\documentclass[twocolumn]{aastex631}

\usepackage{amsmath}
\usepackage{csvsimple}
\usepackage{tabularray}
\usepackage{xcolor}
\usepackage{fourier} 
\usepackage{array}
\usepackage{makecell}
\usepackage{ulem}
\usepackage{graphicx}
\usepackage{rotating}
\usepackage{romannum}

\begin{document}

\title{Radial Gradients and Intrinsic Scatter in MaNGA Galaxies}

\correspondingauthor{Tathagata Pal}
\email{tathagata.pal@nasa.gov}

\author[0000-0002-3077-4037]{Tathagata Pal}
\affiliation{Astrophysics Science Division \\
NASA Goddard Space Flight Center\\
8800 Greenbelt Rd.\\
MD 20771}

\author[0000-0003-1388-5525]{Guy Worthey}
\affiliation{Washington State University \\
Department of Physics and Astronomy, Webster Hall\\
100 Dairy Road Room 1245, Pullman\\
WA 99164}

\begin{abstract}

We derive stellar population parameters and radial gradients within 0.65 R$_e$ for spatially resolved spectra of 2968 early-type galaxies from MaNGA, spanning stellar velocity dispersions ($\sigma$) of 50–340 km s$^{-1}$. Light weighted mean age, C, N, Na, Mg and Fe abundances are obtained by inverting metallicity–composite stellar population models with isochrones that respond in $T{\rm_{eff}}$ to individual element abundance changes.

Globally, log age increases ($\sim$0.53 dex per decade), [Fe/H] declines slightly ($\sim$–0.06 dex per decade), and [X/Fe] for light elements rises (0.19–0.37 dex per decade for log $\sigma<$ 2.0 but steepens to nearly double slope for log $\sigma>$ 2.0) with log $\sigma$. [Fe/H] peaks at log $\sigma$$\approx$2.0 and falls on either side. Light element [X/Fe] anticorrelates with [Fe/H] ($\sim$–0.1 dex per decade). Astrophysical scatter is largest in low-$\sigma$ galaxies, especially for Fe and N.

Internally, age gradients are nearly flat in low-$\sigma$ galaxies and slightly negative in high-$\sigma$ systems ($\sim$–0.04 dex per decade). [Fe/H] radial gradients steepen from –0.06 dex per decade to –0.15 dex per decade across $\sigma$, while light elements (except Na) show $\sim$–0.03 dex per decade gradients. Scatter in gradients peaks in high-$\sigma$ galaxies, most strongly for Fe ($\sim$0.23 dex), suggesting comparable numbers of inside-out and outside-in formation.

A near-zero ($\sim$–0.03) age and light-element gradient plus mild [Fe/H] gradients supports hierarchical merging for ETG evolution. Simulations match the observed age structure, slope change at log $\sigma$$\approx$2.0, and flat gradients, though they overpredict absolute abundances.


\end{abstract}


\section{Introduction} \label{sec:intro}

\subsection{Galaxy Evolution}

The cosmic evolution of the chemical properties of stellar populations, along with that of the interstellar and intergalactic medium, offers valuable insights into the processes that drive galaxy formation and evolution. A wide range of models and cosmological simulations have been created to explore the evolution of the baryonic matter, star formation processes, and the characteristics of galaxies throughout cosmic history, including in the local universe \citep{2005MNRAS.361..776S, 2014MNRAS.444.1518V, 2015ARA&A..53...51S, 2015MNRAS.446..521S, 2019ComAC...6....2N}. The abundances of certain heavy elements relative to hydrogen constrain these models. The metal content of galaxies, often traced via [Fe/H], reflects heavy elements added by supernovae and other stellar sources such as nucleosynthetically-enriched mass loss and neutron star mergers, minus the amount of metals removed through mechanisms such as outflows and stripping \citep{2004ApJ...613..898T, 2008MNRAS.385.2181F, 2011MNRAS.417.2962P, 2015BAAA...57...49D, 2017MNRAS.469.4831C}. Heavy element abundances are also affected by dilution caused by inflows of pristine gas. Iron is a key element in understanding the evolution of galaxies \citep{2005MNRAS.363....2K, 2011MNRAS.416.1354D, 2014ApJ...783...45S} but various other chemical elements are formed over different timescales by different types of stars, and these elements might also offer insights. The hope is that their relative abundances can be translated into the star formation history and chemical evolution within a galaxy \citep{1979ApJ...229.1046T, 1996ASPC...98..529M, 2003MNRAS.339..897T, 2013ARA&A..51..457N, 2017ApJ...837..183W}.

Analyzing chemical abundances on spatially resolved scales, often summarized as radial gradients, may provide further insight into the processes that have shaped the growth and assembly of galaxies \citep{1994ApJ...420...87Z, 2010MNRAS.401..852R, 2013ApJ...766...17L, 2017MNRAS.466.4731G, 2017MNRAS.464.3597L, 2020MNRAS.491.2137E}. Studies on [Fe/H] and elemental abundance gradients also offer information on whether galaxies formed ``inside-out'' or ``outside-in'' \citep{1997ApJ...477..765C, 2004cmpe.conf..171G, 2011ApJ...729...16K,2015A&AT...29....9S}, and elucidate the role of internal processes such as galactic fountains, stellar migration, and radial gas inflows \citep{1986ApJ...303...39D, 2002MNRAS.336..785S, 2008MNRAS.387..577O, 2008ApJ...675L..65R, 2008ApJ...684L..79R, 2009A&A...504...87S, 2015MNRAS.446..299M}. 

The formation history and evolution of galaxies remain a subject of considerable debate. Two prominent scenarios provide convenient bookends for discussion. The \textit{monolithic collapse} hypothesis suggests that galaxies form from the dissipation of large, pre-galactic gas clouds, leading to the formation of all their stars in one fairly brief event \citep{1974MNRAS.166..585L, 1976MNRAS.176...31L, 1999MNRAS.302..537T, 2010MNRAS.407.1347P}. According to this model, galaxies are predicted to form via ``outside-in'' formation, where star formation in the outermost regions ceases earlier due to the earlier onset of galactic winds in areas with shallower gravitational potential wells and have strong metallicity gradients \citep{2006ApJ...638..739P}. Conversely, the \textit{hierarchical merger} hypothesis proposes that large galaxies form through the merging of smaller subunits \citep{1993MNRAS.264..201K, 1998MNRAS.294..705K, 2007MNRAS.375....2D}. In this scenario, star formation is an ongoing process that depends on the gas content of the progenitor galaxies. Galaxies formed through the merging of smaller units tend to flatten existing [Fe/H] gradients \citep{1980MNRAS.191P...1W, 2009A&A...499..427D}. In contrast, galaxies formed through dissipative processes are expected to have steep to moderate gradients in [Fe/H] and other heavy element abundances \citep{1976MNRAS.176...31L, 1985ApJ...298..486C}. There have also been studies where the fundamentals of both of these theories are shown to be partially true \citep{2003A&A...407..423M, 2005ApJ...632L..61O, 2017MNRAS.466.4731G}. An overview of both of these theories can be found in \cite{2015A&AT...29....9S}. 

\subsection{Models for the Integrated Light of Stellar Populations}

The most efficient way to extract information from galaxy spectra is through stellar population analysis, which provides details on age, [Fe/H], and individual element abundances for the galaxy in question. For such analysis, we can employ either the full spectrum fitting technique \citep{2009A&A...501.1269K, 2014ApJ...780...33C, 2018ApJ...854..139C} or absorption index measurements \citep{1998ApJS..116....1T, 2011AJ....141..184S, 2012MNRAS.421.1908J}. There is no fundamental difference between the two approaches. It is merely a question of at which stage empirical adjustments are imposed. Because synthetic stellar spectra match empirical spectra poorly \citep{2007MNRAS.381.1329M}, but because control over individual elements has become essential, some form of forced matching between synthetic and observed stellar spectra is required. The use of indices, adopted for this work, allow an average over many stellar spectral libraries at a variety of instrumental resolutions prior to the empirical zeropointing, at cost of operating within a predefined set of diagnostic indices.

Stellar population analysis works at all redshifts. It is an important tool in analyzing the chemical compositions of the galaxies that hold clues to the galaxy formation and evolution over many epochs of time \citep{2011Ap&SS.331....1W, 2013ARA&A..51..393C}. It is possible that formation time-scales in galaxies can be measured via the ratio of $\alpha$-elements to that of heavy metals (Fe-peak elements) \citep{1999MNRAS.302..537T}, although the high-mass initial mass function (IMF) may potentially also alter it \citep{1992ApJ...398...69W,2014ApJ...783...20W}. Abundances of other lighter metals like carbon (C) and nitrogen (N) put constraints on star formation timescale and also gas inflow \citep{2003MNRAS.339...63C, 2012MNRAS.421.1908J, 2019ApJ...874...93B}. Spectral analysis of galaxies have also been shown to be effective in constraining the slope in the low-mass IMF of galaxies \citep{2015ApJ...798L...4M, 2017ApJ...841...68V, 2018MNRAS.477.3954P}. 

The majority of existing studies has concentrated on detailed stellar population analysis of Early-Type Galaxies (ETGs). This focus is due to the fact that ETGs can be more accurately modeled using simple star formation histories, as they typically exhibit no ongoing star formation \citep{2003MNRAS.344.1000B, 2010MNRAS.404.1639V, 2014ApJ...783...20W} and thus allow analysis of Balmer features that are otherwise susceptible to emission line fill-in. ETGs are predominantly composed of older stars, with minimal or no current star formation activity. In favorable cases, at least, they have had only modest changes over much of cosmic time, and present themselves as good laboratories for studying the long-term evolution of stellar populations. 

\subsection{A Focus on Stellar Population Gradients}

The first science focus for this paper is on the light-weighted age, [Fe/H], and C, N, sodium (Na), and magnesium (Mg) abundances for the galaxies and their dependence on stellar velocity dispersion ($\sigma$) and [Fe/H]. We also explore radial gradients in the same list of parameters and their $\sigma$ dependence. Finally, we compute the intrinsic scatter ($\delta_i$) in all the quantities and discuss the trends.

The study of gradients in various parameters has long been used as a mean to understand the formation and evolution of galaxies. Two of the first studies available in literature on color gradients were done by \cite{1977egsp.conf.....T} and \cite{1990ApJ...359L..41F}. In the early 1990s, a number of pioneering studies were published on line strength gradients, by \cite{1989egss.book.....P, 1990MNRAS.245..217G}, \cite{1992A&A...258..250B}, \cite{1993MNRAS.262..650D}, and others. The first mention of an [Fe/H] gradient (as opposed to a line strength gradient unattached to the solar-zeroed [Fe/H] standard) that we found in literature is by \cite{1993MNRAS.265..553C} where they measured line strengths of different features from 42 ETGs to infer an [Fe/H] gradient. These studies did not necessarily use any stellar population model to infer age, [Fe/H], or element abundances but they pioneered the field of gradient studies. Later on with the development reliable stellar population models, a large number of studies had been completed on spatial gradients in age, [Fe/H], and element abundances within both ETGs and Late Type Galaxies (LTGs) \citep{2003A&A...407..423M, 2005MNRAS.361L...6F, 2005MNRAS.362..857P, 2007A&A...474..763J, 2009ApJ...691L.138S, 2010MNRAS.401..852R, 2010MNRAS.408...97K, 2017A&A...607A.128G, 2017MNRAS.466.4731G, 2018MNRAS.477.3954P, 2018MNRAS.476.1765L, 2019MNRAS.489..608F, 2020MNRAS.495.2894D, 2021ApJ...923...65F, 2021MNRAS.502.5508P, 2023MNRAS.526.1022L}. These studies generally agree that ETGs and the bulges of spiral galaxies display negative [Fe/H] gradients, while showing small positive or flat gradients in both age and the $\alpha$/Fe ratio. This ``globalness'' of age and $\alpha$ throughout the body of the galaxy is an example of a useful constraint to be matched by galaxy formation models.

The emergence of Integral Field Unit (IFU) surveys for galaxies opens up the scope of using high-quality data to investigate gradients in stellar population parameters further. Such surveys include the SAURON and ATLAS-3D project \citep{2001MNRAS.326...23B, 2004MNRAS.352..721E, 2011MNRAS.413..813C}, CALIFA project \citep{2012A&A...538A...8S}, SAMI project \citep{2012MNRAS.421..872C}, and MaNGA project \citep{2015ApJ...798....7B}. IFU spectroscopy allows us to get galaxy spectra from different parts of the galaxy enabling detailed radial studies. Although these surveys provide us with a large volume of data, most of the studies to date are either limited by sample size (the largest number of individual galaxies studied to date is by \cite{2017MNRAS.466.4731G} where they studied around 700 ETGs and \cite{2019MNRAS.489..608F} at about 500 ETGs) or the studies stack spectra \citep{2018MNRAS.477.3954P}. Stacking suppresses information on individual galaxies. Other studies treat each galaxy individually but are restricted in sample size \citep{2015ApJ...807...11G, 2018MNRAS.478.4464A, 2021ApJ...923...65F}. In this work, we consider 2968 individual unique galaxies from the Mapping Nearby Galaxies at Apache Point Observatory (MaNGA, \cite{2015ApJ...798....7B}) for a detailed study of the age, [Fe/H], and element abundances along with their corresponding gradients.

\subsection{A Focus on Exploiting Improved Models}

It is important to state at the outset a couple of important recent improvements in the models that separate these results from previous work. First, the isochrones upon which the models rest now vary with individual element abundances \citep{2022MNRAS.511.3198W}. This enables an effect we term ``magnesium amplification'' that applies to Mg and other elements as well. Addition of Mg, alone, to the chemical mixture in a stellar population, will add opacity to the stars, increase their radii, and cool them. Cooler stars have stronger Mg lines. So, not only will Mg line strengths increase due to the increased number of absorbers, Mg lines will also increase in strength due to arising in cooler stars. Since inclusion of this effect is new to the field starting with this work, our results will not necessarily agree with previous estimates of quantities such as [Fe/H], [C/Fe], or [Mg/Fe]. Second, we incorporate an abundance distribution function (ADF) instead of a delta function in abundance \citep{2014MNRAS.445.1538T}. It is important to do this because metal-poor populations are brighter in the blue, and thus dilute metallic lines found in that spectral region. In the red, the balancing favors metal-rich populations more. We parameterize by the peak (mode) of the ADF, not by, for example, the $V$-weighted average abundance, or, for example, the mass-weighted average abundance. This introduces modest, well-behaved offsets between results derived from a simple stellar population (SSP) and this work, as illustrated in \cite{2014MNRAS.445.1538T}.

\subsection{A Focus on Astrophysical Scatter}

The influence of uncertainties in different phases of stellar evolution and IMF on derived stellar population parameters like the stellar masses, mean ages, metallicities, star formation histories, and other has been well documented by \cite{2009ApJ...699..486C} (see also \cite{2010ApJ...708...58C} and \cite{2010ApJ...712..833C}). A lot of fundamental work was also done in 1990s on the effect of intrinsic scatter in ETGs \citep{1992ApJ...384...43G, 1993MNRAS.265..731G}. \cite{2011MNRAS.417.1787F} quantified intrinsic scatter in scaling relations for nearby elliptical galaxies.

However, intrinsic, astrophysical scatter in age, [Fe/H], and elemental abundances goes unstudied in recent literature. To improve this situation, we treated each MaNGA galaxy in our study individually (not using any kind of stacking algorithm).

This article is arranged as follows: $\S$\ref{sec:data} explains the source and type of data used in this work, $\S$\ref{sec:analysis} contains the details of the analysis performed along with brief explanation of the stellar population model used for this work, $\S$\ref{sec:results} elucidates all of our results, $\S$\ref{sec:discussion} discusses the implications of our results and comparison with previous works, and also discusses our results on the intrinsic scatter. $\S$\ref{sec:conclusion} concludes the article with reiterating our main results and forecasting future work, and an appendix gives a library of graphical comparisons with the TNG-100 cosmological evolution simulation, part of tables elucidating all the indices used in this work and result of the inversion program \textsc{compfit}.

\section{Data} \label{sec:data}

In this section, we describe the data sources and key features relevant to our analysis. Specifically, we employ stellar population modeling using the MaNGA Data Release 17 (DR17) IFU to derive stellar age, [Fe/H], elemental abundance patterns, and their corresponding spatial gradients.

\subsection{MaNGA Data}\label{subsec:manga}

The Mapping Nearby Galaxies at Apache Point Observatory (MaNGA) survey, part of the Fourth Generation Sloan Digital Sky Survey (SDSS-IV, \cite{2017AJ....154...28B}), is designed to map the internal kinematics and chemical composition of nearby galaxies. Utilizing IFU spectroscopy, MaNGA obtains spatially resolved spectra across a large number of positions within each galaxy \citep{2015ApJ...798....7B}. The IFUs are connected to the BOSS spectrographs \citep{2013AJ....146...32S}, which are mounted on the Sloan 2.5-meter telescope at Apache Point Observatory \citep{2006AJ....131.2332G}.

For this work, we use the final and most comprehensive data release, DR17 \citep{2022ApJS..259...35A}, which contains IFU spectroscopy for 10,010 unique galaxies. These spectra cover a wavelength range of 3000 \AA to 10000 \AA with a spectral resolution of R $\sim$ 2000.

The MaNGA Data Analysis Pipeline (DAP) provides spatially stacked spectra in the form of datacubes \citep{2019AJ....158..231W, 2021AJ....161...52L}, along with derived properties such as stellar kinematics (velocity and dispersion), nebular emission line fluxes, and spectral indices \citep{2019AJ....158..160B}. For this analysis, we subtract the DAP-provided nebular emission lines from each spectrum prior to further modeling.

\begin{figure}[ht!]
\plotone{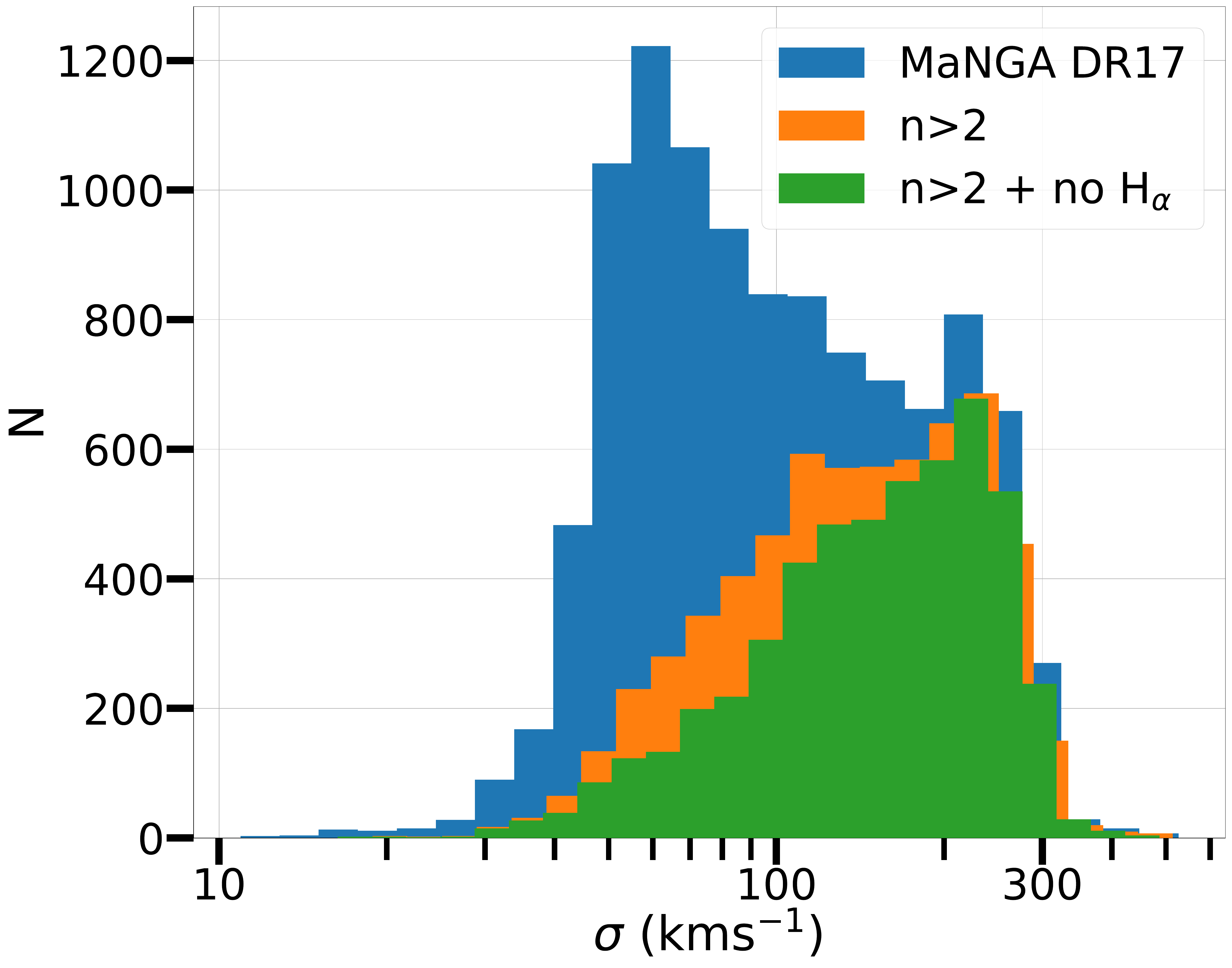}
\caption{The distribution of all galaxies from the MaNGA DR17 (blue), galaxies with $n>$2 (orange), and galaxies with $n>$2 \& no H$_\alpha$ emission (green) as a function of velocity dispersion ($\sigma$). 
\label{fig:gal_dist}}
\end{figure}

As our focus is on ETGs, we restrict the sample to systems with elliptical morphology. It is well established that elliptical galaxies exhibit high S\'ersic index values ($n$, \cite{1963BAAA....6...41S}). Thus, from the full set of 10,010 unique galaxies in MaNGA DR17, we select those with $n > $2, yielding a subsample of 6,273 galaxies. Fig.~2 of \cite{2022MNRAS.509.4024D} and Fig.~12 of \cite{2019MNRAS.483.2057F} show the distribution of MaNGA galaxies in terms of S\'ersic index for DR17 and DR15 respectively. The number distribution of all these 6273 galaxies is shown in Fig.~\ref{fig:gal_dist} (in orange) along with all MaNGA DR17 galaxies (in blue) and our sample with $n > $2 \& without any H$_\alpha$ emission (in green).

\subsection{Further Filtering}\label{subsec:filter}

Following the selection of 6,273 galaxies from MaNGA DR17, we applied an additional cut to exclude systems with detectable H$_\alpha$ emission, as our analysis focuses exclusively on ETGs with minimal or no ongoing star formation. Specifically, we retained only those galaxies for which the integrated H$_\alpha$ emission line flux is less than 15\% of the observed flux within the same spectral interval. Both quantities were computed within the H$_\alpha$ index bandpass, defined as 6548 \AA $< \lambda <$ 6578 \AA \citep{1998ApJ...496..808C}, using the DAP datacube products. This criterion reduced the sample to 5,164 galaxies.

All wavelengths referenced here are in the rest-frame. To convert from observed to rest-frame wavelengths, we use the following equation:

\begin{align}\label{eqn:restframe}
\lambda_{\mathrm{rest}} = \frac{\lambda_{\mathrm{observed}}}{(1+z)(1 + \frac{v}{c})}
\end{align}

In Eqn.~\ref{eqn:restframe}, $z$ denotes the galaxy redshift, $v$ is the internal stellar velocity at the sampled location within the galaxy, and $c$ is the speed of light. We associated a global $\sigma$ to measure the stellar velocity dispersion at 0.5 R$_e$. To extrapolate $\sigma$ to other radial bins, we utilize median radial gradient trends derived from \cite{JJ_thesis}. We validated this choice by reducing the gradient amplitude by 50\% and re-computing the stellar population parameters; the results showed no significant deviation. However, when applying individual $\sigma$ gradient profiles from \cite{1981seng.proc...27I}, we observed substantial differences in the derived quantities. Given that \cite{JJ_thesis} provides $\sigma$ gradients for a statistically larger ETG sample than the limited dataset in \cite{1981seng.proc...27I}, we adopt the former for our analysis. Although MaNGA DAP products offer annular measurements of $\sigma$ directly, we opt instead for the external trend to mitigate potential degeneracies between line strength and $\sigma$, which may affect the robustness of abundance measurements.

\begin{figure}[ht!]
\plotone{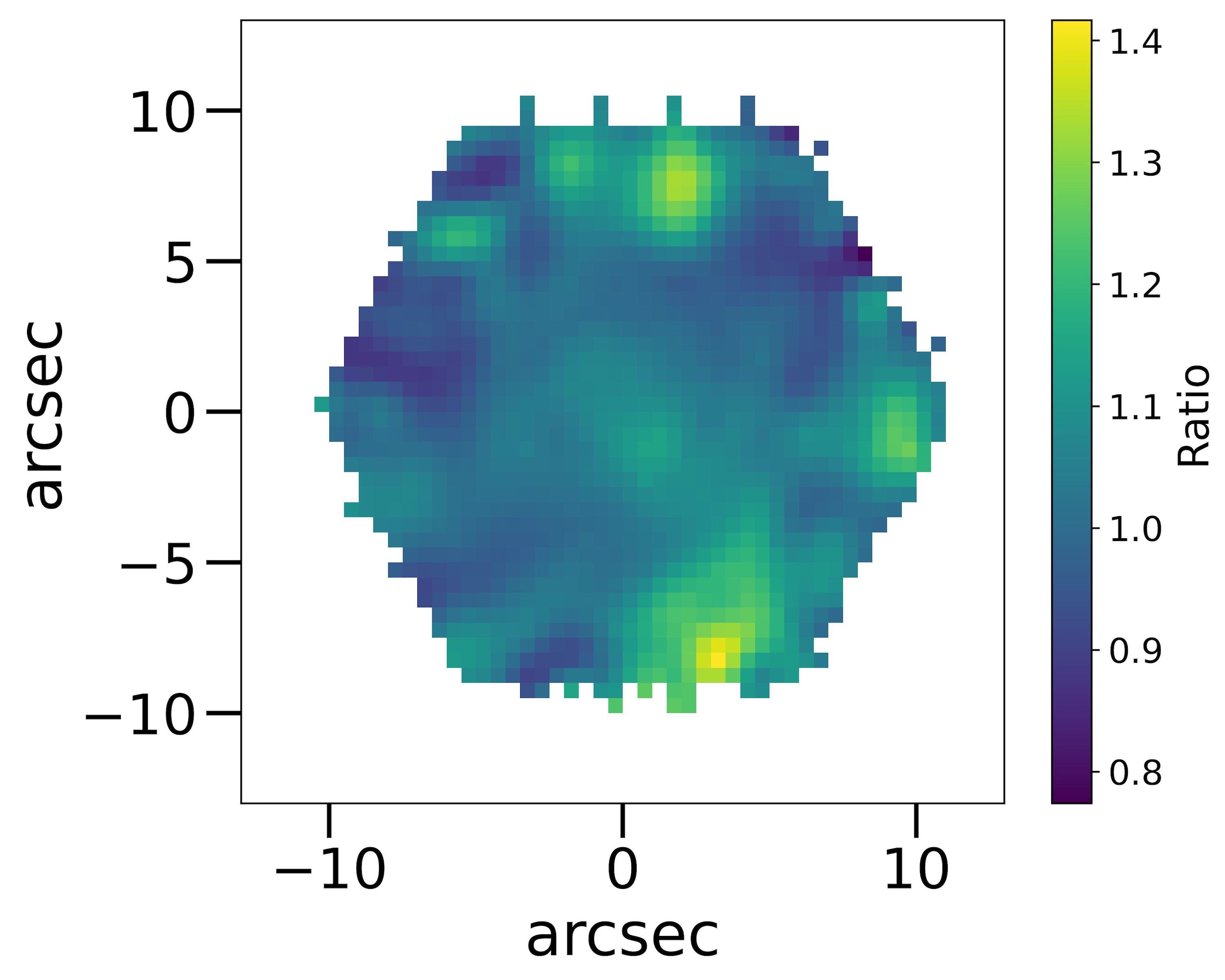}
\caption{Position of the MaNGA IFU with respect to the center of the MaNGA galaxy 1-245451. The colorbar shows the ratio of mean flux within H$_\alpha$ index band and mean flux in the continuum. 
\label{fig:h_alp_ratio}}
\end{figure}

Fig.~\ref{fig:h_alp_ratio} shows the ratio of mean fluxes within the H$_\alpha$ index band and the continuum for the MaNGA galaxy 1-245451. The red and blue continuum are defined to be 6600\AA$<\lambda<$6640\AA\ and 6420\AA$<\lambda<$6455\AA\ respectively \citep{1998ApJ...496..808C}. The average of flux within these two continuum bands is taken to be the mean flux within the continuum band. Since the ratio is close to one at the location of the galaxy (see Fig.~\ref{fig:annuli}), it shows that the algorithm that we opted to reject galaxies with H$_\alpha$ emission, is working optimally. All the wavelengths are converted to restframe wavelength using Eqn.\ref{eqn:restframe}.

\subsection{Creating Annuli}\label{subsec:annulus}

The MaNGA IFU provides us with spectroscopic data from different regions of the galaxy and the MaNGA DAP then processes this spatial and spectroscopic information into a 3D datacube. Two axes of this datacube hold positional information while the third axis gives spectroscopic information. 

We divide each galaxy into 10 annular regions starting from center to 1.0R$_e$ with an increment of 0.1R$_e$, where R$_e$ is the effective radius of the galaxy, synonymous to the half-light radius of the galaxy for S\'ersic functions. The following equation is used to define the ellipse in the positional space of the galaxy.

\begin{multline}\label{eqn:ellipse}
    \frac{[(x-h)cos\theta+(y-k)sin\theta]^2}{a^2}+ \\
    \frac{[(x-h)sin\theta-(y-k)cos\theta]^2}{b^2} = 1
\end{multline}

In Eqn.~\ref{eqn:ellipse}, $x$ \& $y$ are the RA and Dec respectively of the spaxel (spatial pixel) in consideration, $h$ \& $k$ are the RA and Dec respectively of the center of the galaxy, $\theta$ is the angle of inclination (East of North), $2a$ \& $2b$ are the semi-major and semi-minor axes respectively of the galaxy. In this equation, $a$ is used synonymously with R$_e$. The $b$ is obtained from the $b/a$ ratio for the galaxy in question. All of $\theta$, $a$, and $b/a$ come from the S\'ersic profile as provided by the MaNGA DAP. It is to be noted that this system of dividing the galaxy in several annular regions is motivated from the the work of \cite{2018MNRAS.477.3954P}. 

\begin{figure}[ht!]
\plotone{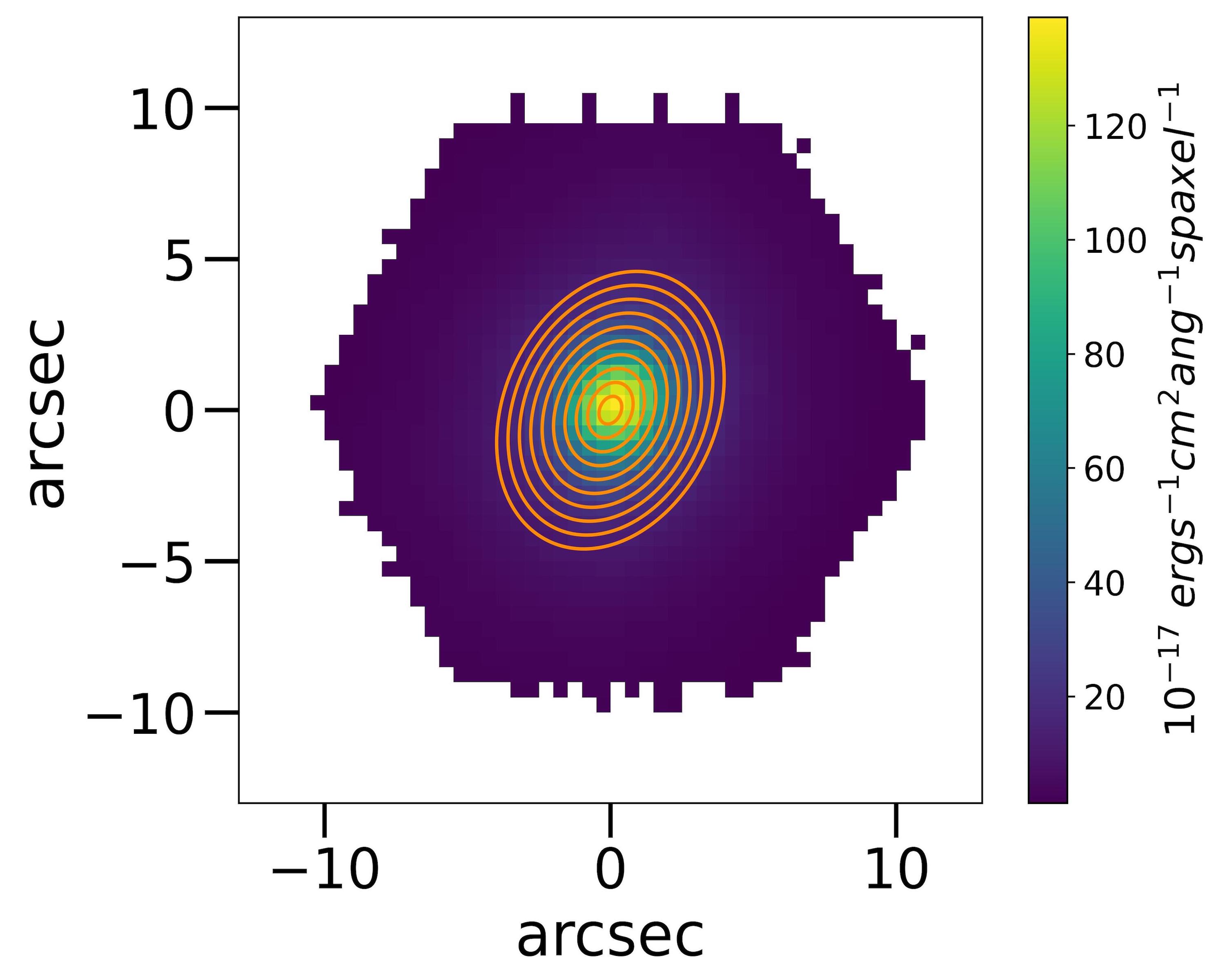}
\caption{The map of H$_\alpha$ flux within the H$_\alpha$ index band (in absolute scale) for the MaNGA galaxy 1-245451. The orange ellipses are calculated using Eqn.\ref{eqn:ellipse} with parameters specific to this galaxy.
\label{fig:annuli}}
\end{figure}

\begin{figure}[ht!]
\plotone{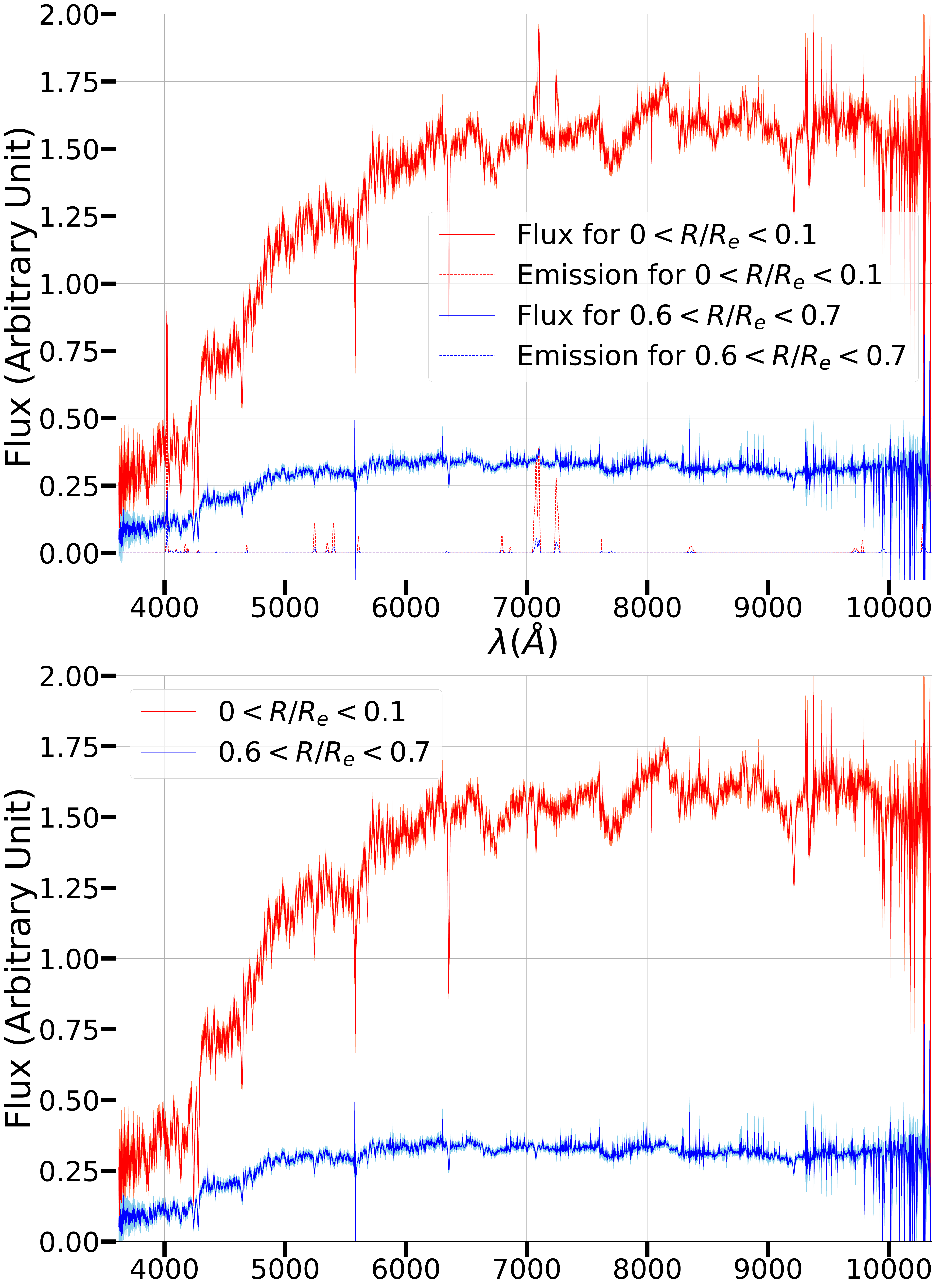}
\caption{Top panel: Mean flux (red/blue solid line) and emission line spectra (red/blue dashed line) from the innermost  and 7$^{th}$ annuli for the MaNGA galaxy 1-245451. Bottom panel: Emission line subtracted spectra for the same galaxy (red: innermost annulus and blue: 7$^{th}$ annulus). The error shading shows the 1-$\sigma$ uncertainty in the measured fluxes.
\label{fig:spec}}
\end{figure}

Fig.~\ref{fig:annuli} illustrates the annulus scheme used in this work. The representative spectrum for each annulus is the average spectrum of all spaxels within that annulus. The mean spectrum per annulus is computed by averaging the spectra of all constituent spaxels. The MaNGA DAP provides inverse variance estimates for each spaxel; we calculated the error for each annulus as the mean of these inverse variances across all spaxels within the annulus. MaNGA DAP also supplies masks for bad spaxels, which we apply when computing annular averages. The final representative spectrum for each annulus is obtained by subtracting the average emission-line spectrum from the average flux spectrum. Fig.~\ref{fig:spec} presents two example spectra from distinct galaxy regions. The top panel displays mean fluxes (solid lines) from the innermost (red) and 7$^{th}$ (blue) annuli alongside their corresponding emission line spectra (dashed lines). The bottom panel shows the emission line-subtracted spectra for these annuli. Shaded regions represent 1-$\sigma$ uncertainties, computed as the square root of the inverse of the DAP-provided errors (inverse variance) at each wavelength. The selected galaxies do not exhibit prominent emission line features. Notably, we do not employ any stacking scheme, as our goal is to probe scatter in measured quantities by treating each galaxy individually. This approach entails limitations. First, spectra beyond 9500\AA\ are extremely noisy, preventing reliable index strength measurements in that region. Second, beyond the 7$^{th}$ or 8$^{th}$ annulus, flux levels decline significantly, further complicating index measurements. To mitigate these issues, we limit subsequent analyses to the 7$^{th}$ annulus for each galaxy (counting the galaxy center as the 0$^{th}$ annulus, this corresponds to up to the 6$^{th}$ annulus). Furthermore, in element abundance analysis, we restrict ourselves to elements with indices below 9500\AA, and we omit IMF analysis based on the Wing-Ford band at $\lambda\lambda$9900.

\section{Analysis} \label{sec:analysis}

In this section, we outline the analysis pipeline used in this work. All analyses were performed on individual spectra from each annulus of a given galaxy. A key objective is to investigate the intrinsic scatter in derived properties such as age, [Fe/H], and chemical abundances (see $\S$\ref{subsec:diss_scatter}). To that end, galaxies are treated individually without stacking. The increased noise in the spectra is addressed by the strategies described in Section~\ref{subsec:annulus}.

\subsection{Spectral Analysis}\label{subsec:spec_analysis}

Stellar population parameters were determined by fitting stellar population models to spectra. This can be achieved either by fitting the entire spectrum \citep{2009AN....330..960K, 2017MNRAS.466..798C, 2018ApJ...854..139C} or by focusing on specific absorption line features \citep{2000AJ....119.1645T, 2007ApJS..171..146S, 2010MNRAS.404.1775T}. In this work, we adopt the latter method and focus on using the strongest absorption features.

\begin{table*}[]
\centering
\caption{The most influential index definitions used in this work. The acronyms are as follows: IB=Index Band, BC=Blue Continuum, RC=Red Continuum. ``Start" and ``End" denote the start or end wavelengths of the passbands in \AA. The ``Unit" column specifies if the accepted unit of the index is \AA\ or magnitudes.}
\label{table:index}
\begin{tabular}{ccccccccc}
\hline\hline
Parameter & Feature & IB Start & IB End & BC Start & BC End & RC Start & RC End & Unit \\ \hline
Age + [Fe/H]     & Fe4383   & 4369.125 & 4420.375 & 4359.125 & 4370.375 & 4442.875 & 4455.375 & \AA \\ 
         & Fe4531 &  4514.25 & 4559.25 & 4504.25 & 4514.25 & 4560.50 & 4579.25 & \AA \\ 
         & C$_2$4668 & 4634 & 4720.25 & 4611.5 & 4630.25 & 4742.75 & 4756.5 & \AA \\ 
         & H$_\beta$ & 4847.875 & 4876.625 & 4827.875 & 4847.875 & 4876.625 & 4891.625 & \AA \\ 
         & Fe5270 & 5245.65 & 5285.65 & 5233.15 & 5248.15 & 5285.65 & 5318.15 & \AA \\ 
         & Fe5335 & 5312.125 & 5352.125 & 5304.625 & 5315.875 & 5353.375 & 5363.375 & \AA \\ 
         & Fe5406 & 5387.50 & 5415.00 & 5376.25 & 5387.50 & 5415.00 & 5425.00 & \AA \\ 
         & H$\gamma_A$ & 4319.75 & 4363.50 & 4283.50 & 4319.75 & 4367.25 & 4419.75 & \AA \\
        & H$\delta_F$ & 4091.00 & 4112.25 & 4057.25 & 4088.50 & 4114.75 & 4137.25 & \AA \\ 
        & Ni3667 & 3655.10 & 3679.40 & 3630.40 & 3646.60 & 3685.40 & 3705.30 & \AA \\ \hline
{[}Na/R{]} & NaD & 5876.875 & 5909.375 & 5860.625 & 5875.625 & 5922.125 & 5948.125 & \AA \\ \hline
{[}Mg/R{]} & Mg$_1$ & 5069.125 & 5134.125 & 4895.125 & 4957.625 & 5301.125 & 5366.125 & mag \\ 
         & Mg$_2$ & 5154.125 & 5196.625 & 4895.125 & 4957.625 & 5301.125 & 5366.125 & mag \\   
         & Mg $b$ & 5160.125 & 5192.625 & 5142.625 & 5161.375 & 5191.375 & 5206.375 & \AA \\    \hline   
{[}N/R{]} & CN$_1$ & 4142.125 & 4177.125 & 4080.125 & 4117.625 & 4244.125 & 4284.125 & mag \\ 
         & CN$_2$ & 4142.125 & 4177.125 & 4083.875 & 4096.375 & 4244.125 & 4284.125 & mag \\ 
         & CNO3862 & 3840.3 &  3883.4 & 3768.1 & 3812.3 & 3896.4 & 3916.2 & \AA \\ 
         & CNO4175 & 4129.4 & 4219.8 & 4082.5 & 4123.3 & 4243.3 & 4284.2 & \AA \\  \hline
{[}C/R{]} & C$_2$4668 & 4634 & 4720.25 & 4611.5 & 4630.25 & 4742.75 & 4756.5 & \AA \\ 
         & Mg$_1$ & 5069.125 & 5134.125 & 4895.125 & 4957.625 & 5301.125 & 5366.125 & mag \\ 
         & CO4685 & 4626.40 & 4743.30 & 4557.30 & 4589.50 & 4805.10 & 4835.30 & \AA \\ 
         & CO5161 & 5154.3 & 5167.3 & 5108.4 & 5138.8 & 5188.4 & 5202.0 & \AA \\ 
         \hline
\end{tabular}
\end{table*}

In $\S$\ref{subsec:manga}, we describe the specific data products used from the MaNGA DAP \citep{2019AJ....158..231W}. The DAP currently provides three binning schemes: SPX, which fits each individual spaxel with $g$-band S/N $\geq$ 1; VOR10, which applies Voronoi binning to achieve $g$-band S/N $\geq$ 10 \citep{2003MNRAS.342..345C}; and HYB10, a hybrid of the two. We adopt the SPX scheme and do not apply any additional binning. Stellar kinematics are derived by fitting the data with templates from the MILESHC stellar library, which comprises 42 representative spectra constructed through hierarchical clustering of the MILES library \citep{2006MNRAS.371..703S}. We use the $\sigma$ from this continuum fit as a proxy for velocity dispersion at 0.5 R$_e$. Emission lines are modeled using templates from the MASTAR library \citep{2022ApJS..259...35A, 2019ApJ...883..175Y}. These emission lines are subtracted from the observed IFU spectra prior to further analysis.

In this work, we specifically analyzed the light weighted age, [Fe/H] in lockstep with other heavy species, and individual chemical abundances of C, N, Na, and Mg. The reason for choosing these specific elements is mostly the ease with which the abundances of these elements can be measured. The indices corresponding to these elements are prominent and are not wrought with degeneracies. Also, we needed to keep in mind that we do not pick elements that have indices beyond 9500\AA. Several absorption spectral features are used to determine the variables in question using the model described in Section \ref{subsec:model}. The strengths of the absorption features used are calculated using Equations 2 and 3 in \cite{1994ApJS...94..687W}. Selected index definitions are given in Table \ref{table:index}. The definitions are the standard ones from \citet{1994ApJS...94..687W,1997ApJS..111..377W,2005ApJ...627..754S}. For a given annulus in a given galaxy, up to 70 out of a list of 80 indices were used to calculate age and abundance parameters. A full list of the 80 indices is given in Appendix \ref{appen:feature_table}. Table \ref{table:index} shows the indices that most strongly affect the underlying variable for a typical old galaxy. To assess this, the throw of the index is calculated and then put into units of the observational uncertainty [$(\Delta I/0.3 {\rm \ dex\ change\ in\ species\ }Z) / \sigma$]. The throw of any given index will of course change with observational uncertainty and with location in the age-metallicity plane, but in general the ETGs we study here could have been analyzed with \textit{only} the Table \ref{table:index} indices with little loss of fidelity compared to the full set. Throw values are generally between 2 and 6. Two indices appear in the list twice. Mg abundance affects Mg$_1$ strongly, but C abundance also has an effect. C$_2$4668 is useful for measuring C abundance and also happens to be extraordinarily metallicity sensitive. The ``R" in Table \ref{table:index} stands for any generic heavy element. For the main results of this work, we have varied Fe in lockstep with other Fe-peak and heavier elements. In this paper, therefore, ``R'' can be used interchangeably with ``Fe.'' For example, [Fe/H] = [R/H] and [Mg/Fe] = [Mg/R].

\begin{figure}[ht!]
\plotone{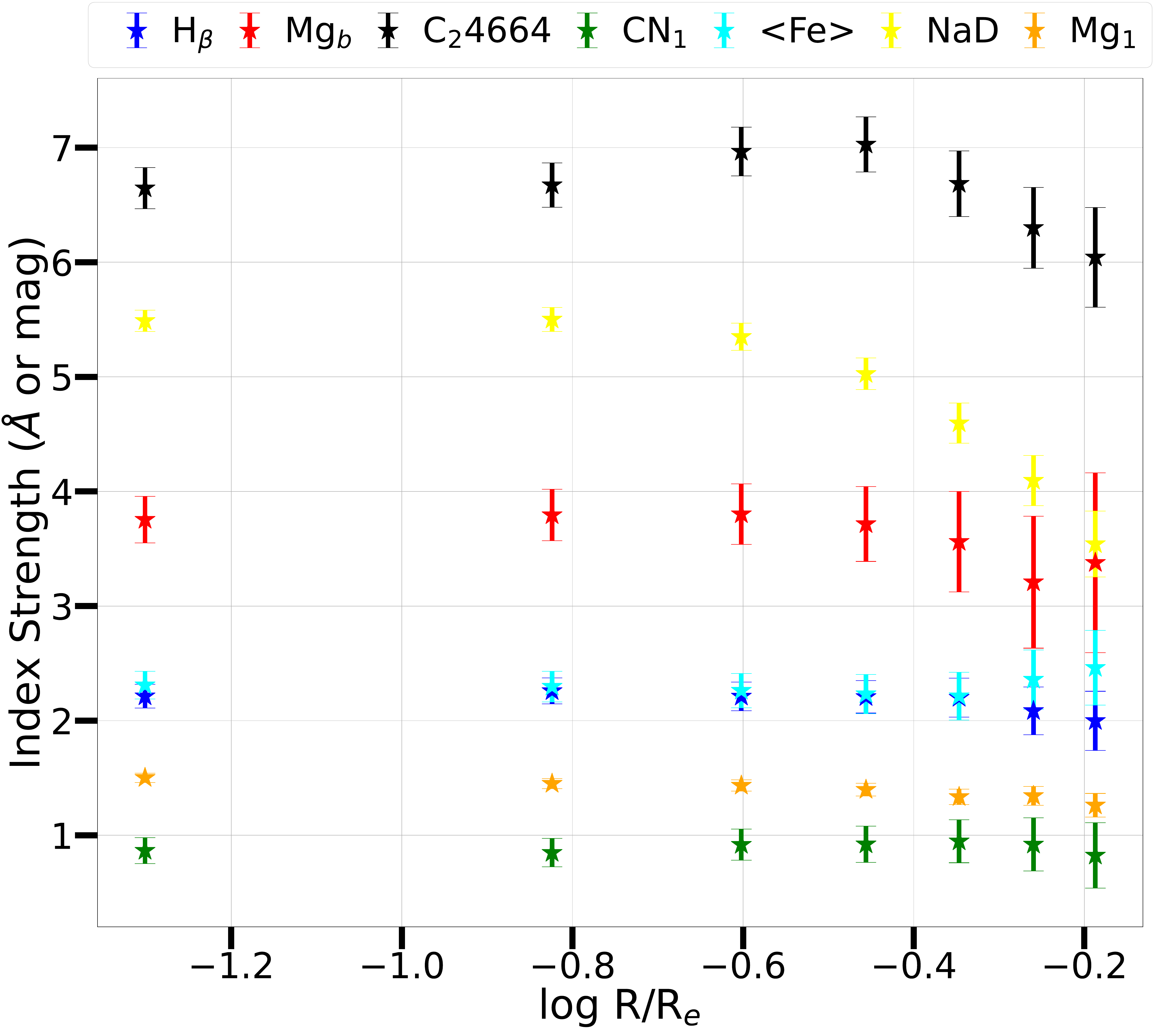}
\caption{The strengths and uncertainties of selected features are displayed for MaNGA galaxy 1-25451. Values are shown for the 1$^{\rm st}$ through 7$^{\rm th}$ annulus, inclusive. For graphical scaling, CN$_1$ is multiplied by 20, and Mg$_1$ is multiplied by 15.
\label{fig:indices}}
\end{figure}

Fig.~\ref{fig:indices} shows selected index strengths (in \AA\ for atomic features and in magnitude for molecular features) of major indices used in this work for the MaNGA galaxy 1-25451. For purposes of display, the original measured strengths of CN$_1$ and its error is multiplied by 20 and the same is done for Mg$_1$ but multiplied by 15. A substantial number of galaxies were excluded from further analysis after several annuli produced unreliable index measurements, reducing the final sample from 5,164 to 2,968 galaxies.

The uncertainty associated with each measurement comes from the variance supplied with the MaNGA DAP data at each pixel. Under the assumption that the pixel errors are truly random, we calculate the error in the index strength using eqn. 33 of \citet{1998A&AS..127..597C}.  

\subsection{Composite Model Grid and Parameter Estimation}\label{subsec:model}

Naturally, the results of galactic spectral analysis depend on the stellar population model used. The stellar population model utilized for this work is based on \cite{2014ApJ...783...20W} (hereafter \textit{Worthey} model) with the foundational framework established by \cite{1994ApJS...95..107W}. These models use stellar evolutionary isochrones combined with a stellar initial mass function (IMF) to estimate the distribution of stars within the log L versus log T$_{\rm eff}$ diagram. Fluxes are assigned to each star bin, along with empirical estimates of the absorption feature indices.

For any stellar population model three major ingredients are: stellar spectral library, set of isochrones, and initial mass function (IMF). Because synthetic spectra match real ones poorly, spectral libraries usually start with empirical observations with the spectral changes wrought by elemental abundance changes computed synthetically and added later. We adopt this approach. Our empirical library incorporates several stellar spectral libraries: 1. a cross-correlated variant of the Lick spectra \citep{2014A&A...561A..36W}, 2. the MILES spectral library \citep{2006MNRAS.371..703S}, 3. the Indo-US library \citep{2004ApJS..152..251V}, 4. the IRTF spectral library \citep{2003PASP..115..362R}, and 5. the ELODIE spectral library \citep{2001A&A...369.1048P}. These libraries are smoothed or back-corrected to a common 200 km s$^{-1}$ resolution before indices are measured. Polynomial fitting functions model index behavior as a function of T$_{\rm eff}$, log($g$), and [Fe/H]. Element sensitivity was added using three spectral synthesis codes as in \cite{2009ApJ...694..902L}.

Element sensitivity in $T_{\rm eff}$ was added to the \cite{2008A&A...482..883M} isochrones according to the formula of \cite{2022MNRAS.511.3198W}. We adopted a Kroupa IMF \citep{2001MNRAS.322..231K}. To generate an integrated-light index strength, index strengths are calculated for each ``star'' along the isochrone, and decomposed into an index flux and a continuum flux (weighted by the star's luminosity at the wavelength of interest and the number of stars in that bin). A running total of the flux and continuum values are then summed over the isochrone. After the summation, the resulting pair of fluxes (continuum and index passbands) is reformulated into a final index value for the single stellar population (SSP). The models used here are not SSPs. They are composite models that incorporate an abundance distribution function (ADF) \citep{2014MNRAS.445.1538T, 2005ApJ...631..820W} that matches the Milky Way and other local galaxies. The models are thus metallicity-composite, but still with a single burst age. In this work we adopt the `normal' width ADF with FWHM = 0.65 dex from \cite{2014MNRAS.445.1538T}. When we quote [Fe/H] results, they refer to the peak metallicity in the asymmetric ADF, not the mass-weighted mean, light-weighted mean, or SSP-equivalent value.

With a grid of composite models in hand, we employed an inversion program ({\sc compfit}, (to {\sc fit} {\sc comp}osite stellar population models) to the observations. Somewhat less sophisticated versions of {\sc compfit} were employed in \citet{2014ApJ...783...20W,2022MNRAS.511.3198W}, and \citet{2023MNRAS.518.4106W}.
{\sc compfit} smooths the model grid, then uses hand-picked age- and metallicity-sensitive indices to derive an average (single) age and peak metallicity. Because of parameter space limitations in the underlying isochrone grid, the peak metallicity is constrained to lie in the range $-0.6 < $ [R/H] $ < 0.4$, where R stands for a generic heavy element. In this paper, Fe is left in lockstep with the other heavy elements, so that [Fe/R] = [Fe/H] - [R/H] happens to be always zero, and [Fe/H] = [R/H] are conveniently interchangeable in these pages. For purposes of deciding on age and ADF peak abundance, we employed particularly age-sensitive features (H$\delta_F$ and H$\beta$) or strongly heavy-element sensitive definitions (Ni3667, Fe4383, Fe4531, C$_2$4668, Fe5270, Fe5335, and Fe5406).

In a predetermined order, COMPFIT examines various features that are diagnostic of the abundance pattern, weighed by their own sensitivity. With a trial abundance mixture in hand, the program then iterates, improving the age, metallicity, and, separately, chemical element mixture until the model optimally matches each set of galaxy indices. The "predetermined order" maximizes efficiency compared to relatively blind techniques such as Markov Chain Monte Carlo or Particle Swarm Optimization by starting with unambiguous elements like Na, which affects primarily the Na D index, and proceeds to elements that affect many indices. Grid discretization involving age and metallicity exists, since this portion of the algorithm tests all parts of the parameter space, but the space is parceled into bins. The binning is fine enough so that it is not readily apparent in figures involving age and metallicity. Abundance ratios are linearized, so they do not exhibit discretization.

\begin{figure*}[ht!]
\plotone{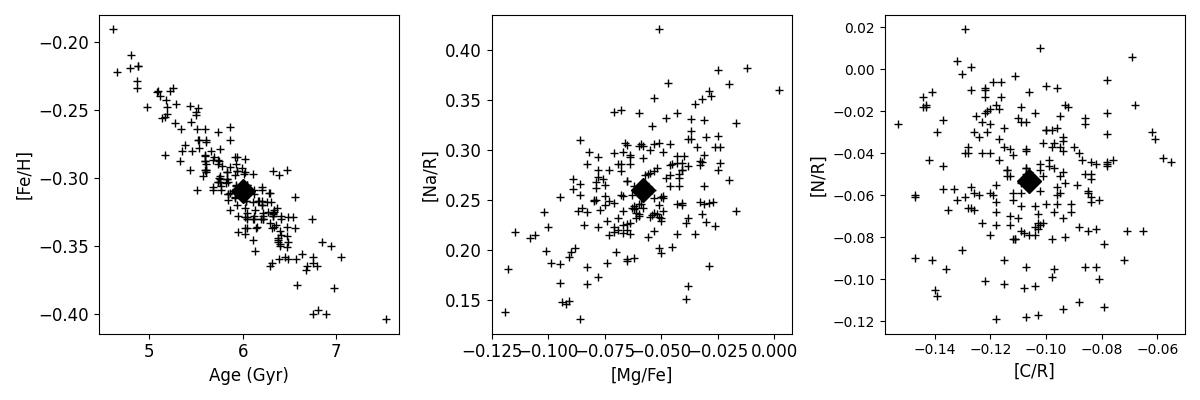}
\caption{Example Monte Carlo parameter inversions for $\sigma\sim70$ galaxy 11009-12702, innermost annulus. The bold diamond represents the nominal solution, while other points represent parameter solutions from altered versions of 11009-12702.
\label{fig:smallgmc}}
\end{figure*}

\begin{figure*}[ht!]
\plotone{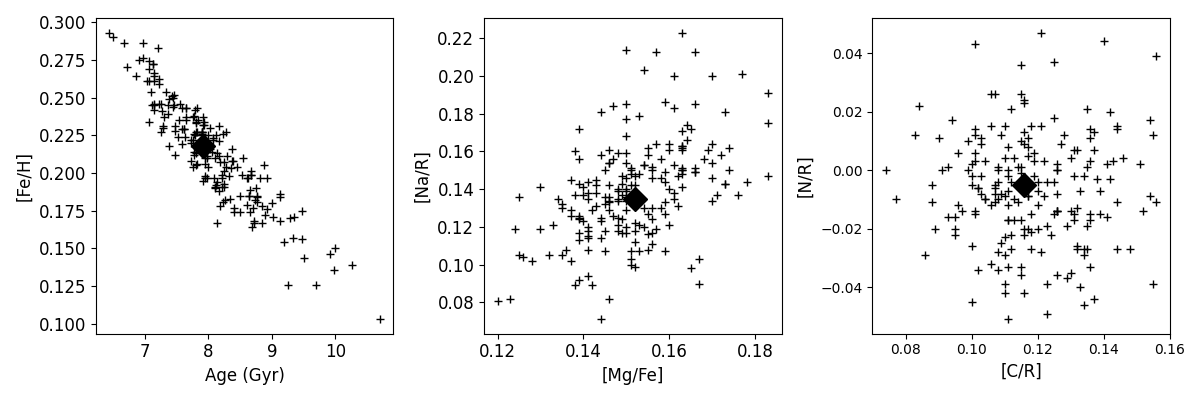}
\caption{Example Monte Carlo parameter inversions for $\sigma\sim200$ galaxy 11965-9101, innermost annulus. Symbols as in Fig. \ref{fig:smallgmc}.
\label{fig:largegmc}}
\end{figure*}

Once a solution for optimal age and chemical mixture is obtained, the analysis is repeated for copies of the original galaxy index set whose index values have been altered randomly within the observational error envelope. Typically, 50 such Monte Carlo realizations are computed to give output uncertainties. Degeneracies are apparent, especially in the age-metallicity plane, as seen in Figs. \ref{fig:smallgmc} and \ref{fig:largegmc}. As seen in these figures, the ratios of individual elements generally do not correlate with each other.

Here are a few notes on how to interpret the inverted-model results. The final abundance for each element is listed here as the peak (mode) of the ADF, not any kind of SSP-equivalent abundance. Light elements (C, N, O, Si, Na) are assumed to be related to Fe and $Z$ in a single ratio for each solution (that is, no trend of [$\alpha$/Fe] with Fe or $Z$ was assumed). Furthermore, due to the subtlety of the spectra features affected by Si and O, inclusion of these elements gives noisy solutions. Therefore, these two elements were tied to Mg, which strongly affects the spectrum. That is, [Mg/Fe]=[Si/Fe]=[O/Fe] always. This matters because isochrone temperatures vary with O and Si abundance, and there is a strong O-age degeneracy \citep{2022MNRAS.511.3198W}. Elements not mentioned, such as Ni, Cr, Pd, or U, were similarly tied in lockstep to Fe. We refer to such unnamed elements by the symbol `R.' Helium abundance has a non-adjustable trajectory with $Z$ determined by the isochrone set. This is out of pragmatic necessity, since He cannot be measured in the galaxies studied here, but should be mentioned since it puts an irreducible systematic error into the results.

\section{Results} \label{sec:results}

This section presents the results for stellar age, [Fe/H], and the abundances of C, N, Na, and Mg, along with their radial gradients. For the purposes of this study, we refer to the single-burst-equivalent mean age, [Fe/H], [C/R], [N/R], [Na/R], and [Mg/R] collectively as stellar population parameters. In $\S$\ref{subsec:abundances}, we show the variation of these parameters as a function of R/R$_e$, out to R/R$_e$ = 0.7. Average values of the parameters, binned by velocity dispersion ($\sigma$), are reported in Table~\ref{table:bins_vd}, while Table~\ref{table:bins_met} presents the same parameters binned by [Fe/H]. In both cases, the tabulated values correspond to the mean measurements from the 3$^{\mathrm{rd}}$ annulus of galaxies within each bin. Section~\ref{subsec:gradients} presents the results on the radial gradients of the stellar population parameters.

\subsection{Stellar Population Parameters}\label{subsec:abundances}

Stellar population parameters were derived using {\sc compfit}, as described in $\S$\ref{subsec:model}. Uncertainties were estimated by performing 50 Monte Carlo (MC) realizations per annulus for each galaxy. In each realization, the measured, corrected index strengths were perturbed within a Gaussian envelope of 0.7 SD (Standard Deviation). The 0.7 factor was chosen to suppress outlying solutions and improve error robustness in some cases. The standard deviation across the 50 realizations was adopted as the parameter uncertainty for each annulus. While [C/R], [N/R], [Na/R], and [Mg/R] solutions are treated as linear and unbounded, age and [Fe/H] can fall near the model grid edges—specifically, [Fe/H] near 0.4 and age near 16 Gyr. For solutions at or near these boundaries, MC-derived errors were linearly inflated, up to a factor of two, to account for the absence of solutions beyond the grid limits.

\begin{table*}[]
\centering
\caption{Bins in $\sigma$ used in this work. For each $\sigma$ bin, the total number of galaxies in that bin and their population parameters are also tabulated. Note that the values tabulated are mean of values from the 3$^{rd}$ annulus for the galaxies within the $\sigma$ bin.}
\label{table:bins_vd}
\begin{tabular}{cccccccc}
\hline\hline
log $\sigma$ Range & N & Mean log age & Mean [Fe/H] & Mean [C/R] & Mean [N/R] & Mean [Na/R] & Mean [Mg/R] \\[-3pt] 
(km/s) & & (Gyr) & (dex) & (dex) & (dex) & (dex) & (dex) \\[5pt] \hline
{[}1.6, 2{]}       & 432  & 0.55$\pm$0.17  & 0.11$\pm$0.17  & -0.08$\pm$0.07  & -0.15$\pm$0.15  & -0.23$\pm$0.14  & -0.14$\pm$0.10  \\ \hline
{[}2, 2.2{]}       & 699  & 0.67$\pm$0.15  & 0.19$\pm$0.14  & 0.04$\pm$0.06   & -0.03$\pm$0.11  & -0.05$\pm$0.15  & 0.00$\pm$0.07   \\ \hline
{[}2.2, 2.34{]}    & 930  & 0.77$\pm$0.15  & 0.15$\pm$0.13  & 0.12$\pm$0.06   & 0.06$\pm$0.10   & 0.05$\pm$0.13   & 0.08$\pm$0.07   \\ \hline
{[}2.34, 2.45{]}   & 755  & 0.87$\pm$0.12  & 0.09$\pm$0.13  & 0.19$\pm$0.07   & 0.14$\pm$0.11   & 0.14$\pm$0.12   & 0.14$\pm$0.07   \\ \hline
{[}2.45, 2.53{]}   & 152  & 0.95$\pm$0.10  & 0.03$\pm$0.13  & 0.24$\pm$0.07   & 0.23$\pm$0.10   & 0.23$\pm$0.12   & 0.19$\pm$0.07   \\ \hline
\end{tabular}
\end{table*}

\begin{table*}[]
\centering
\caption{Bins in [Fe/H] used in this work. For each [Fe/H] bin, the total number of galaxies in that bin and mean population parameters are tabulated. Note that the values tabulated are mean of values from the 3$^{rd}$ annulus of the galaxies within the [Fe/H] bin.}
\label{table:bins_met}
\begin{tabular}{cccccc}
\hline\hline
[Fe/H] Range & N & Mean [C/R] & Mean [N/R] & Mean [Na/R] & Mean [Mg/R] \\[-3pt] 
(dex) & & (dex) & (dex) & (dex) & (dex) \\[5pt] \hline
[-0.38, -0.16] & 111 & 0.13$\pm$0.18 & 0.05$\pm$0.20 & 0.05$\pm$0.24 & 0.07$\pm$0.17  \\ \hline
[-0.15, 0.06]  & 705 & 0.12$\pm$0.14 & 0.03$\pm$0.18 & 0.04$\pm$0.21 & 0.06$\pm$0.14  \\ \hline
[0.06,0.28]    & 1696 & 0.09$\pm$0.10 & 0.04$\pm$0.15 & 0.02$\pm$0.18 & 0.05$\pm$0.12  \\ \hline
[0.28, 0.5]    & 456 & 0.05$\pm$0.09 & 0.02$\pm$0.13 & -0.03$\pm$0.18 & 0.02$\pm$0.11  \\ \hline
\end{tabular}
\end{table*}

\subsubsection{Age}\label{subsubsec:age}

\begin{figure}[ht!]
\plotone{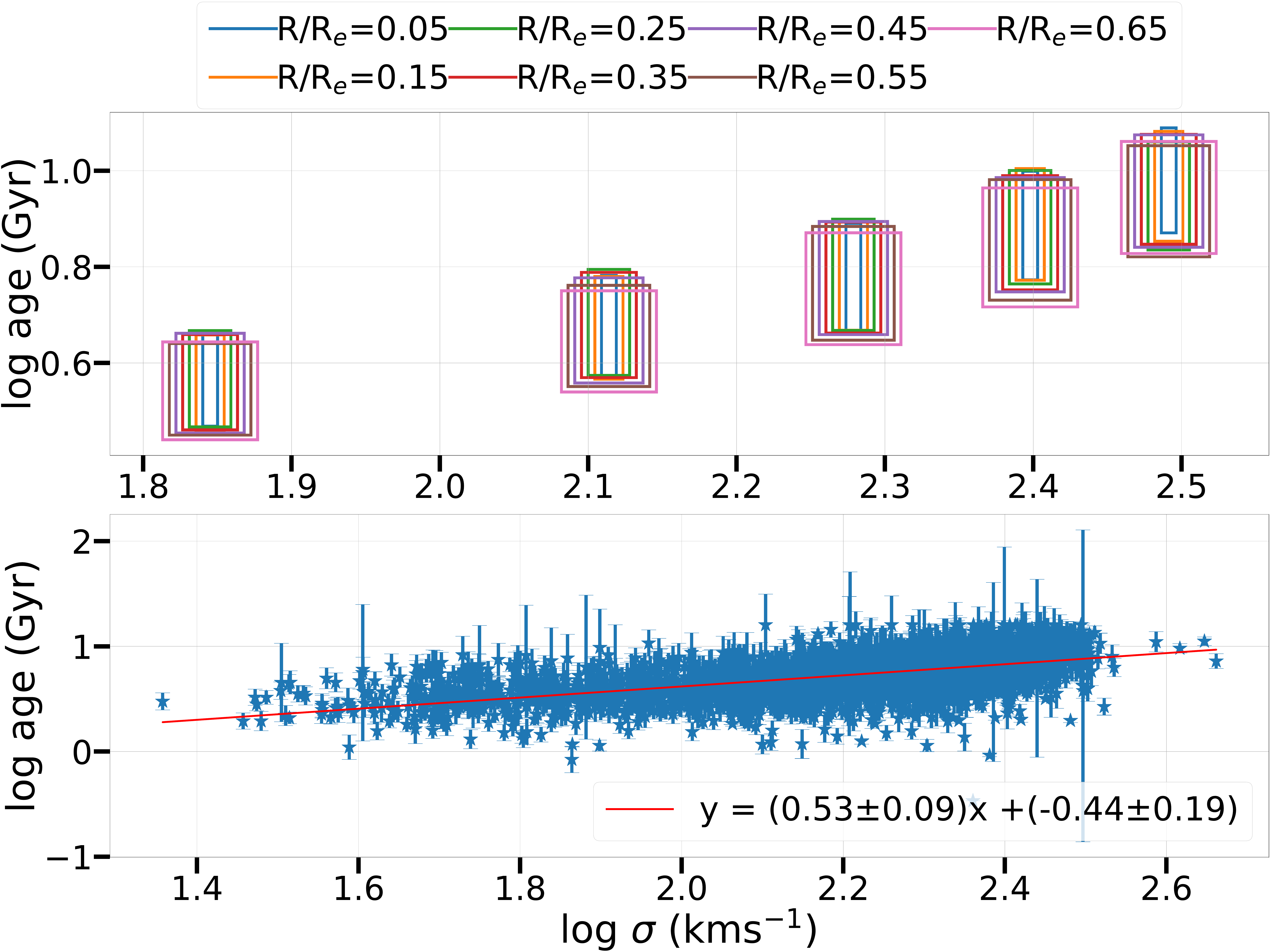}
\caption{The (log of) mean age is plotted against log of $\sigma$. In the top panel, the center of the rectangular box corresponds to the mean age for each $\sigma$ bin ($\sigma$ bins are described in Table \ref{table:bins_vd}) color coded by annulus radius. The vertical extent of each box corresponds to $\pm$average error for a given annulus and the width increases with increasing annulus value.} The legend on top lists the middle value for each radial bin. The bottom panel shows the mean age (with errorbars) of each galaxy's 3$^{\rm rd}$ annulus. The legend describes the best fit line with `y' as the mean age and `x' as the log of $\sigma$.
\label{fig:age_vd_rad_with_3ann}
\end{figure}

Since mass is proportional to $\sigma^2$ but scales linearly with galaxy size \citep{1992ApJ...399..462B}, $\sigma$ serves as an imperfect yet sufficient proxy for galaxy mass. Fig.~\ref{fig:age_vd_rad_with_3ann} shows the variation of log(mean age [Gyr]) with $\sigma$. Ages derived in this work are, roughly, $U$-band-weighted mean ages. In the top panel, the mean ages (represented by the center of the rectangles) across $\sigma$ bins (color-coded by annulus) are plotted against the log of $\sigma$, where the $\sigma$ values represent bin centers as defined in Table~\ref{table:bins_vd}. It is to be noted that the height of each rectangular box in top panel of Fig.~\ref{fig:age_vd_rad_with_3ann} corresponds to error for a given annulus at a given $\sigma$ whereas the width of the box, although arbitrarily chosen, tracks radius. The bottom panel displays the age at the 3$^{\rm rd}$ annulus for each galaxy, which we adopt as a representative value, more so than the central region or outermost, noisier annuli. A positive trend is apparent, indicating that galaxies with higher $\sigma$ tend to be older. The slope of the best-fit line (shown in red in the bottom panel) is 0.53 dex per log $\sigma$. The error bars in the top panel reflect the average parameter uncertainty across all galaxies within a $\sigma$ bin for each annulus, while in the bottom panel they correspond to the uncertainty in age at the 3$^{\rm rd}$ annulus of each individual galaxy.

\subsubsection{[Fe/H]}\label{subsubsec:met}

\begin{figure}[ht!]
\plotone{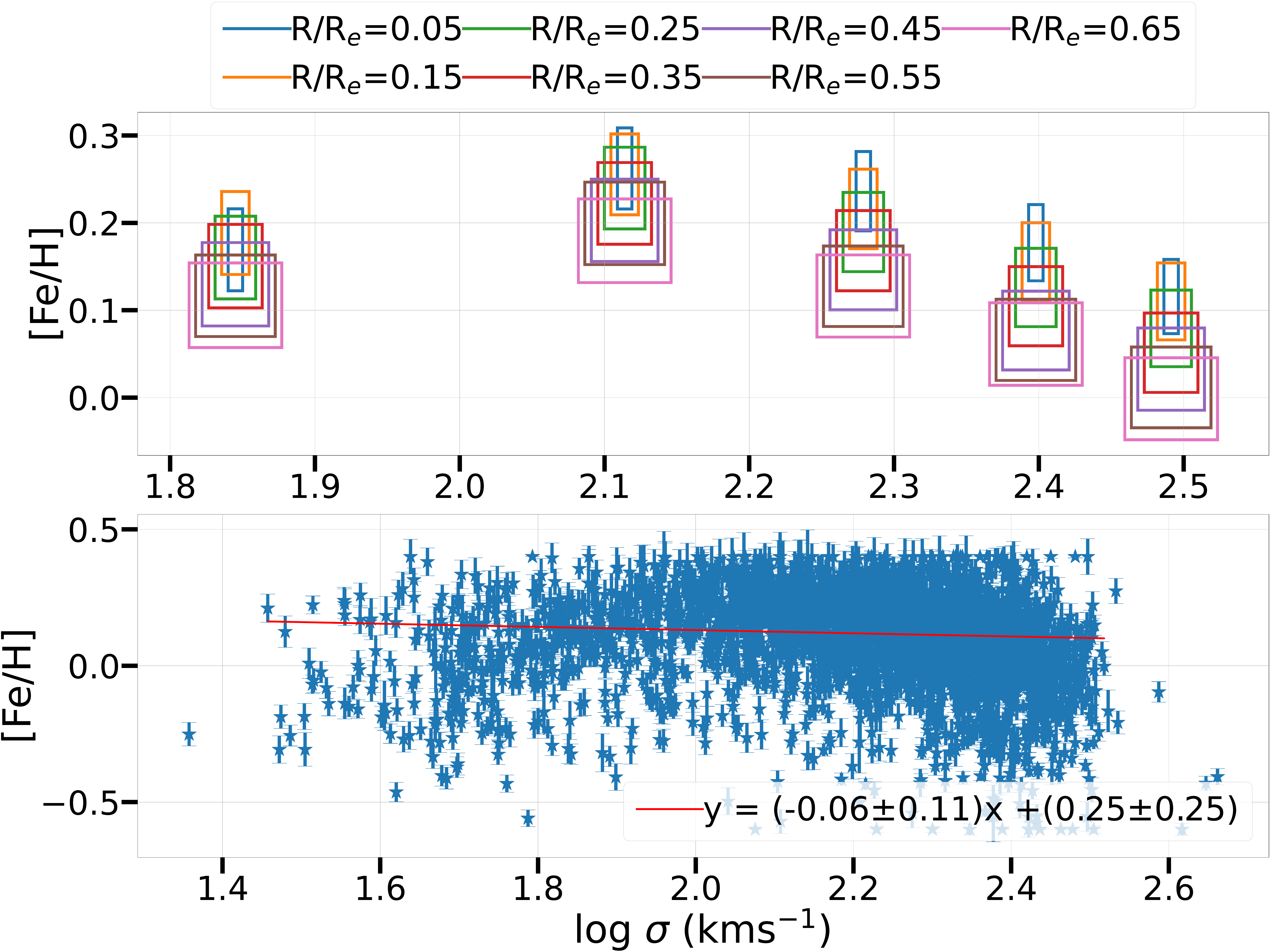}
\caption{Mean [Fe/H] is plotted against log $\sigma$. In the top panel, the center of the rectangular box corresponds to the mean value for each $\sigma$ bin color coded by annulus. The height of the box corresponds to $\pm$average error in [Fe/H] for a given annulus and the width increases arbitrarily with increasing annulus number.} The legend on top lists the middle value for each radial bin. The bottom panel shows the mean [Fe/H] (with errorbars) of each galaxy for its 3$^{\rm rd}$ annulus. The legend describes the best fit line with `y' as the [Fe/H] and `x' as the log of $\sigma$.
\label{fig:met_vd_rad_with_3ann}
\end{figure}

Fig.~\ref{fig:met_vd_rad_with_3ann} shows how the [Fe/H] values change with $\sigma$. In the top panel, the mean [Fe/H] (represented by the center of the rectangles) across $\sigma$ bins (color-coded by annulus) are plotted against the log of $\sigma$. Here also the height of each rectangular box in the top panel corresponds to $\pm$average error for a given annulus whereas the width of the box increases with increasing annulus value. The bottom panel shows the [Fe/H] at the 3$^{\rm rd}$ annulus for each individual galaxy. We observe a decreasing trend of [Fe/H] with $\sigma$ from the 2$^{\rm nd}$ $\sigma$ bin onward. When analyzed individually, we find a slightly negative slope (-0.06 dex per log $\sigma$) overall for the [Fe/H] vs log($\sigma$) graph in bottom panel of Fig.~\ref{fig:met_vd_rad_with_3ann}, though the two-segment character is still visible in the data, with an inflection point at roughly log $\sigma \approx$ 2.0 or $\sigma \approx$ 100 km s$^{-1}$.

\subsubsection{C, N, Na, and Mg}\label{subsubsec:abun}

Having examined the trends of age and [Fe/H] with local velocity dispersion, we now turn to the elemental abundances of C, N, Na, and Mg in our sample of 2968 MaNGA galaxies. 

\begin{figure*}[ht!]
\plotone{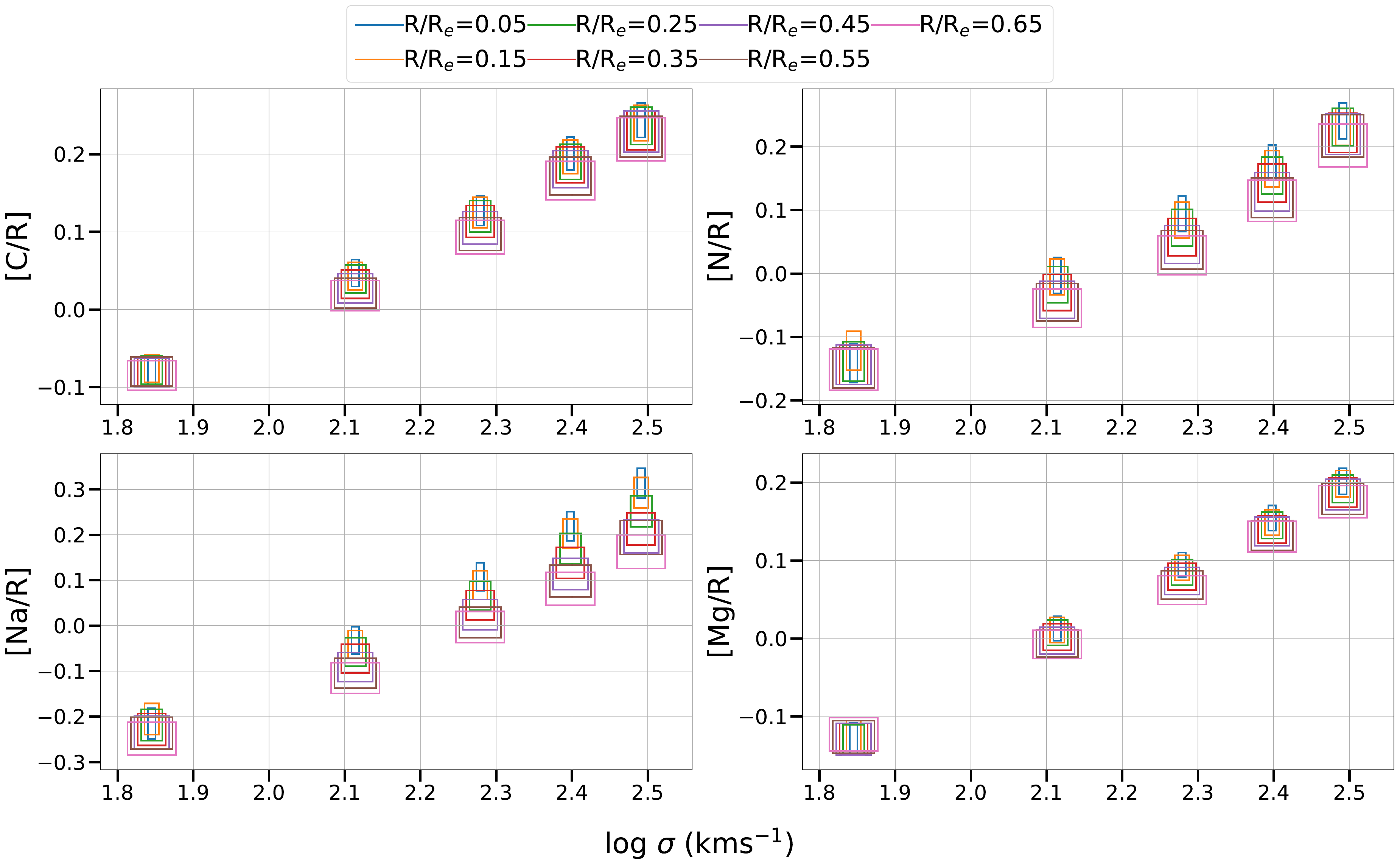}
\caption{The mean elemental abundances for C, N, Na, and Mg plotted against log of $\sigma$. As in Figs. \ref{fig:age_vd_rad_with_3ann} and \ref{fig:met_vd_rad_with_3ann}, the center of the rectangular box corresponds to mean value and the error is represented by the height of the box.} The annuli are color coded and annulus midpoints are listed in the legend on top of the figure. 
\label{fig:abun_vd_rad}
\end{figure*}

The galaxies are grouped according to their $\sigma$ (at the 3$^{\rm rd}$ annulus) according to Table~\ref{table:bins_vd} and the average abundance values in each $\sigma$ bin are plotted in Fig.~\ref{fig:abun_vd_rad} (color-coded by annulus). Note that the center of the rectangular box represents the average abundance value while the height represents the corresponding error. The errors are obtained by taking the average of the error of each measurement within a $\sigma$ bin for every annulus. Given the greater abundance of light elements relative to Fe-peak elements, this figure clearly illustrates that the mean abundance rises with increasing $\sigma$, even though the Fe abundance alone does not (Fig.~\ref{fig:met_vd_rad_with_3ann}). As $\sigma$ is correlated with galaxy mass, it can be safely concluded that massive ETGs exhibit significantly elevated abundances of C, N, Na, and Mg compared to their less massive counterparts.  

Note once again that `R' stands for generic heavy element. While {\sc compfit} is capable of treating Fe separately from other heavy elements, for the runs that we report on here, Fe has not been allowed to vary by itself, so any [X/R] is equivalent to [X/Fe]. 

\begin{figure*}[ht!]
\plotone{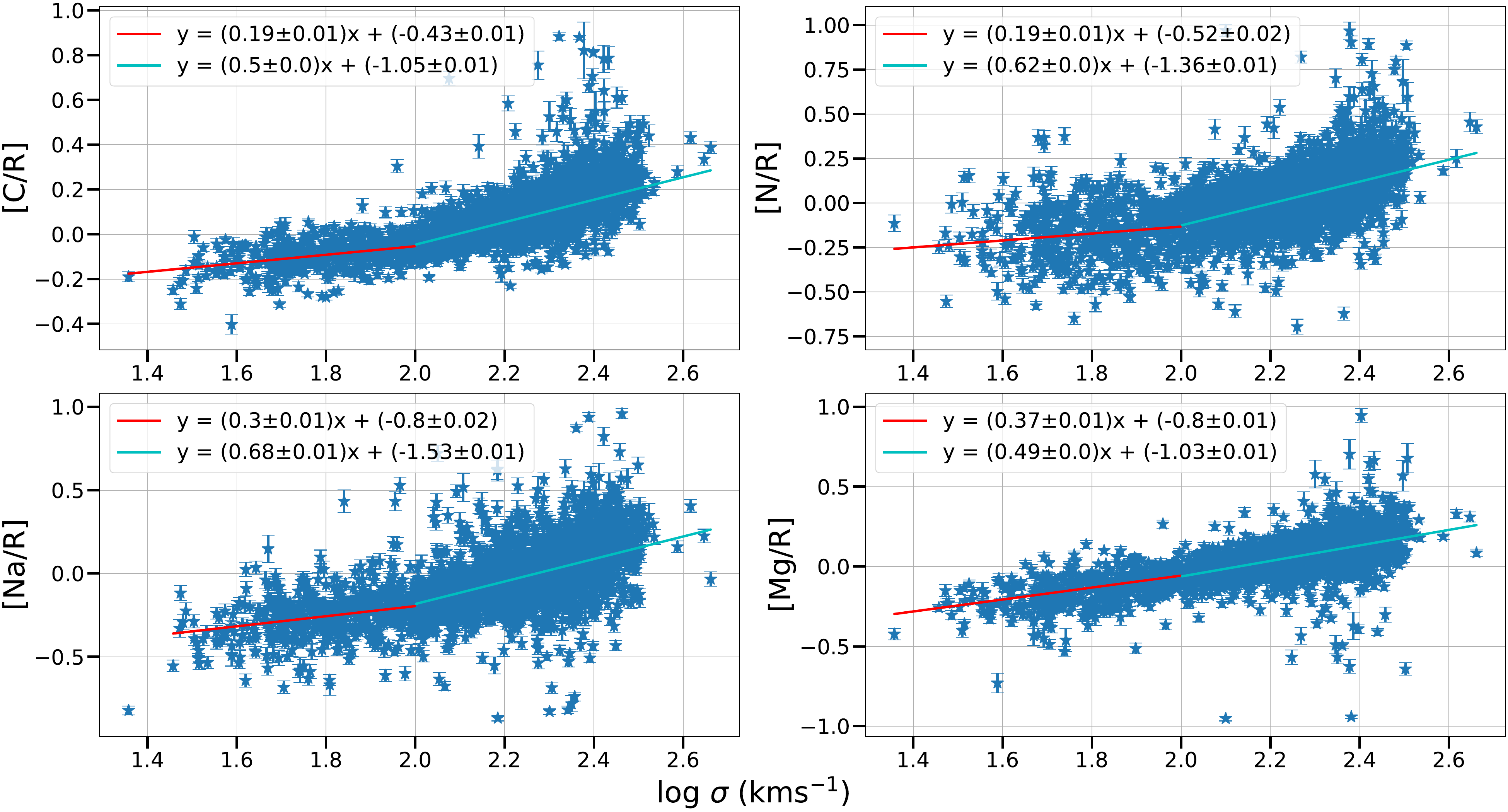}
\caption{The elemental abundances for the 3$^{\rm rd}$ annulus of each galaxy are plotted against $\sigma$. The error bars come from the 50 Monte Carlo realizations as described in the text. The trendlines show the best fit to the data for log($\sigma$)$<$2.0 (in red) and log($\sigma$)$>$2.0 (in cyan). The best fit equations are shown as legend with `y' being the elemental abundance and `x' being the log($\sigma$).
\label{fig:abun_vd_3ann}}
\end{figure*}

A notable finding emerges when examining elemental abundances at the individual galaxy level rather than in groups. As depicted in Fig.~\ref{fig:abun_vd_3ann}, a consistent trend of increasing abundance with $\sigma$ is observed, mirroring the pattern in Fig.~\ref{fig:abun_vd_rad}. However, a distinct change in slope is evident at log($\sigma$)=2.0 for all four elements analyzed in this study. Analysis of the trend line equations in Fig.~\ref{fig:abun_vd_3ann} reveals that the slope increases by a factor of 2.6 for C, 3.26 for N, 2.26 for Na, and 1.3 for Mg for log $\sigma>$2.0 compared to log $\sigma<$2.0. This transition at log($\sigma$)=2.0 corresponds to the downward inflection in Fe abundance, shown as a `kink' in Fig.~\ref{fig:met_vd_rad_with_3ann}. These observations suggest that mid-sized to large galaxies may follow distinct evolutionary pathways compared to dwarf elliptical galaxies.

\subsubsection{Binned by [Fe/H]}\label{subsubsec:binnedbyFe}

In addition to examining the relationship between elemental abundances and $\sigma$, we investigated their variation with respect to [Fe/H]. As [Fe/H] is derived from the {\sc compfit} inversion program, we incorporate the associated uncertainties in [Fe/H] when charting the variation of other parameters with [Fe/H].

\begin{figure}[ht!]
\plotone{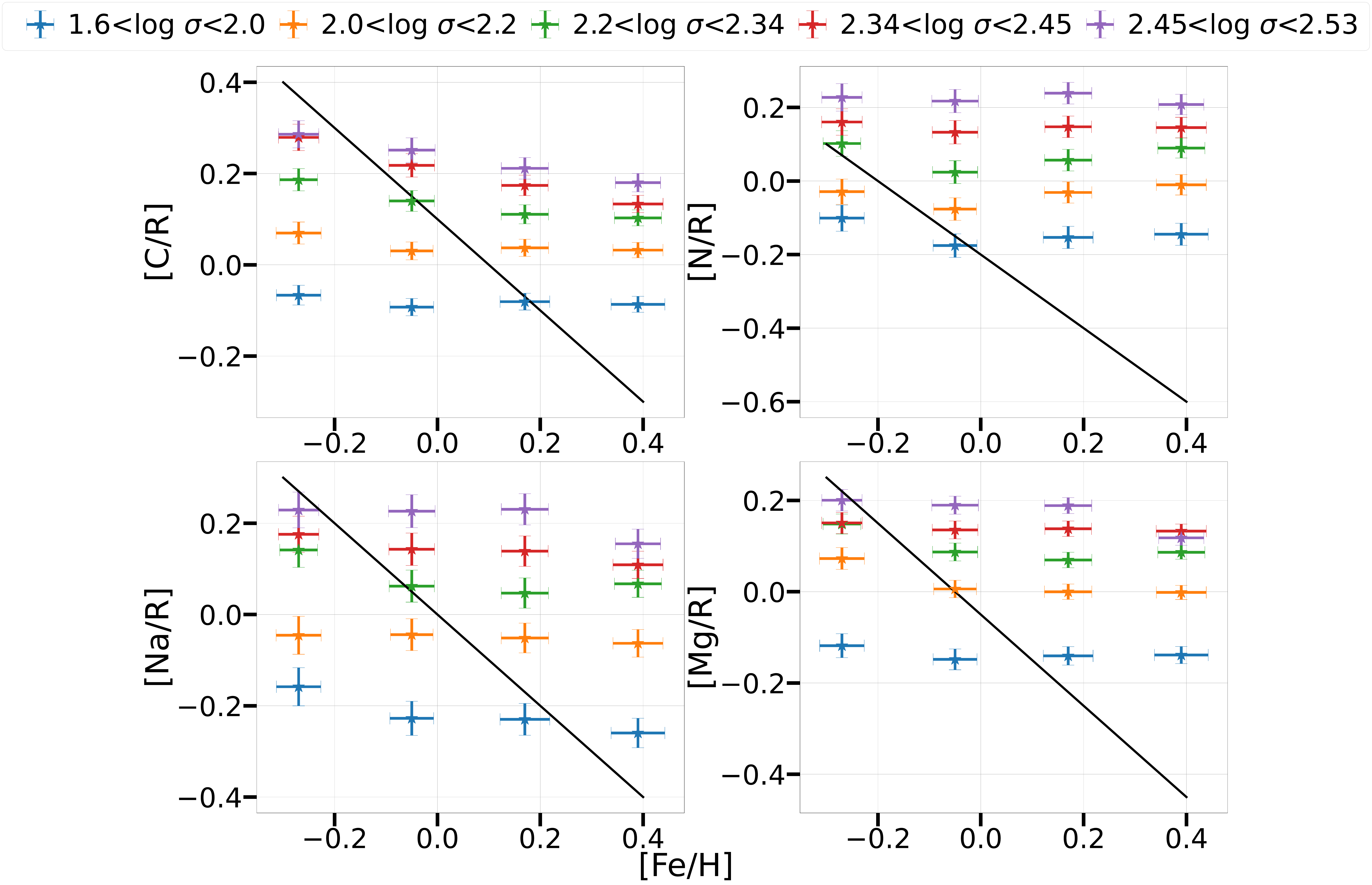}
\caption{Change of light elements' abundance plotted against [Fe/H]. The $\sigma$ bins are color coded and shown in the legend on top. Note that the galaxies are grouped by $\sigma$ and [Fe/H] at the 3$^{rd}$ annulus. There are no large galaxies for the highest $\sigma$ and [Fe/H] bin, thus causing absence of data point in the above figure for that bin. Lines of slope $-1$ (black solid line) indicate the behavior if the light element is held at constant abundance while Fe alone increases.
\label{fig:abun_met_vd}}
\end{figure}

\begin{figure}[ht!]
\plotone{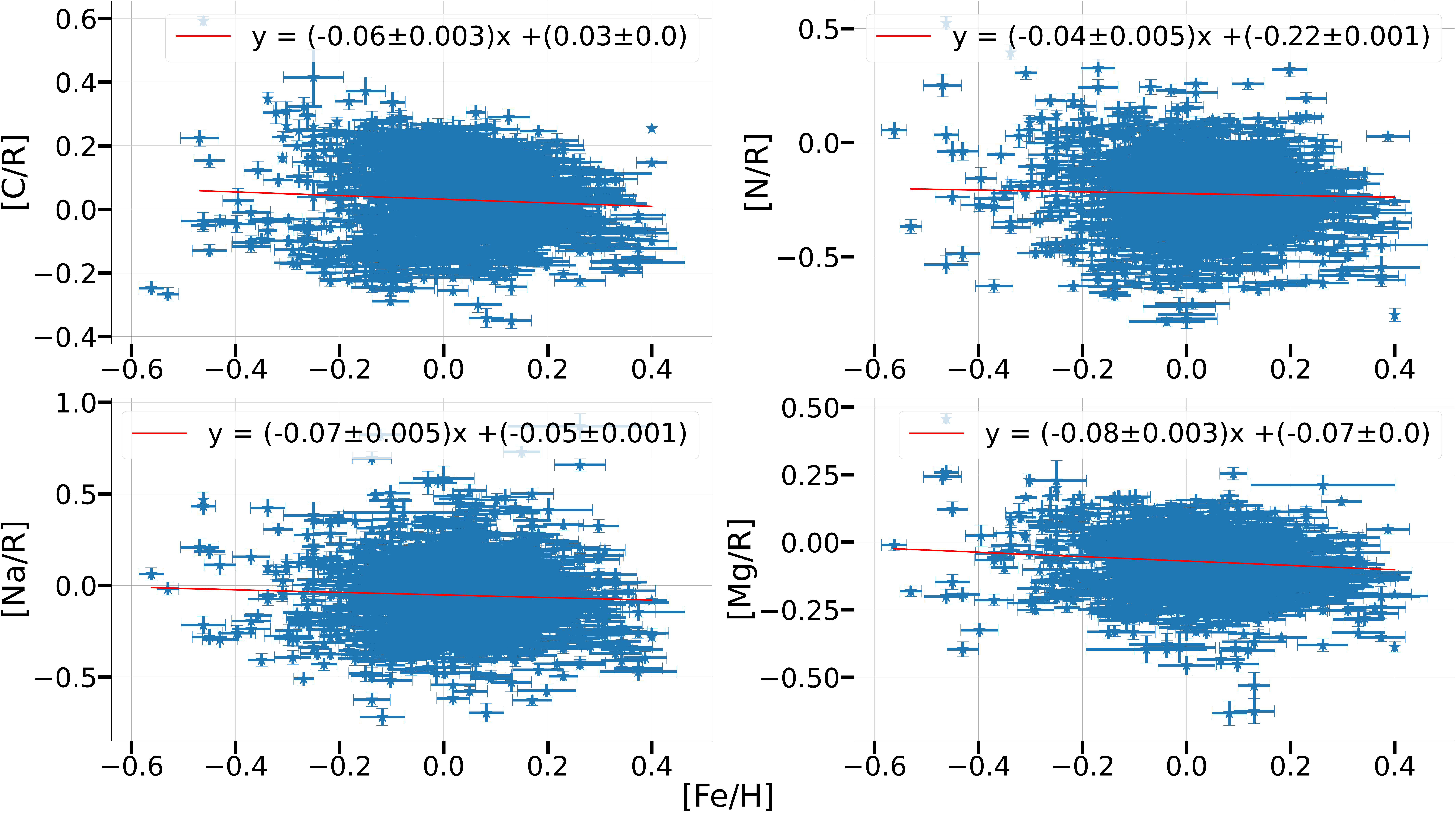}
\caption{Same as Fig.~\ref{fig:abun_vd_3ann} but with [Fe/H] as the independent variable.
\label{fig:abun_met_3ann}}
\end{figure}

Fig.~\ref{fig:abun_met_vd} presents a plot of elemental abundances as a function of [Fe/H]. In this figure, we plotted the average abundance value across each [Fe/H] bin color-coded by $\sigma$. The error bars represent the mean of uncertainties within each [Fe/H] bin across all $\sigma$ bins, reflecting a typical single-galaxy error. A general decreasing trend in light element abundances is observed with increasing [Fe/H]. Trend lines with a slope of $-1$ (solid black lines in Fig.~\ref{fig:abun_met_vd}) illustrate the expected behavior if light element abundances remain constant while Fe increases, as would occur if Type \Romannum{1}a supernovae were the sole contributors. The observed shallower decline in light element abundances suggests ongoing enrichment of these elements at higher [Fe/H]. This trend is further quantified in Fig.~\ref{fig:abun_met_3ann}, where the elemental abundances at the 3$^{\rm rd}$ annulus of each galaxy are plotted against [Fe/H]. Linear trend lines fitted to the data exhibit slightly negative slopes ($\sim$ -0.1 dex per decade) for all elements (see Fig.~\ref{fig:abun_met_3ann}). A slope of zero would indicate lockstep enrichment with Fe, whereas a slope of $-1$ would suggest no enrichment. The observed slopes are significantly closer to the lockstep scenario. \cite{2010A&A...522A..32R} documents the behavior of various elements with respect to [Fe/H]. Within the [Fe/H] regime of -0.6$<$[Fe/H]$<$0.4, our findings for C, N, Na, and Mg very closely resonates with the findings of \cite{2010A&A...522A..32R}.

The consistency of the findings can be verified by comparing Figures \ref{fig:met_vd_rad_with_3ann}, \ref{fig:abun_vd_3ann}, and \ref{fig:abun_met_3ann}. Fig.~\ref{fig:met_vd_rad_with_3ann} illustrates a decreasing trend in [Fe/H] with increasing $\sigma$, indicating that more massive galaxies exhibit lower [Fe/H]. Conversely, Fig.~\ref{fig:abun_vd_3ann} demonstrates that more massive galaxies possess higher abundances of light metals. Consequently, it is anticipated that light metal abundances would decrease with increasing [Fe/H], a trend confirmed in Fig.~\ref{fig:abun_met_3ann}. Notably, the change in slope observed when plotting light metal abundances against $\sigma$ disappears when these abundances are plotted against [Fe/H], at least for N and Mg. This distinction is significant, as $\sigma$ reflects galaxy structure, whereas elemental abundances are indicative of chemical evolution processes.

\subsection{Gradients in Stellar Population Parameters}\label{subsec:gradients}

This section presents the results of our radial gradient analysis. Radial gradients are computed in logarithmic space by dividing the stellar population parameters (age, [Fe/H], [C/R], [N/R], [Na/R], and [Mg/R]) by the logarithm of the normalized radius, log(R/R$_e$). Thus, the gradient represents the change in a given parameter (on a logarithmic scale) per unit change in log(R/R$_e$). The associated uncertainties in the radial gradients are derived from the covariance matrix of the paired quantities. Table \ref{table:bins_vd_grad} provides the details of $\sigma$ binning, alongside the mean radial gradients for each bin for age, [Fe/H], and the abundances of C, N, Na, and Mg.

\begin{table*}[]
\centering
\caption{Table showing mean radial gradients in log(age), [Fe/H], and abundances for C, N, Na, and Mg for each $\sigma$ bin used in this work.}
\label{table:bins_vd_grad}
\begin{tabular}{cccccccc}
\hline\hline
log $\sigma$ Range & N & Mean $\nabla_r$log age & Mean $\nabla_r$[Fe/H] & Mean $\nabla_r$[C/R] & Mean $\nabla_r$[N/R] & Mean $\nabla_r$[Na/R] & Mean $\nabla_r$[Mg/R] \\[-3pt] 
(km/s) & & & & & & & \\[5pt] \hline
{[}1.6, 2{]}     & 430  & 0.00$\pm$0.12  & -0.07$\pm$0.14 & -0.01$\pm$0.07 & 0.00$\pm$0.16  & -0.03$\pm$0.16 & 0.00$\pm$0.09 \\ \hline
{[}2, 2.2{]}     & 699  & -0.04$\pm$0.13 & -0.08$\pm$0.13 & -0.02$\pm$0.05 & -0.05$\pm$0.09 & -0.08$\pm$0.13 & -0.02$\pm$0.05 \\ \hline
{[}2.2, 2.34{]}  & 924  & -0.02$\pm$0.14 & -0.12$\pm$0.15 & -0.03$\pm$0.07 & -0.06$\pm$0.12 & -0.12$\pm$0.18 & -0.03$\pm$0.07 \\ \hline
{[}2.34, 2.45{]} & 747  & -0.05$\pm$0.18 & -0.13$\pm$0.17 & -0.02$\pm$0.10 & -0.05$\pm$0.16 & -0.13$\pm$0.21 & -0.03$\pm$0.17 \\ \hline
{[}2.45, 2.53{]} & 151  & -0.05$\pm$0.16 & -0.15$\pm$0.17 & -0.02$\pm$0.08 & -0.02$\pm$0.14 & -0.16$\pm$0.18 & -0.02$\pm$0.08 \\ \hline
\end{tabular}
\end{table*}

\subsubsection{Age gradients}\label{subsubsec:age_grad}

\begin{figure}[ht!]
\plotone{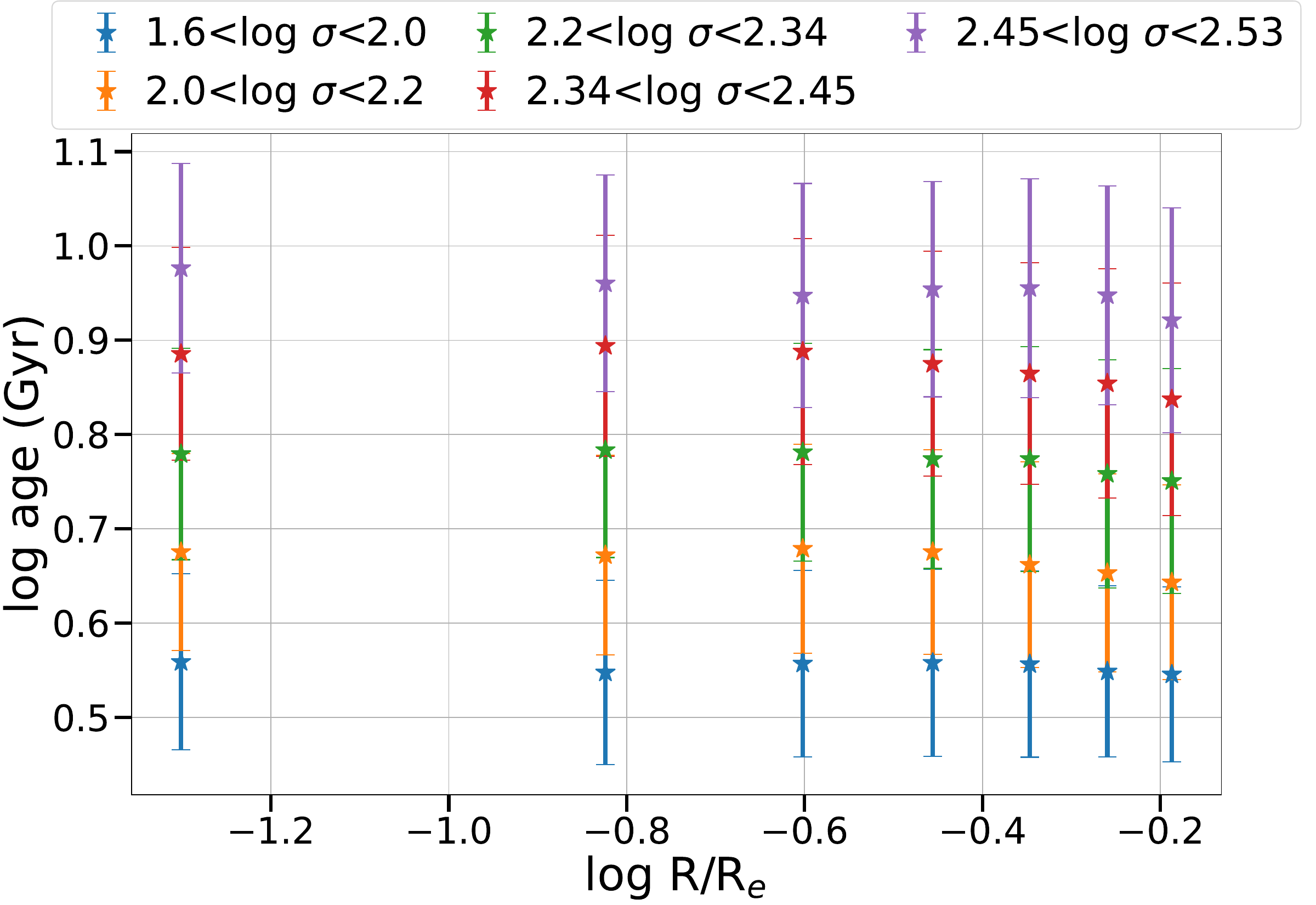}
\caption{The mean age of each $\sigma$ bin is plotted against log(R/R$_e$) with errorbars. The legend shows the extent of each $\sigma$ bin.
\label{fig:age_rad_vd}}
\end{figure}

This section focuses on the radial gradients in age across galaxies within specific $\sigma$ bins, as detailed in Table~\ref{table:bins_vd_grad}. In Fig.~\ref{fig:age_rad_vd}, the mean age for each $\sigma$ bin is plotted against log R/R$_e$ (color-coded by $\sigma$ bins), with error bars representing the mean uncertainty within each bin. On average, radial age gradients are flat or slightly negative ($\sim$-0.04 dex per log(R/R$_e$) across all $\sigma$ bins, though significant scatter indicates that many galaxies have younger outskirts, while a comparable number are younger at their centers. Only the highest $\sigma$ bin, with limited statistical reliability, consistently shows older centers and younger outskirts.

\begin{figure}[ht!]
\plotone{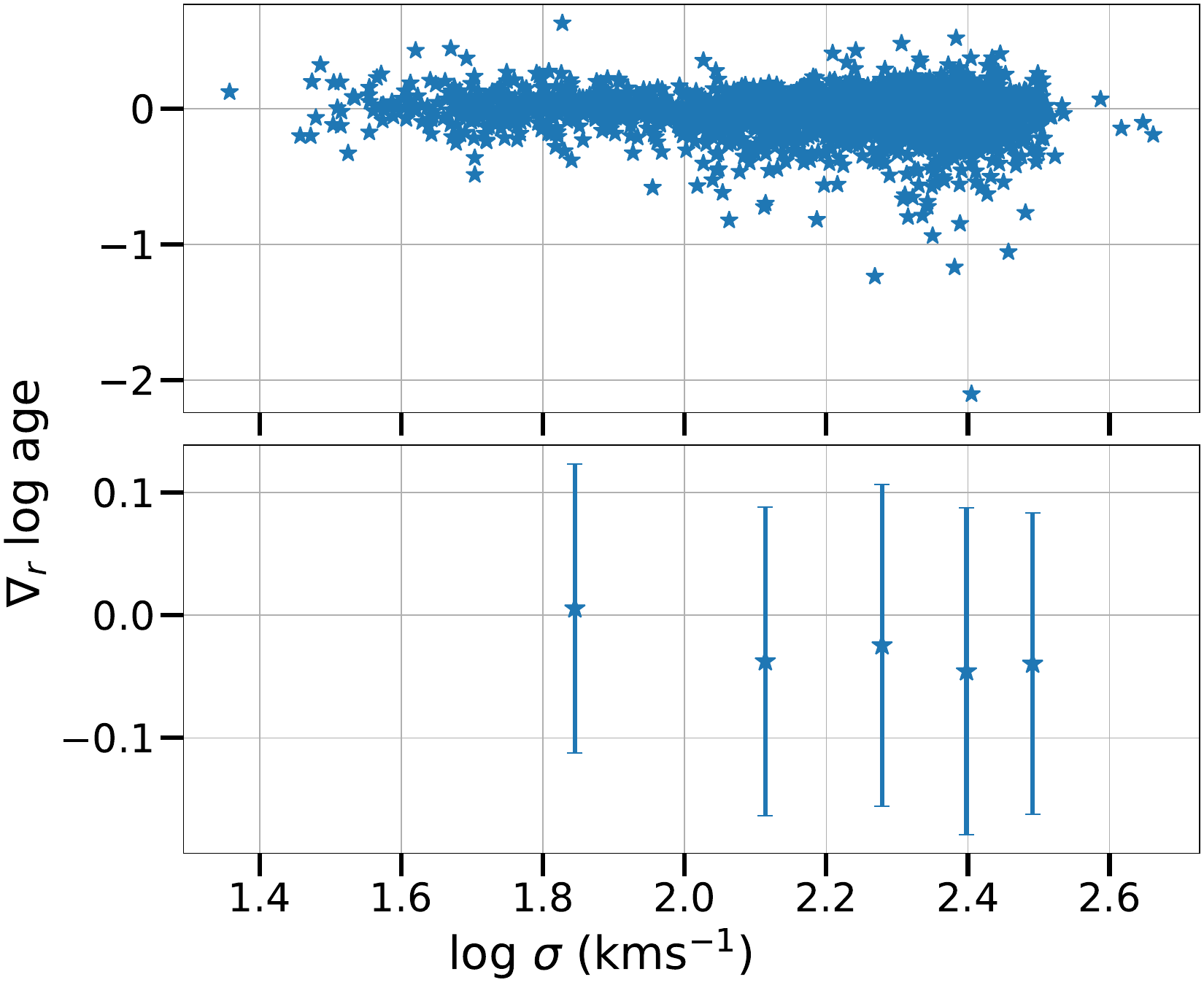}
\caption{The age gradients for individual galaxy are plotted against $\sigma$ in the top panel. In the bottom panel, the galaxies are grouped according to their $\sigma$ (at the 3$^{\rm rd}$ annulus) and average gradient values are plotted against midpoint of the $\sigma$ bin with errorbars indicating typical error for one galaxy within a $\sigma$ bin.
\label{fig:gradage_vd}}
\end{figure}

The top panel of Fig.~\ref{fig:gradage_vd} shows radial age gradient values for individual galaxies as a function of $\sigma$, with most values clustering near zero. The bottom panel plots the mean radial gradient for each $\sigma$ bin against the bin's central $\sigma$ value. The lowest $\sigma$ bin exhibits a near-zero mean radial gradient, while other bins maintain a near-constant gradient of -0.04 dex/log(R/R$_e$). Notably, the gradient error for each bin is derived from the mean of individual radial gradient uncertainties within the bin, reflecting a typical single-galaxy error, rather than the standard deviation of the bin.

Several methodological nuances warrant clarification regarding Figures \ref{fig:age_rad_vd} and \ref{fig:gradage_vd}. Mean values in Fig.~\ref{fig:age_rad_vd} and the bottom panel of Fig.~\ref{fig:gradage_vd} exclude outliers identified using the robust Z-score method, with data points exceeding a Z-score of 3.0 classified as outliers. Discrepancies between the gradients in Fig.~\ref{fig:age_rad_vd} and the mean radial gradients in the bottom panel of Fig.~\ref{fig:gradage_vd} arise from different computational approaches. In Fig.~\ref{fig:age_rad_vd}, the gradient is computed after calculating the mean age (post-outlier removal) for each $\sigma$ bin, whereas in Fig.~\ref{fig:gradage_vd}, radial gradients are first calculated for individual galaxies before averaging. As the gradient of the mean does not generally equal the mean of the gradients, and different data points are excluded as outliers, the results are not identical. These considerations also apply to the pairs Fig.~\ref{fig:met_rad_vd} and Fig.~\ref{fig:gradmet_vd}, as well as Fig.~\ref{fig:abun_rad_vd} and Figures \ref{fig:gradC_vd}, \ref{fig:gradN_vd}, \ref{fig:gradNa_vd}, and \ref{fig:gradMg_vd}.

\subsubsection{[Fe/H] gradients}\label{subsubsec:met_grad}

\begin{figure}[ht!]
\plotone{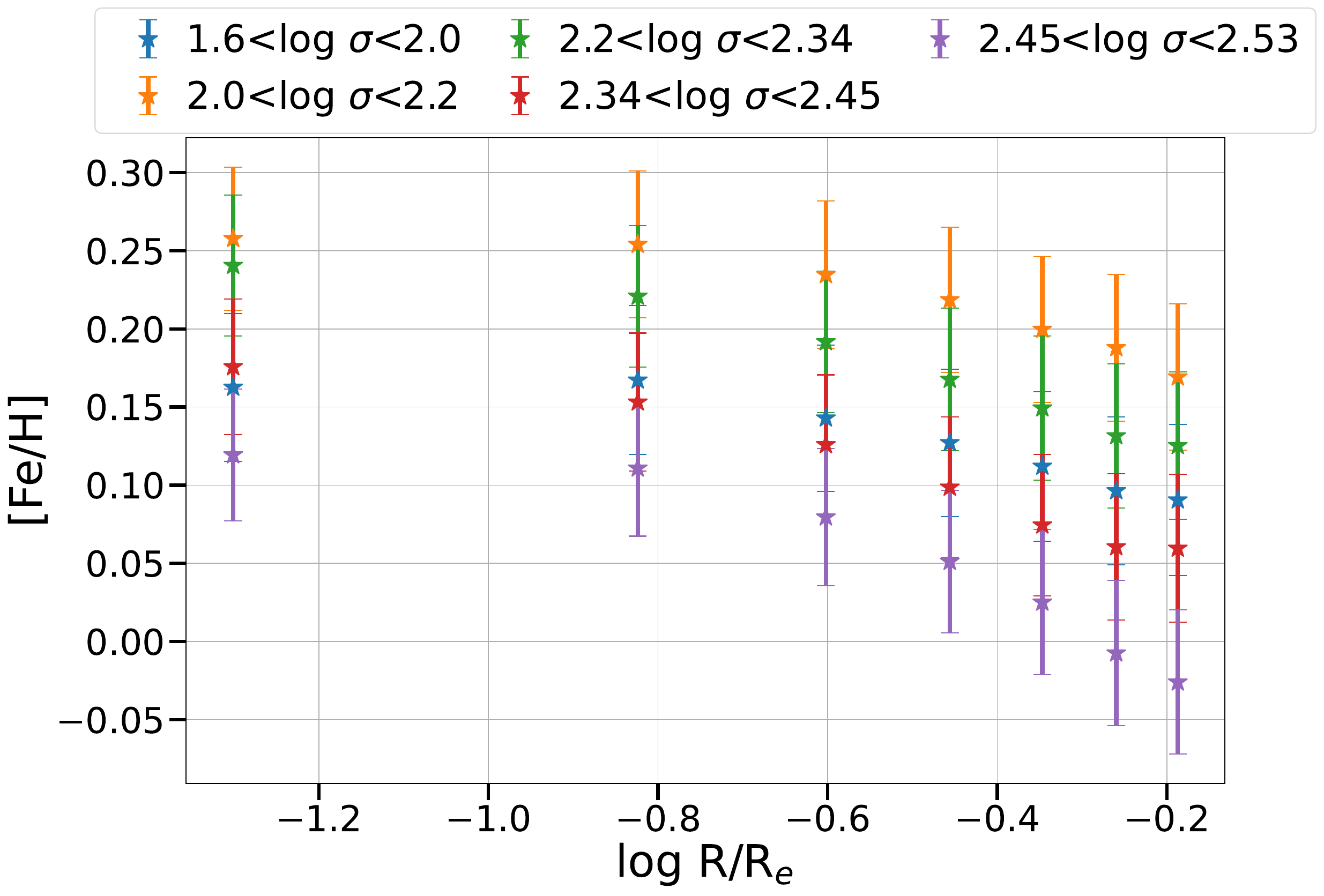}
\caption{Same as Fig.~\ref{fig:age_rad_vd} but with [Fe/H] on the y-axis. 
\label{fig:met_rad_vd}}
\end{figure}

\begin{figure}[ht!]
\plotone{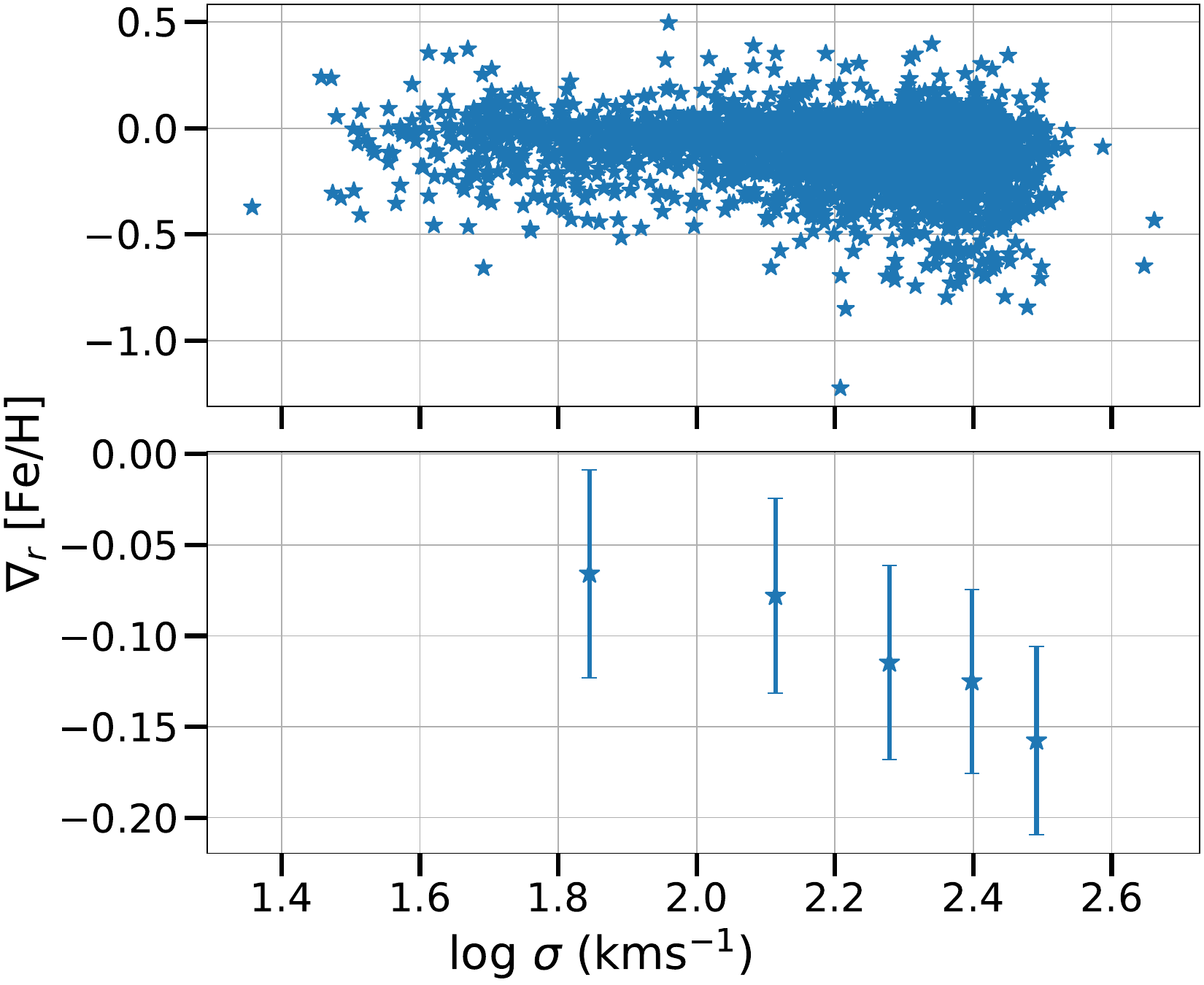}
\caption{Same as Fig.~\ref{fig:gradage_vd} but with [Fe/H] on the y-axis.
\label{fig:gradmet_vd}}
\end{figure}

Fig.~\ref{fig:met_rad_vd} shows the variation of mean [Fe/H] across different $\sigma$ bins (color-coded) as a function of radius, using log(R/R$_e$) as the radial proxy. At a glance, a pronounced negative radial gradient is consistently observed across all $\sigma$ bins. Fig.~\ref{fig:gradmet_vd} shows the radial gradient values in dex per decade. The top panel of Fig.~\ref{fig:gradmet_vd} plots individual radial gradients against their corresponding $\sigma$ values. Consistent with $\S$\ref{subsubsec:age_grad}, galaxies are grouped into five $\sigma$ bins, and the mean gradient values for each bin are plotted against $\sigma$ in the bottom panel of Fig.~\ref{fig:gradmet_vd}. The error bars represent the mean of individual uncertainties within each $\sigma$ bin.

Compared to $\S$\ref{subsubsec:age_grad}, the radial gradients in [Fe/H] have higher negative values and smaller intrinsic scatter (we will  show in $\S$\ref{subsec:diss_scatter} that most of the scatter is astrophysical, not observational in origin). In Fig.~\ref{fig:gradmet_vd} the highest $\sigma$ bin shows a gradient value of -0.16 dex/log(R/R$_e$) which implies that the [Fe/H] drops by 0.16 dex when we move from the innermost annulus to the 5$^{th}$ annulus. The fact that the smallest galaxies, which may or may not be merger remnants, have the shallowest [Fe/H] gradients while the largest galaxies, which are definitely shaped primarily by mergers, have the strongest gradients is a point for discussion (see $\S$\ref{subsec:discuss_grad}).

\subsubsection{C, N, Na, and Mg gradients}\label{subsubsec:abun_grad}

\begin{figure*}[ht!]
\plotone{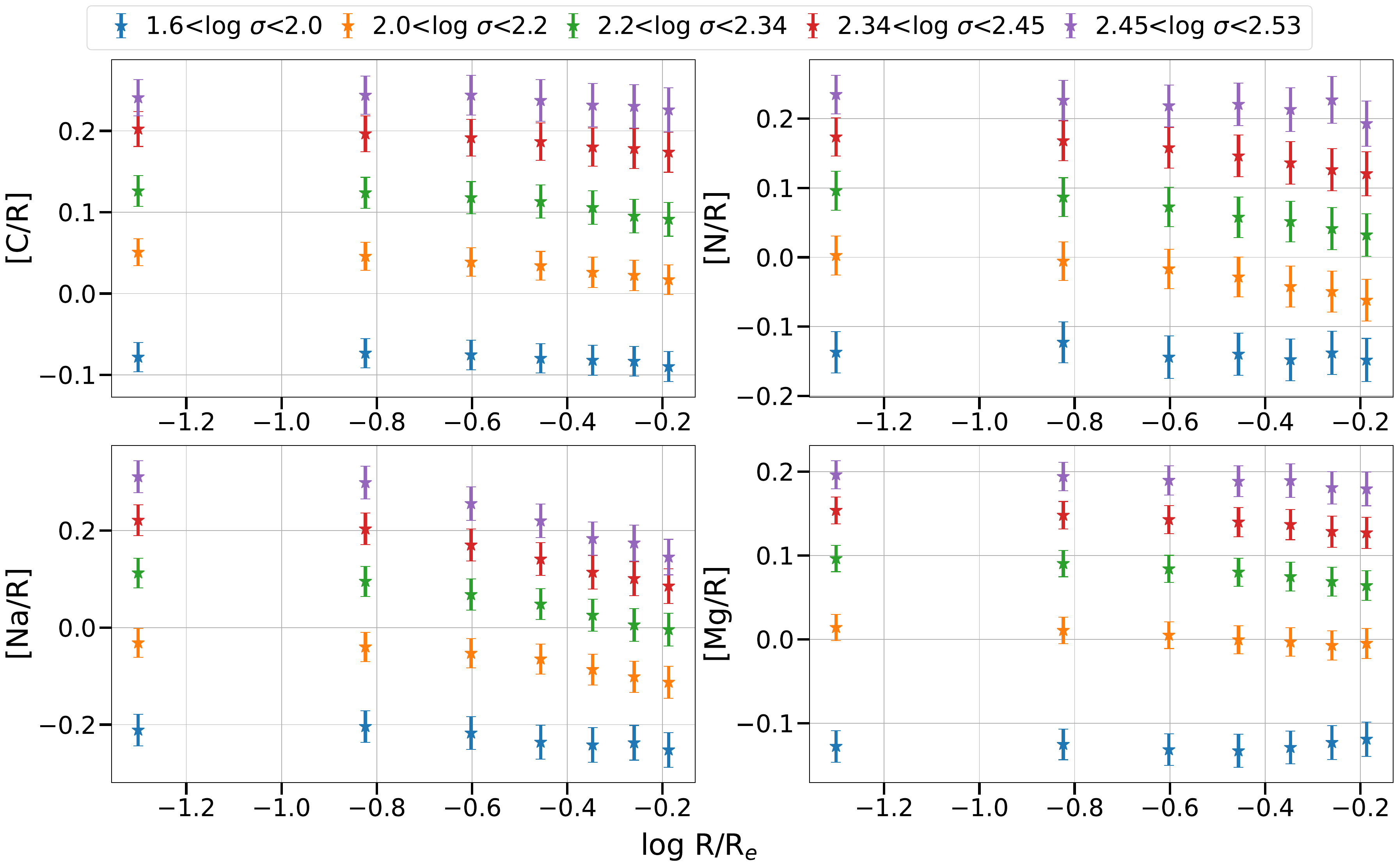}
\caption{The abundance of all the elements are plotted against log(R/R$_e$). The galaxies are grouped according to their $\sigma$ values in five different $\sigma$ bins (color coded) with errorbars. The legend lists the extent of each $\sigma$ bin, following Table \ref{table:bins_vd}. 
\label{fig:abun_rad_vd}}
\end{figure*}

Fig.~\ref{fig:abun_rad_vd} shows how [X/R] is changing radially, in five different $\sigma$ bins. The errorbars associated with each bin is the mean of errors in radial gradient values for the galaxies within that bin. The figure shows once again that the abundances increase with increasing $\sigma$ value (see also Fig.~\ref{fig:abun_vd_rad}) at all radial values. Na appears to have the highest gradients for all $\sigma$ values, though due to axis scaling this might most easily be seen in Table \ref{table:bins_vd_grad}. Another point worth mentioning here is that apart from Na, all the other light elements considered in this work show $\sim$-0.03 dex/log R/R$_e$ radial gradient value across all the $\sigma$ bins (very similar to the radial gradient in age).

\begin{figure}[ht!]
\plotone{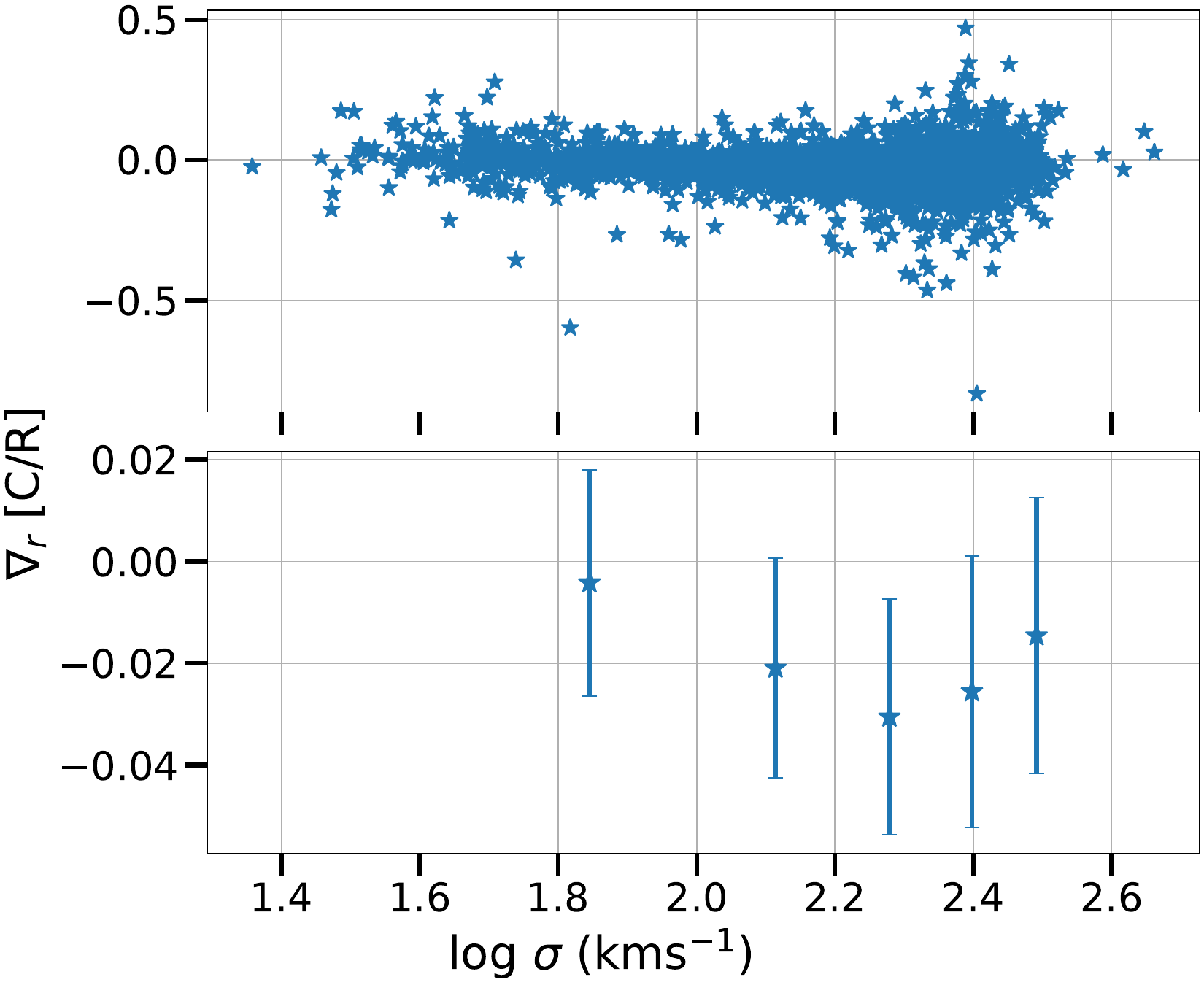}
\caption{The top panel shows the calculated radial gradients in C plotted against $\sigma$ for all the galaxies. The bottom panel shows the mean gradient value for each $\sigma$ bin. The errorbars in the bottom panel comes from the mean errors in radial gradient for all the galaxies within a certain $\sigma$ bin. Errorbars are not shown on the top panel in order to avoid crowding in the figure.
\label{fig:gradC_vd}}
\end{figure}

\begin{figure}[ht!]
\plotone{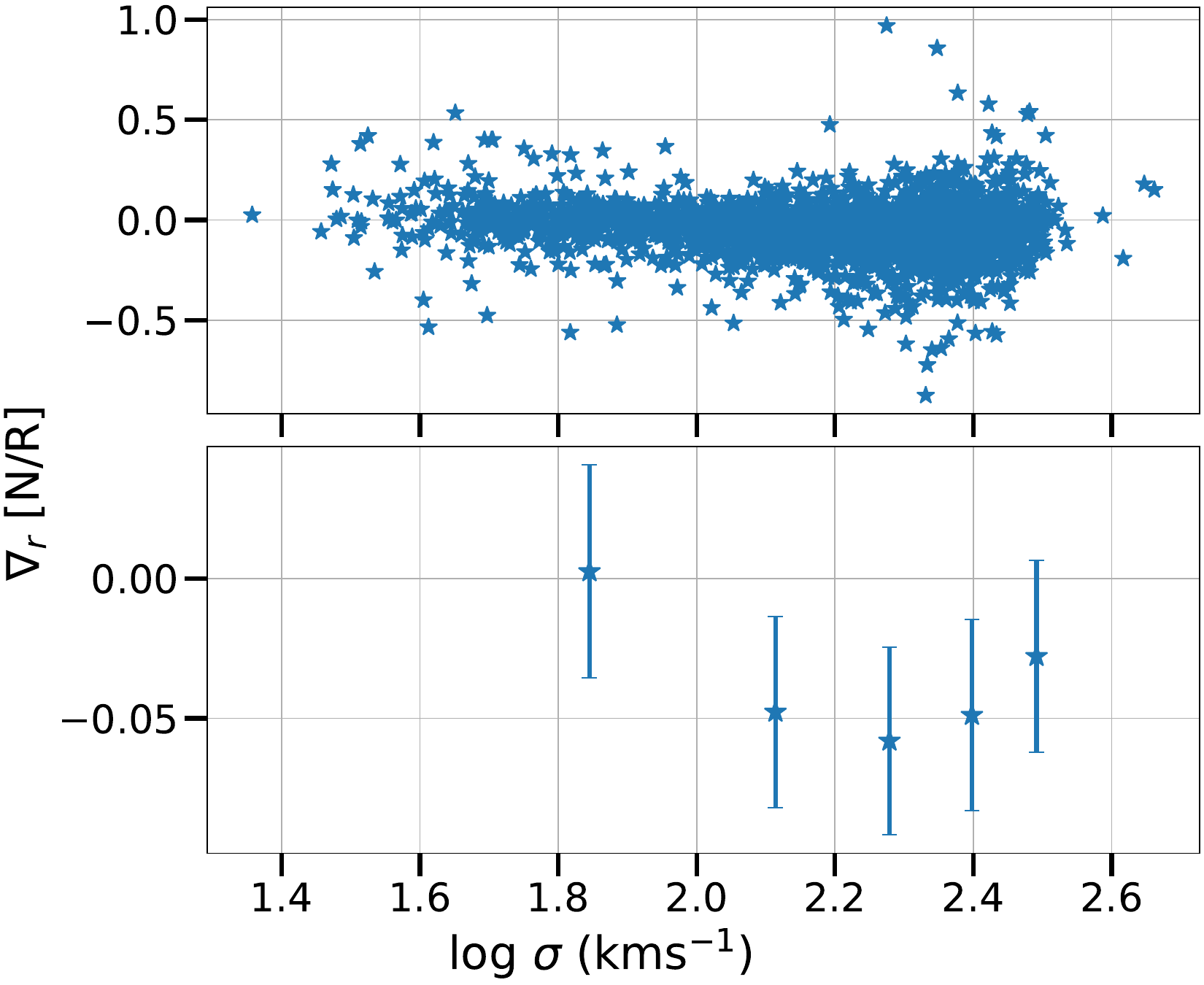}
\caption{Same as Fig.~\ref{fig:gradC_vd} but for N.
\label{fig:gradN_vd}}
\end{figure}

\begin{figure}[ht!]
\plotone{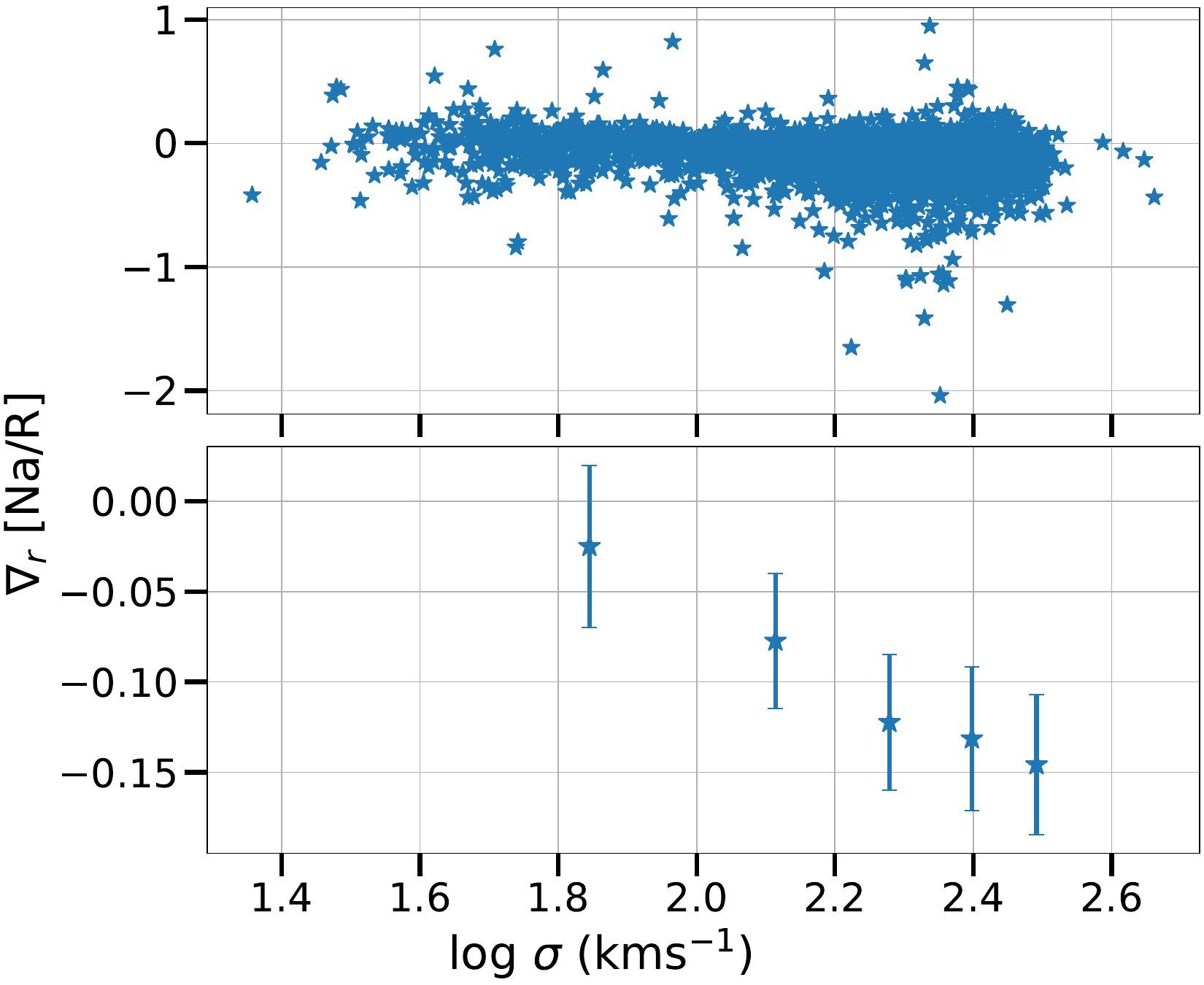}
\caption{Same as Fig.~\ref{fig:gradC_vd} but for Na. 
\label{fig:gradNa_vd}}
\end{figure}

\begin{figure}[ht!]
\plotone{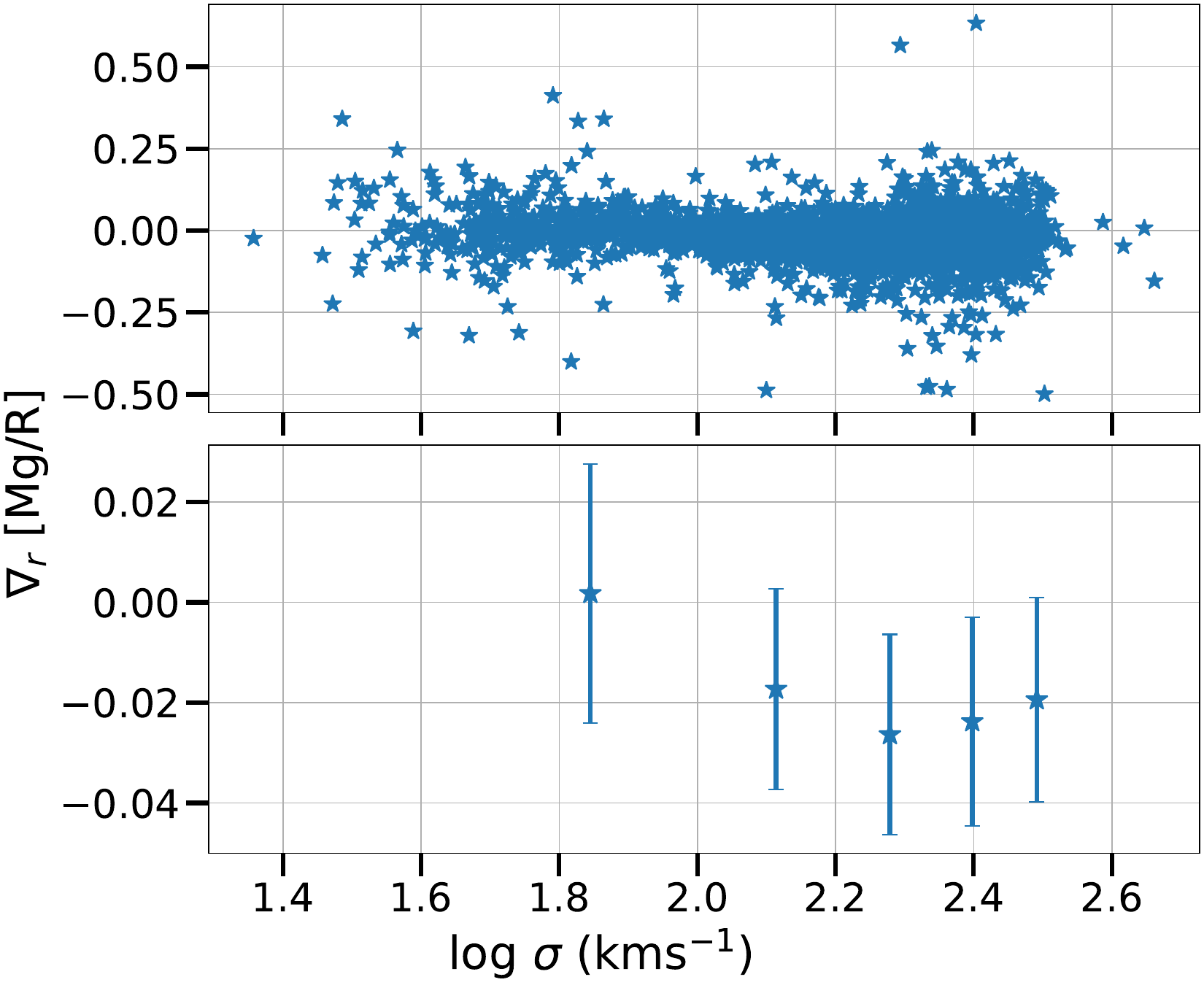}
\caption{Same as Fig.~\ref{fig:gradC_vd} but for Mg. 
\label{fig:gradMg_vd}}
\end{figure}

Figures \ref{fig:gradC_vd}, \ref{fig:gradN_vd}, \ref{fig:gradNa_vd}, and \ref{fig:gradMg_vd} shows the variation of radial gradients for C, N, Na, and Mg, respectively, with respect to $\sigma$. The top panels illustrate individual galaxy radial gradients, while the bottom panels present the mean radial gradients for each $\sigma$ bin. For C, N, and Mg, the mean radial gradients are approximately zero ($\sim$-0.03) across all $\sigma$ bins, showing minimal dependence on $\sigma$. In contrast, Na exhibits a strong dependence of radial gradient on $\sigma$, with both a pronounced variation in gradient and higher absolute gradient values at high $\sigma$ values (Fig.~\ref{fig:gradNa_vd}). Specifically, in the highest $\sigma$ bin of Figure~\ref{fig:gradNa_vd}, a mean negative gradient of 0.15 dex/log R/R$_e$ indicates a decrease in Na abundance by 0.15 dex from the galaxy center to the 5$^{\rm th}$ annulus. Low-$\sigma$ galaxies exhibit near-zero gradients for individual light element abundances, despite predominantly negative [Fe/H] gradients.

\section{Discussion} \label{sec:discussion}

In this work, we derived stellar population parameters (mean light-weighted age, ADF peak [Fe/H], [C/R], [N/R], [Na/R], and [Mg/R]) and their radial gradients for 2968 individual MaNGA galaxies. Here, we review and amplify key results for (1) stellar population parameter trends, (2) gradient trends, and (3) astrophysical scatter. Additionally, in $\S$\ref{subsec:diss_illus} we compare our results with TNG, a galaxy evolution simulation. We compare key results with previous work and mention some broader implications for future focus. 

\subsection{Stellar Population Parameters}\label{subsec:discuss_abun}

\begin{figure}[ht!]
\plotone{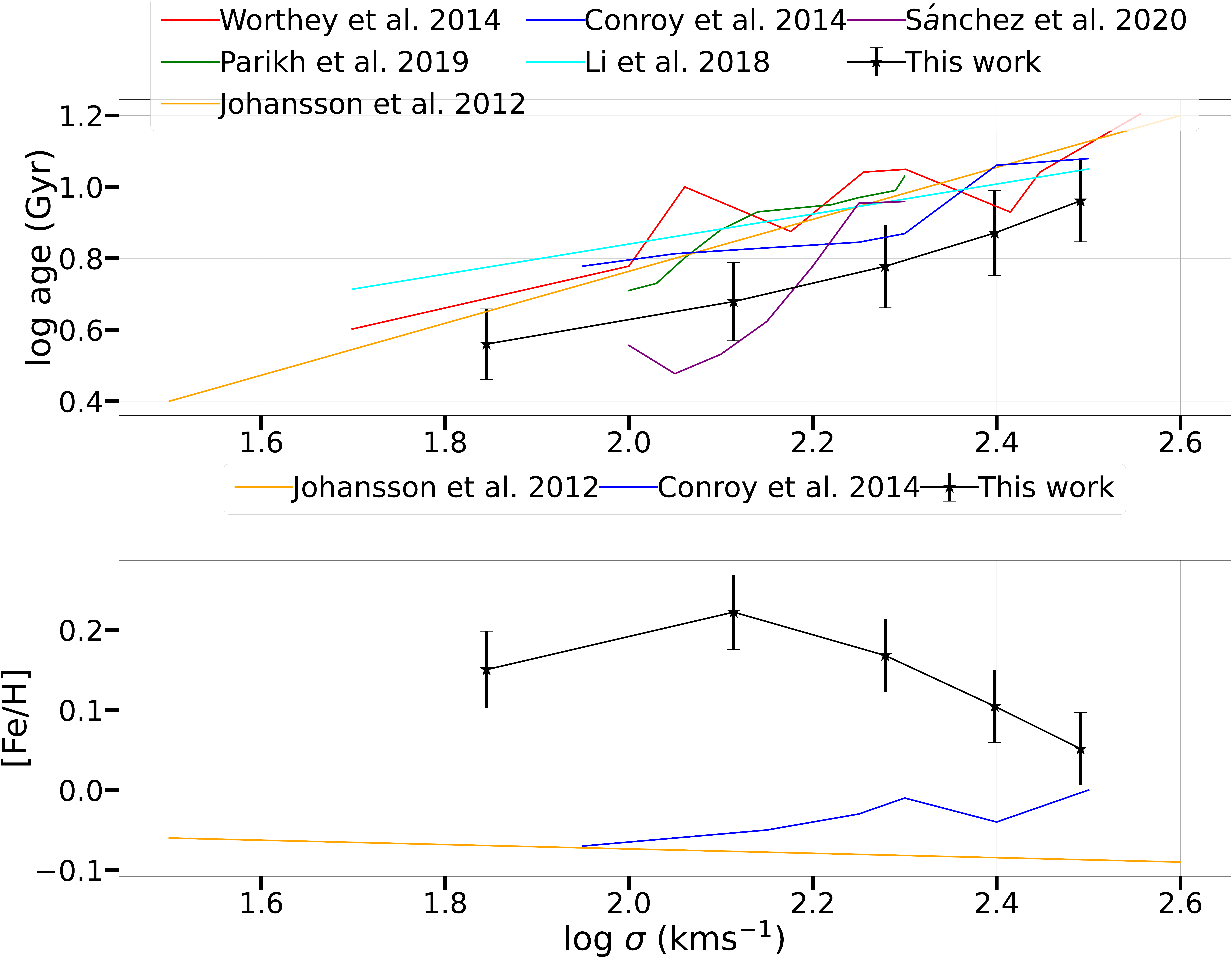}
\caption{Comparison of our results with previous work for log $\sigma$ - log age (top panel) and log $\sigma$ - [Fe/H] (bottom panel) relations. Note that the age plotted here for this work is the mean of age obtained from the 3$^{rd}$ annulus of each galaxy for each $\sigma$ bin.
\label{fig:age_met_lit_comp}}
\end{figure}

Fig.~\ref{fig:age_vd_rad_with_3ann} illustrates the relationship between galaxy age and its $\sigma$. The bottom panel shows the individual age at the third annulus, while the center of each rectangular box in the top panel presents the mean age within each $\sigma$ bin. A positive correlation is evident, with age increasing as a function of galaxy size, as indicated by the slope of 0.53 dex per decade in the bottom panel of Fig.~\ref{fig:age_vd_rad_with_3ann}. In the top panel of Fig.~\ref{fig:age_met_lit_comp}, we compare our log $\sigma$--log age relationship with prior studies. Within the margin of error, our gradient of 0.53 dex per decade aligns closely with values reported in the literature: 0.55 by \cite{2014ApJ...780...33C}, 0.70 by \cite{2014ApJ...783...20W}, 0.73 by \cite{2012MNRAS.421.1908J}, and 0.43 by \cite{2018MNRAS.476.1765L}. However, our result is lower than the 1.07 reported by \cite{2019MNRAS.483.3420P} and 1.1 by \cite{2023MNRAS.526.1022L}. The discrepancy with \cite{2023MNRAS.526.1022L} likely arises from their use of the full MaNGA DR17 galaxy sample without distinguishing between elliptical and spiral galaxies, while \cite{2019MNRAS.483.3420P} employed stacking to derive stellar population parameters, which may account for the observed differences. Also \cite{2020MNRAS.495.2894D} reported a significantly stronger age--$\sigma$ relationship across most $\sigma$ bins, a trend not observed in our analysis. These findings are noteworthy because of the volatile oxygen-age degeneracy \cite{2022MNRAS.511.3198W} along with our assumption that [Mg/R] = [O/R]; with such a large increase in model flexibility, there was no \textit{a priori} reason that the age scale would remain unchanged. Furthermore, \cite{2005ApJ...621..673T} demonstrate that the trend of larger galaxies exhibiting older ages holds across varying environmental densities. These observations indicate that massive ETGs formed their stars earlier in the universe's history and underwent rapid quenching of star formation. This early and efficient shutdown of star formation in massive systems points to feedback processes that were more effective in halting star formation at earlier times. Such findings are consistent with the "cosmic downsizing" scenario \citep{1996AJ....112..839C, 2005ApJ...619L.135J}, in which more massive galaxies complete their star formation sooner than less massive ones. \cite{2019ApJ...872...50L} has shown that the number of MaNGA galaxies experiencing `inside-out' quenching increases with halo mass. This challenges the predictions of simple hierarchical models, which posit a more gradual buildup of massive galaxies through mergers over time.

An ingredient currently missing from the models is blue straggler stars. Inclusion of these bluer stars will increase the inferred ages, likely by Gyr \citep{2000AJ....119.1645T}. The ages as reported here are by no means the last word on the subject.

Shifting our focus to [Fe/H], Fig.~\ref{fig:met_vd_rad_with_3ann} showed how [Fe/H] varies as a function of $\sigma$ for both individual galaxies (3$^{\rm rd}$ annulus; bottom panel) and $\sigma$ bins (center of rectangle; top panel). In conjunction with Fig.~\ref{fig:age_vd_rad_with_3ann}, we can also conclude that older galaxies tend to have slightly lower [Fe/H]. We emphasize again that due to innovations in our modeling (isochrones that shift in temperature for individual elements and metallicity-composite base models) our results should not necessarily track previously published results. That said, \cite{2000AJ....120..165T} echoes our finding that for a given $\sigma$, older galaxies are found to have low [Fe/H]. In the bottom panel of Fig.~\ref{fig:age_met_lit_comp}, we compare our log $\sigma$ - [Fe/H] relation with that of \cite{2012MNRAS.421.1908J} (orange line) and \cite{2014ApJ...780...33C} (blue line). Neither of these works report a `kink' in this relation. Although \cite{2012MNRAS.421.1908J} show a flat to negative dependence of [Fe/H] with $\sigma$, the opposite trend is shown by \cite{2014ApJ...780...33C}.

The behavior of decreasing (or flat) [Fe/H] with $\sigma$ can be accommodated within current galaxy evolution theory. A possible factor that might reduce [Fe/H] in galaxies is a high star formation rate (SFR) of the galaxy \citep{2010A&A...519A..31L} followed by swift quenching before Type \Romannum{1}a products contribute. \cite{2008MNRAS.385..147D} have shown that galaxies with higher stellar mass tend to have higher SFR within a redshift ($z$) range of 0-2. The higher SFR would logically produce stronger outflows within a galaxy that further reduces its [Fe/H] \citep{2016A&A...588A..41C}, although a deeper gravitational potential well would work against that. If star formation occurs in stronger bursts in more massive (and thus older) ETGs, then massive ETGs formed stars early in the Universe and then swiftly quenched star formation, on average. Thus, high SFR in high $\sigma$ ETGs plus stronger quenching winds lead to less chemical evolution and slightly lower [Fe/H] compared to low $\sigma$ galaxies, yielding a slightly negative trend in Fig.~\ref{fig:met_vd_rad_with_3ann}. These high $\sigma$ galaxies have higher light element abundances (Fig.~\ref{fig:abun_vd_rad}), but one could posit that the time delay between Type \Romannum{2} supernova enrichment and delayed Type \Romannum{1}a supernova enrichment makes enriched gas produced by the latter more susceptible to ejection by galactic winds.

In Fig.~\ref{fig:abun_vd_rad} we see a trend of increasing abundance with $\sigma$ for all light elements explored in this work (C, N, Na, and Mg). Previous work on elemental abundances in ETGs shows a very similar trend \citep{2021MNRAS.502.5508P, 2012MNRAS.421.1908J, 2014ApJ...780...33C, 2007ApJ...671..243G}. Here we briefly discuss each element separately as each of them has different origin.

Nitrogen primarily originates from the CNO cycle in stars, utilizing carbon, while carbon is predominantly produced via the triple alpha process, with supernovae contributing to both elements. Stellar winds in late evolutionary stages enrich the interstellar medium with these elements. Carbon production dominates in lower-mass stars (1--3 M$_\odot$), whereas nitrogen requires higher-mass stars (4--8 M$_\odot$) \citep{2003MNRAS.339...63C}. In Milky Way stars, [C/Fe] remains near zero across [Fe/H], while [N/Fe] increases with metallicity \citep{2018AJ....156..126J}. Our analysis reveals nitrogen under-abundance by approximately 0.1 dex relative to carbon and solar ratios, consistent with findings by \cite{2014ApJ...780...33C} for carbon and nitrogen abundances and gradients.

Not included in the models explicitly is the effect of first dredge-up on the first-ascent red giant branch. Nitrogen abundance increases by $\sim0.4$ dex relative to C on the upper RGB [see, e.g., Fig 2 of \citet{2016MNRAS.456.3655M}]. If this effect were included, nitrogen's spectral signatures would strengthen and the inferred N abundance would drop even lower in Fig.~\ref{fig:abun_vd_3ann}, putting [N/C] and [N/Fe] significantly low compared to the Milky Way. The effect \textit{is} included implicitly, in that Milky Way stars were used to empirically zero the spectral library, and this effect is built in, masking as a surface gravity effect. That said, and keeping also in mind the N-rich globular cluster system in M31 that contrasts so strongly with our own cluster system \citep{1998PASP..110..888W} and the small dynamic range of current galaxy formation model predictions (Fig.~\ref{fig:ill_man_N}), it is safe to say that N is telling us much about chemical evolution that we are not yet equipped to interpret.

We see very similar abundances trajectories for sodium and magnesium (see Fig.~\ref{fig:abun_vd_rad}). Although these two elements have different nucleosynthetic origins, both Na and Mg are mostly sourced from type \Romannum{2} supernova (SNII) explosions \citep{2001A&A...370..194M, 2000AJ....119.1645T} with significant contributions from the $\alpha$-rich freezout phase of the explosion \citep{1995ApJS..101..181W}, so their nucleosynthetic origin \textit{site} is similar. The `locked-in' state of ETGs in terms of chemical abundances preserve this signature of Na and Mg over time. 

A remarkable finding from this work is shown in Fig.~\ref{fig:abun_vd_3ann}. The effect shows more clearly because we did not stack spectra. What we see is that there is a very prominent increase in slope of chemical abundances when plotted against $\sigma$ for log($\sigma$)$>$2.0. This slope change seems not to have been mentioned in previous literature on this subject. Possibly related, however, is that \citet{2005MNRAS.362...41G} found a transition from low mass to massive galaxies in the mass-metallicity relation at around log($\sigma$)=2.3.  Evidently, low mass galaxies are less efficient in producing light metals and retaining them compared to massive galaxies. Different factors could lead to this, including star formation efficiency, galactic winds, and the IMF. 

Figs.~\ref{fig:abun_met_vd} (grouped by annulus) and \ref{fig:abun_met_3ann} (third annulus per galaxy) depict elemental abundance dependence on [Fe/H]. A slight negative trend is observed, with light metal abundances decreasing as [Fe/H] increases. This trend in ETGs mirrors the Milky Way’s for magnesium but differs qualitatively for [C/R], [N/R], and [Na/R]. Simulations by \cite{timmes_gal_evol} indicate that carbon and nitrogen remain constant over a wide [Fe/H] range, while sodium and magnesium decrease with increasing [Fe/H]. The decreasing trend of light metal abundances with [Fe/H] likely stems from iron production in type \Romannum{1}a supernova explosions (SNIa), which reduces [X/Fe] while increasing [Fe/H]. For magnesium, \cite{10.1093/mnras/sty534} and \cite{2012ApJ...760...71C} report a consistent decrease in [Mg/Fe] with increasing [Fe/H], corroborated by \cite{2001A&A...370.1103A}.

These chemical signatures of increasing light element abundance and decreasing (or flat) [Fe/H] with $\sigma$, coupled with high SFR and strong outflow, are consistent with `inside-out' growth scenario during their formation. Their central regions likely underwent a brief, intense burst of star formation, rapidly enriching them in light metals and quenching soon after, leading to the high [X/Fe] ratios. More information can be obtained by studying the radial gradients in the abundances of iron-peak elements and other light metals.

\subsection{Radial Gradients in Age and Abundances}\label{subsec:discuss_grad}

The radial gradients in different stellar population parameters provide insight into the formation and evolution of galaxies. In this work, in addition to looking at the trends of stellar population parameters with $\sigma$, we also investigated their radial gradients trend with $\sigma$. 

\begin{figure}[ht!]
\plotone{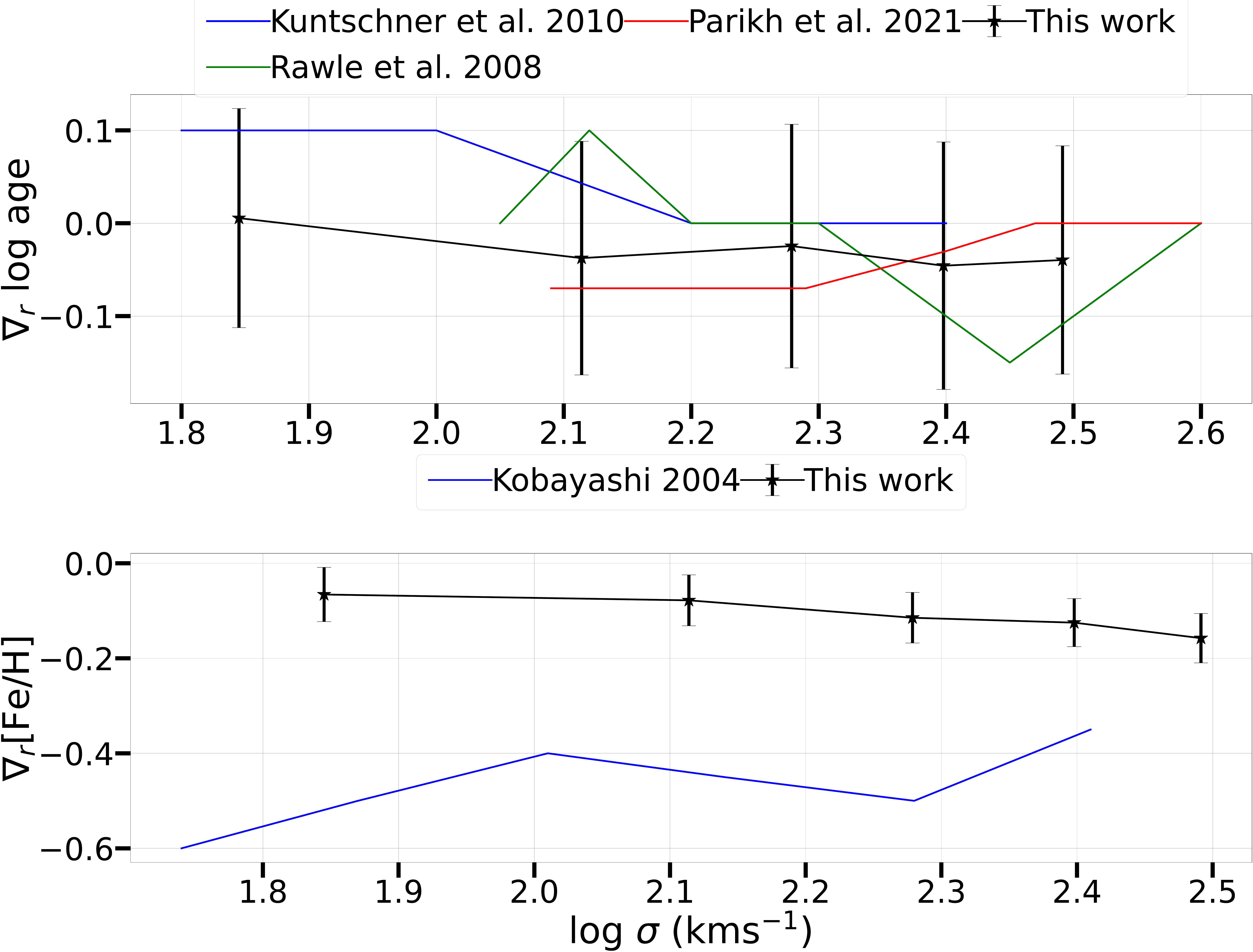}
\caption{Comparison of our results with various other previous work for log $\sigma$ - $\nabla_r$log(age) (top panel) and log $\sigma$ - $\nabla_r$[Fe/H] (bottom panel) relations.
\label{fig:age_met_grad_lit_comp}}
\end{figure}

In Fig.~\ref{fig:age_rad_vd} we show how the mean age for each $\sigma$ bin changes radially from the center of the galaxy to 0.65 R$_e$. We find a modest radial gradient of $\sim$-0.04 log(Gyr)/log(R/R$_e$) in mean age across different $\sigma$ bins, but the scatter is much larger than the mean trend. This result is also validated through Fig.~\ref{fig:gradage_vd}. \cite{2003A&A...407..423M} finds a median value of 0.0 in radial gradient in age for their Coma early-type galaxies. \cite{2007MNRAS.377..759S} reports that 10 out of their 11 galaxies showed zero radial gradient in age. On the top panel of Fig.~\ref{fig:age_met_grad_lit_comp}, we have compared our log $\sigma$ - $\nabla_r$ log age relation with some of the previous work by \cite{2010MNRAS.408...97K}, \cite{2008MNRAS.389.1891R}, and \cite{2021MNRAS.502.5508P}. Our result, showing a nearly flat radial gradient in age, is consistent with these results within the scope of error. Most of the previous studies on radial gradient in age consider ages out to R=R$_e$, but in this work we calculated spatial information to R=0.7 R$_e$. A lack of age gradient might imply that stellar populations in ETGs were formed in bursts that spanned large swaths of galactocentric radius. However, it needs to be remembered that our results apply to the average. Few galaxies conform to the average. Slightly more than half are younger in the outskirts (0.7 R$_e$ for this study) and slightly less than half are younger in the center. Only a minority of galactic accretion events drive gas to galaxy centers, it seems. This result of nearly flat radial gradient in age (on average) indicates that stars likely formed in the inner regions and were subsequently mixed outward via dynamic processes, such as mergers and radial migration. These rapid, violent events in massive galaxies likely homogenized stellar populations across their radii, diminishing initial age gradients.

We find radial gradients in [Fe/H] $\approx$-0.12 [dex per decade, [Fe/H]/log(R/R$_e$)] for the middle $\sigma$ bin with substantial dependence on $\sigma$ (see Fig.~\ref{fig:gradmet_vd}). \cite{2003A&A...407..423M} reports a value of -0.16 for radial gradient in [Z/H] (their Figure 7) and \cite{2021MNRAS.502.5508P} shows an average radial gradient of around -0.19 in [Z/H] (their Figure 13). An unambiguous comparison was not possible between our work and theirs as we have used [Fe/H] as proxy for metallicity and most of the other work use [Z/H] as proxy for metallicity. However, direct comparison with results from \cite{2004MNRAS.347..740K} are shown in Fig. \ref{fig:age_met_grad_lit_comp}. The bottom panel of Fig.~\ref{fig:age_met_grad_lit_comp} shows that our radial gradient in [Fe/H] is on average a factor of two shallower than \cite{2004MNRAS.347..740K}. Earlier estimates averaged a slope of $-0.2$ dex per decade for medium and large ETGS \citep{1999PASP..111..919H}. We see a considerable scatter around zero (see top panel of Fig.~\ref{fig:gradmet_vd}) with values as negative as -0.8 [Fe/H]/log(R/R$_e$) and as positive as 0.4 [Fe/H]/log(R/R$_e$) (not considering outliers). Our trend of radial [Fe/H] gradient with $\sigma$ qualitatively matches with \cite{2009ApJ...691L.138S} at low $\sigma$ values (log($\sigma$)$<$2.0) and with \cite{1999ApJ...513..108B} for higher $\sigma$ values. A negative [Fe/H] gradient favors an `inside-out' mode of chemical evolution in ETGs if star formation and enrichment occur on a timescale similar to the accretion itself. This conclusion is similar to that of \cite{2017MNRAS.466.4731G} where they find similar results with 505 ETGs. The [Fe/H] radial gradient reported here could also be explained as a consequence of hierarchical merging events. If they involve gas, mergers can trigger intense central starbursts, which significantly enrich the core regions with metals leading to a metallicity distribution where the central areas exhibit higher metal content than the outer regions \citep{2004MNRAS.347..740K}. If not subsumed into star formation right away, gas loses energy and angular momentum, funneling into the central regions where concentrated star formation enriches the core with metals more significantly than the outskirts \citep{2014MNRAS.439..990M}. The gradual increase of [Fe/H] radial gradient with $\sigma$ can also be explained by the amplification of these processes for larger galaxies. The monolithic collapse hypothesis, on the other hand, predicts a much higher [Fe/H] gradient \citep{2001ApJ...558..598K, 2010MNRAS.407.1347P} than what is reported here. If monolithic collapses are a common formation channel, then we also require major ``dry'' merger events to explain the much flatter slope in [Fe/H].

Using a linear scale for R/R$_e$, the negative radial gradient in age is 0.04 dex per R/R$_e$ for the lowest $\sigma$ bin and remains constant at an increased value of 0.11 dex per R/R$_e$ for the rest of the $\sigma$ bins. We quantify the goodness of fit for both logarithmic and linear cases using the coefficient of determination (the R$^2$ value). The R$^2$ values for each $\sigma$ bin for the linear case are (in order of increasing $\sigma$): 0.52, 0.93, 0.97, 0.94, and 0.95. On the other hand, for the logarithmic case, the R$^2$ values are: 0.26, 0.70, 0.77, 0.71, 0.85. Thus, in general, the linear case shows better fit. Similar analyses were also performed for the radial gradients in [Fe/H]. The negative radial gradient in [Fe/H] (using linear scale for R/R$_e$) increases from 0.12 dex per R/R$_e$ to 0.23 dex per R/R$_e$ with R$^2$ values of 0.96, 0.97, 0.97, 0.98, and 0.98 for different $\sigma$ bins (in order of increasing $\sigma$). Using the logarithmic scale the R$^2$ values are 0.79, 0.81, 0.91, 0.93, and 0.84, again showing better fit for the linear case. Thus for both age and [Fe/H] we find that a linear physical scale in kpc connects to the spatial dependencies of chemical evolution better than a logarithmic one. We speculate that power-law dependencies such as depth of potential well or dynamical crossing time are less important than turbulence or other processes with a relatively fixed sphere of influence.

Fig.~\ref{fig:abun_rad_vd} shows the radial trends of all of the light elements explored in this work. We discuss each of the elements individually.

Although carbon (C) and nitrogen (N) are not classified as $\alpha$ elements, their radial gradients mirror the qualitative trend of magnesium (Mg). Figs.~\ref{fig:gradC_vd} and \ref{fig:gradN_vd} depict the radial gradients of C and N, respectively, across all $\sigma$ bins. Notably, for massive early-type galaxies (ETGs) with log $\sigma > 2.0$, the radial gradients of C and N exhibit flattening with respect to $\sigma$, albeit with minimal gradient amplitude. This aligns with \cite{2021MNRAS.502.5508P}, who reported nearly flat gradients for both elements across most mass bins, and \cite{2021ApJ...923...65F}, who noted a flat gradient for C and a slightly negative gradient for N. Within uncertainties, our findings are consistent with these studies. Furthermore, like \cite{2021ApJ...923...65F}, we find no significant correlation between C and N radial gradients and [Fe/H] (figure not included).

The strongest gradient values are obtained for sodium. Fig.~\ref{fig:gradNa_vd} shows that the negative radial gradient in Na increases monotonically with $\sigma$. The median negative gradient of around 0.10 dex per decade ([Na/R]/log R/R$_e$) in this work is similar to a number of previously reported values \cite{2017MNRAS.464.3597L, 2021MNRAS.502.5508P}.  \cite{2021ApJ...923...65F} reports strong negative radial gradient in [Na/R] for individual ETGs, but their [Na/R] values are higher than what we report in this work. Given the overall similarity between Mg and Na trends \textit{among} ETGs, the $\sim2$ times stronger Na gradients \textit{within} ETGs are puzzling. 

Let us briefly consider absorption of Na in the cold interstellar material (ISM). The Na D index measures a pair of transitions whose lower state is the ground state, i.e., a resonant feature in the ISM. Absorption along lines of sight out of the Milky Way plane is on the order of tenths of \AA\ \citep{1986A&A...166...83B} compared with observed ETG Na D strengths of $\sim$3 \AA . While more massive ETGs tend to harbor more cool ISM \citep{2010ApJ...725..100W, 2014MNRAS.439.2291W} in the outskirts, the amount of Na would have to be considerable. Also, more absorption in the outskirts would lead to weaker, not stronger, negative radial gradients. All things considered, we judge that cold interstellar Na is not a plausible explanation for the [Na/R] gradient trends in ETGs.

Magnesium (Mg) is considered an $\alpha$-element that is predominantly produced inside massive stars via fusion of helium nuclei. In this work, Mg shows negative radial gradient for all $\sigma$ bins (qualitatively in Fig\ref{fig:abun_rad_vd} and quantitatively in Fig.~\ref{fig:gradMg_vd}). Early work by \cite{1993MNRAS.262..650D} showed that the radial gradient in the Mg$_2$ index has a value of around -0.06 which can be translated roughly to similar radial gradient values in [Mg/R]. Our highest $\sigma$ bin has a mean radial gradient value of around -0.02 dex per decade ([Mg/R]/log R/R$_e$) and gradually decreases to a flat gradient for the lowest $\sigma$ bin. Many previous studies reported a radial gradient value of near -0.06 \citep{2008MNRAS.389.1891R, 2019MNRAS.489..608F, 2021ApJ...923...65F}, whereas \cite{2003A&A...407..423M} and \cite{2021MNRAS.502.5508P} report zero radial gradient for [Mg/R]. In the average ETG, if there is such a thing, the Mg abundance is global; independent of radial distances from the center of the galaxy.

The nearly flat radial gradients of C, N, and Mg in ETGs support a hierarchical assembly scenario, where repeated mergers and accretion events mix stars and gas and spatially homogenize the stellar populations. This contrasts with the steep gradients expected from a simple monolithic collapse. The shallow light element gradients favor mergers as agents to homongenize ETGs and favor their formation as extended, multi-phase events rather than rapid, single-collapse systems.

\subsection{Scatter in Astrophysical Parameters and their Radial Gradients}\label{subsec:diss_scatter}

Once observational effects are subtracted, the scatter of a derived quantity within a group of galaxies reveals the extent of variation within the population allowed by natural evolutionary pathways. With over 2000 galaxies, our sample size should not be a concern, and the extent of natural variation should be well-measured. In this section, we show results of our calculated intrinsic scatter ($\delta_i$) and how they correlate with $\sigma$. In general, we find that observational errors are small enough that almost all of the scatter in previous figures comes from the intrinsic scatter.

Intrinsic scatter ($\delta_i$) is calculated via
\begin{align}
\label{eqn:delta}
    \delta_T^2=\delta_i^2+\delta_o^2 ,
\end{align}
where $\delta_T$ is the total scatter in a derived quantity and $\delta_o$ is the variation in that quantity given the observational scatter. $\delta_T$ for each $\sigma$ bin is the standard deviation in the astrophysical parameter or radial gradient within that $\sigma$ bin. 

Optimal estimation of $\delta_o$ relies on repeated observations of the same target, which MaNGA does not provide. However, MaNGA yields per-spectral-pixel uncertainty, readily convertible to uncertainty per measured index value. We assume that spectral response ripples over 10\AA\ to 100\AA\ scales, a potential error source for some spectrographs, are negligible for MaNGA. To estimate $\delta_o$, we randomly perturbed each index measurement within its Gaussian probability distribution and recalculated {\sc compfit} solutions. The mean was derived from the as-measured solution, with 50 variations (MC realizations)) generated per annulus to approximate the standard deviation for each measurement. For radial gradient measurement errors, we utilized the covariance matrix error obtained during gradient calculations. 

To get to $\delta_o$ from the errors from MC realizations or gradient errors, we use 
\begin{align}\label{eqn:delta_o}
    \delta_o=\sqrt{\frac{1}{\sum_{i=1}^n\Delta_i^{-2}}}\times\sqrt{N} ,
\end{align}
where $\Delta_i$ is the error from MC realizations or gradient errors for a particular galaxy's measurement within a $\sigma$ bin and N is the total number of galaxies within that same $\sigma$ bin. Armed with these formulae we calculated the $\delta_i$ for the age, [Fe / H], C, N, Na, and Mg estimates and their radial gradients for each $\sigma$ bin.

\begin{figure}[ht!]
\plotone{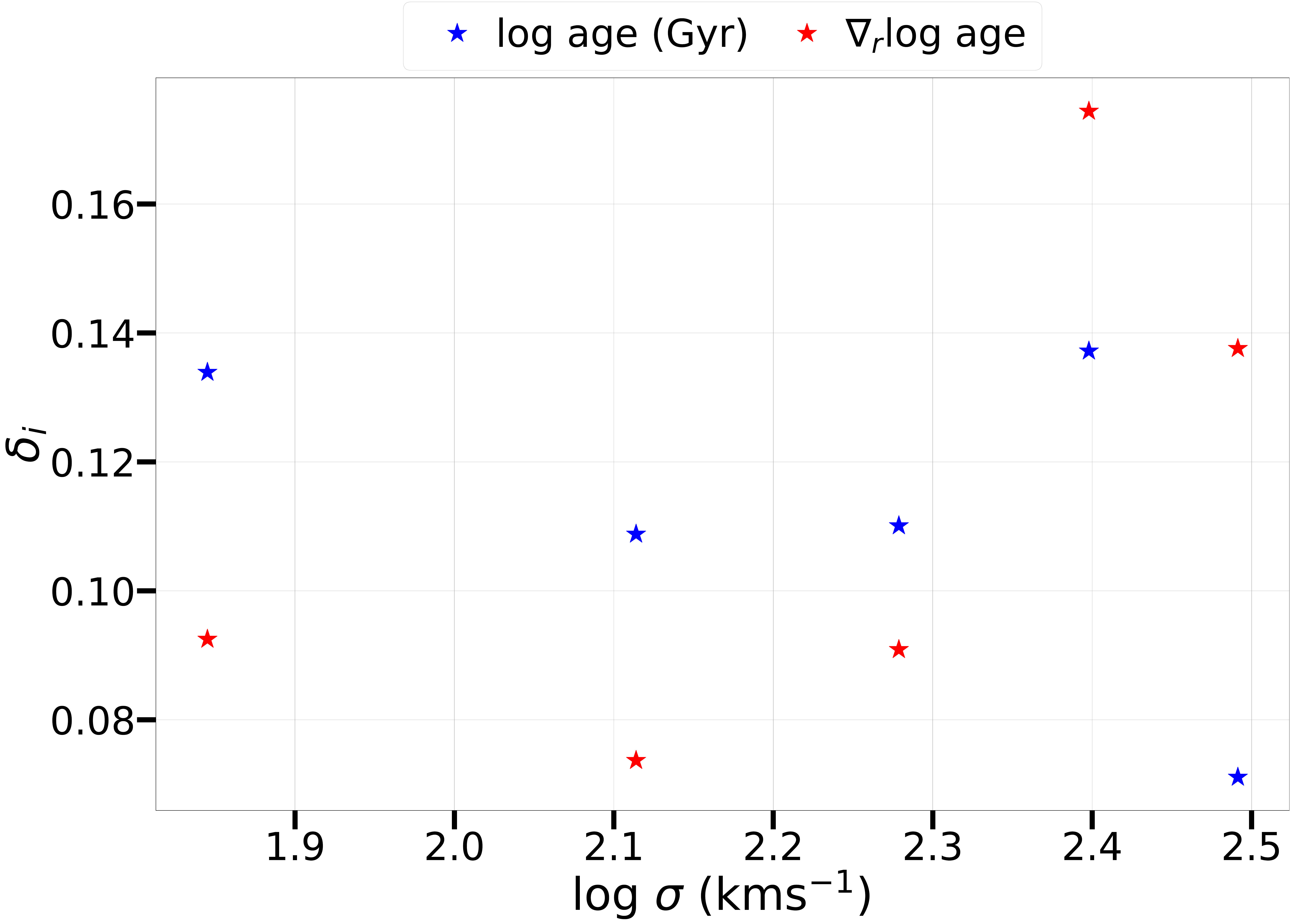}
\caption{The intrinsic scatter ($\delta_i$) in age (blue) and in radial age gradient (red) plotted against $\sigma$. The unit associated with age is dex and that with radial gradient is dex/log(R/R$_e$).
\label{fig:delta_age_gradage}}
\end{figure}

In Fig.~\ref{fig:delta_age_gradage} we show the calculated $\delta_i$ for the age and radial gradient in age. The figure shows a slightly decreasing trend with $\sigma$ for age scatter whereas the opposite is observed for age gradient scatter. The flat gradients in low-$\sigma$ galaxies show little scatter ($\sim$0.09 dex/log(R/R$_e$)), while the almost-flat age gradients in high-$\sigma$ galaxies show slightly larger intrinsic scatter ($\sim$0.15 dex/log(R/R$_e$)). Sliced the other way, low-$\sigma$ galaxies are well-mixed radially, but globally show substantial age scatter while high-$\sigma$ galaxies have small but measurable age gradients, both negative and positive, but globally are older and more homogeneous in their senescence.

The slightly contradictory observation that the most age-homogeneous galaxies also have one of the highest astrophysical spreads in age gradient actually fits well with the scenario where the more massive galaxies are formed by large number of merger events compared to smaller galaxies which are considered to have gone through fewer merger events. \cite{2007ApJ...665..265F} showed that the number of red galaxies (ETGs) has increased from $z\sim$1 to the present day ($z$=0). In order to explain this scenario, the authors posit that migration of galaxies to the red sequence is caused by both quenching and merger events. We also support a more homogeneous age distribution for galaxies with high $\sigma$. In a merger-dominated scenario, the placement of young stars post-merger (nearer to the center or nearer to the edges) might depend on gas content, kinematic initial conditions, and the ages of the merger components. That is, it is plausible that increased merging causes larger scatter in age radial gradients.

At the low-$\sigma$ end, if these galaxies are objects that escaped becoming incorporated into larger units, they are missing an important quenching mechanism (merging) and therefore might be expected to have a large scatter in mean age, as we observe.

Some galaxy formation scenarios posit that a gas collapse forms strong abundance gradients in galaxy subunits, many of which later merge and therefore lessen the gradients. These scenarios predict the weakest gradients should be seen in high-$\sigma$ galaxies. If gas is involved in the mergers, high-$\sigma$ galaxies should also show the highest abundances and youngest ages. None of these predictions are consistent with our results. TNG hierarchical simulations, on the other hand, reproduce the gross age trend with $\sigma$, show the abundance slope changes at log($\sigma$)$\approx$2.0 that we observe, and exhibit flat gradients similar to those we derive, although the abundances predicted by TNG seldom exactly track the observations. The TNG comparison is discussed below in Sec.~\ref{subsec:diss_illus}.

\begin{figure}[ht!]
\plotone{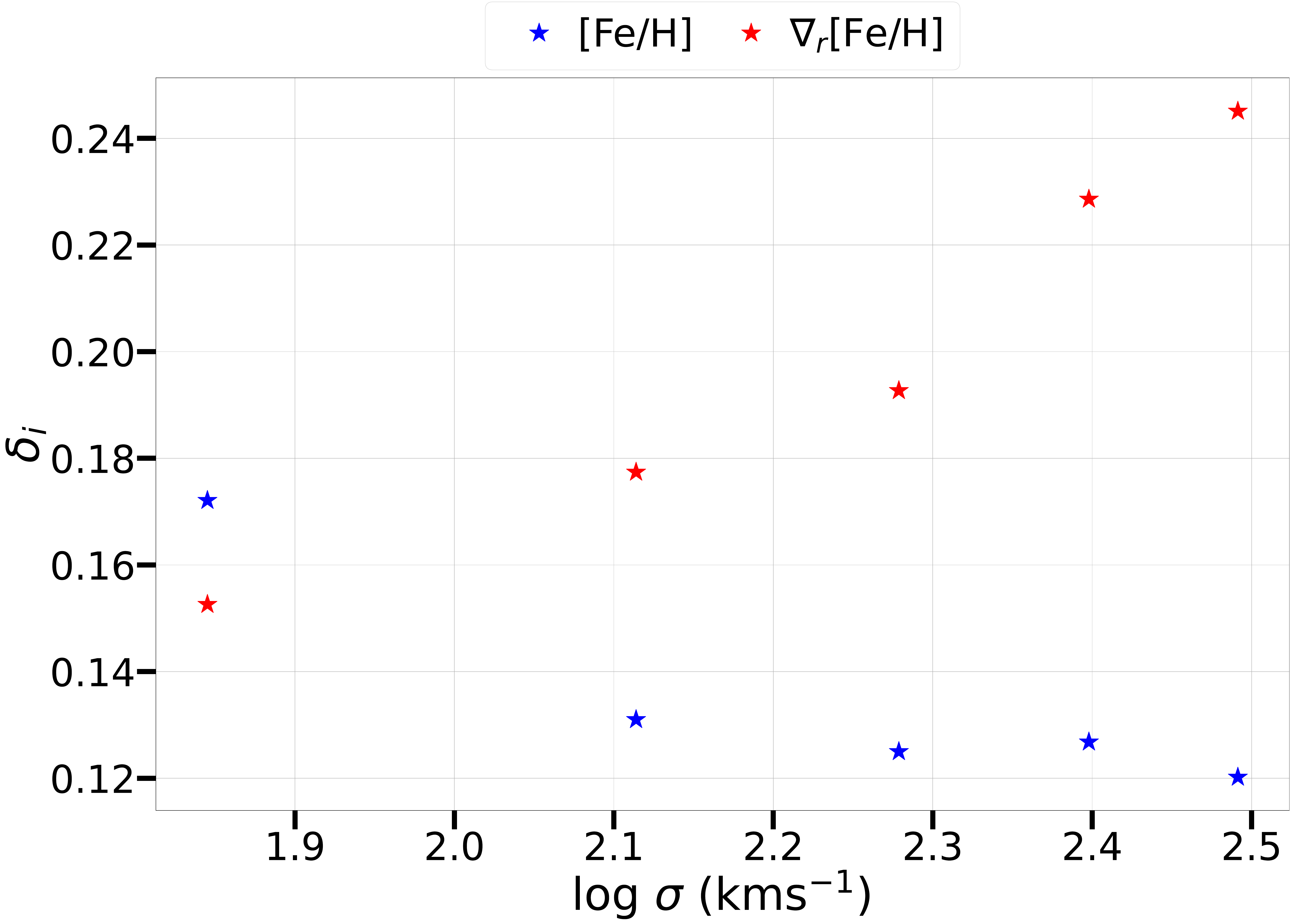}
\caption{The intrinsic scatter ($\delta_i$) in [Fe/H] (blue) and in radial gradient of [Fe/H] (red) plotted with $\sigma$. The unit associated with [Fe/H] is dex and that with radial gradient is dex/log(R/R$_e$).
\label{fig:delta_met_gradmet}}
\end{figure}

Fig.~\ref{fig:delta_met_gradmet} shows the variation of $\delta_i$ in [Fe/H] and corresponding radial gradients for all of our defined $\sigma$ bins. As with mean age, a decreasing trend of $\delta_i$ in [Fe/H] can be seen with increasing $\sigma$, paired with an increase of variation in gradient. The arguments just articulated for support of a $\sigma$-dependent merger likelihood apply in the case of [Fe/H] as well.

\begin{figure}[ht!]
\plotone{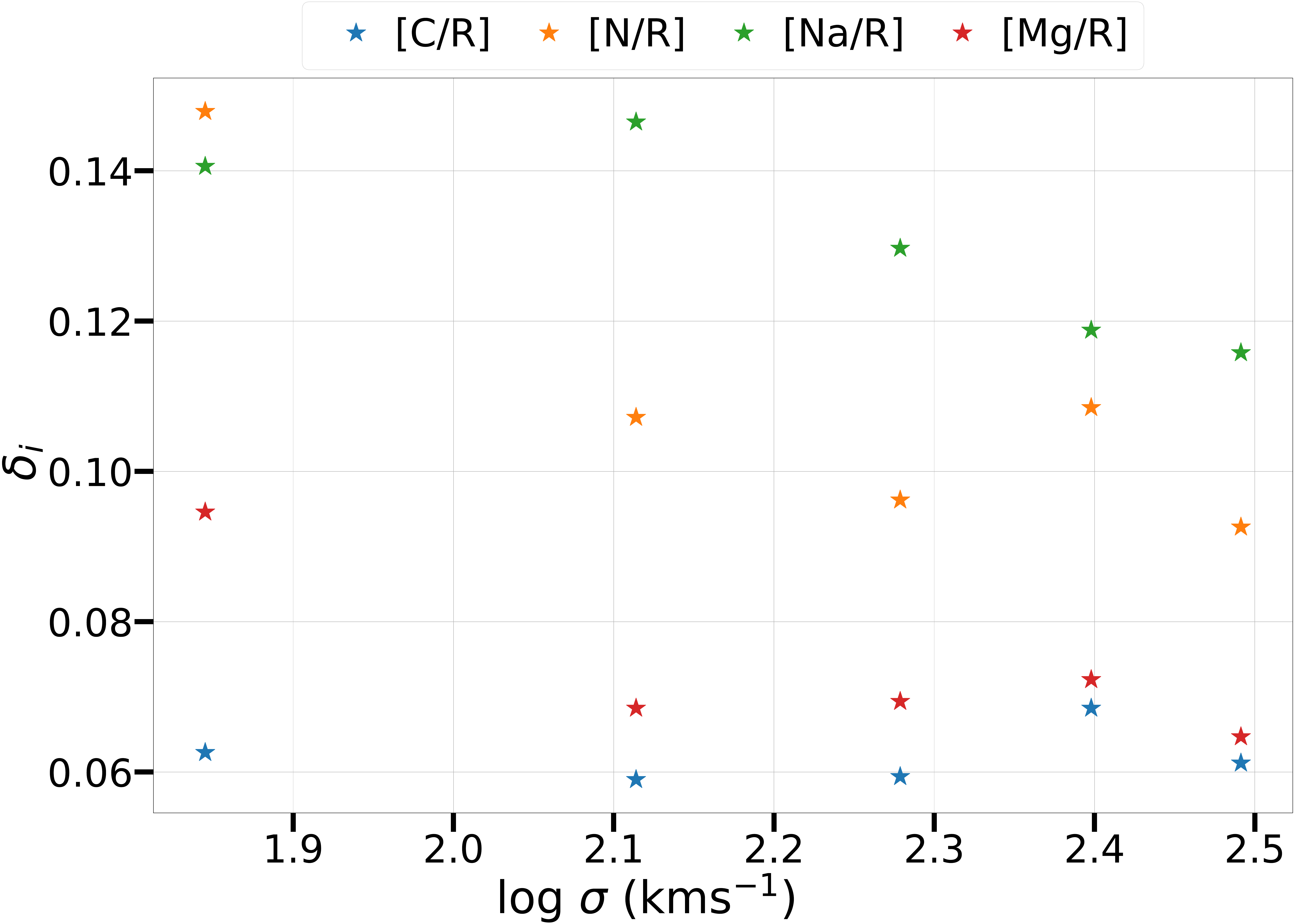}
\caption{The intrinsic scatter ($\delta_i$) in abundances of C (blue), N (orange), Na (green), and Mg(red) plotted with $\sigma$. The unit is dex for $\delta_i$.
\label{fig:delta_abun}}
\end{figure}

Fig.~\ref{fig:delta_abun} shows the $\delta_i$ values in C, N, Na, and Mg abundances. The intrinsic scatter in C abundance is the least, with Mg a close second. Apart from the first $\sigma$ bin, N abundance has nearly constant $\delta_i$ of 0.1 dex for all the other $\sigma$ bins. The high value of $\delta_i$ in N for smaller galaxies (log($\sigma$)$<$2.0) might shed light on the time-scale of chemical enrichment in them. If N from intermediate-mass stars is important, a variation in N abundance implies a variation in formation timescale. It is plausible that these variations would be attenuated in galaxies with richer merger histories. The $\delta_i$ for the Na abundance declines from $\sim$0.14 dex for small galaxies to $\sim$0.12 dex for large ones. Unique to Na is that additional absorption of Na in the interstellar medium (ISM) could increase inferred [Na/R] in some galaxies and increase scatter.   

\begin{figure}[ht!]
\plotone{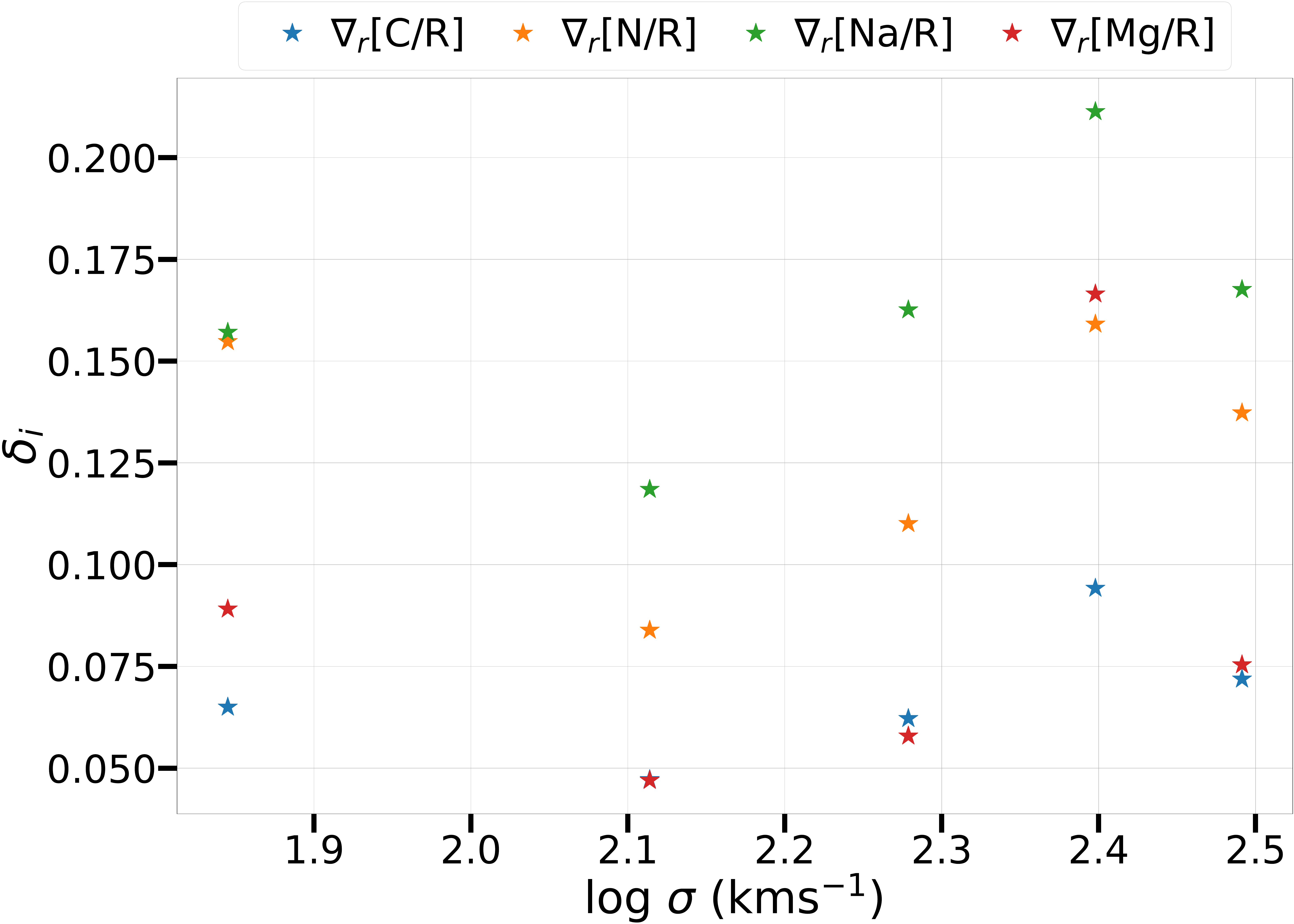}
\caption{The intrinsic scatter ($\delta_i$) in radial gradients for C (blue), N (orange), Na (green), and Mg(red) abundances plotted with $\sigma$. The unit is dex/log(R/R$_e$) for $\delta_i$.
\label{fig:delta_gradabun}}
\end{figure}

The $\delta_i$ values for radial gradients in C, N, Na, and Mg abundances appear in Fig.~\ref{fig:delta_gradabun}. The most interesting result is the increasing trend of $\delta_i$ from the second $\sigma$ bin onwards till the penultimate $\sigma$ bin, reaching a value of almost a factor of two higher compared to the second $\sigma$ bin. Surprisingly, the $\delta_i$ in all the elements for the highest $\sigma$ bin drops to a lower value. Radial gradients in C and Mg have the lowest values in $\delta_i$ followed by N and then Na. This similar trend is also seen in Fig.~\ref{fig:delta_abun}. It is noteworthy that radial gradients in Na have significantly higher $\delta_i$ compared to other elements considered in this work reaching as high as 0.22 dex/log(R/R$_e$) in the fourth $\sigma$ bin. In conjunction with the Fig.~\ref{fig:gradNa_vd}, it can be stated that there is a significant and as-yet-unexplained anomaly regarding Na, which has not been sufficiently addressed in the existing body of literature. ISM absorption may play a small role, but the overall coherence of Na with Mg and other elements argues that ISM absorption is the exception rather than the rule.

Table \ref{table:bins_delta} summarises the results for our $\delta_i$ calculations. The $\delta_i$ values are rounded to two decimal places in the table. 

\begin{figure}[ht!]
\plotone{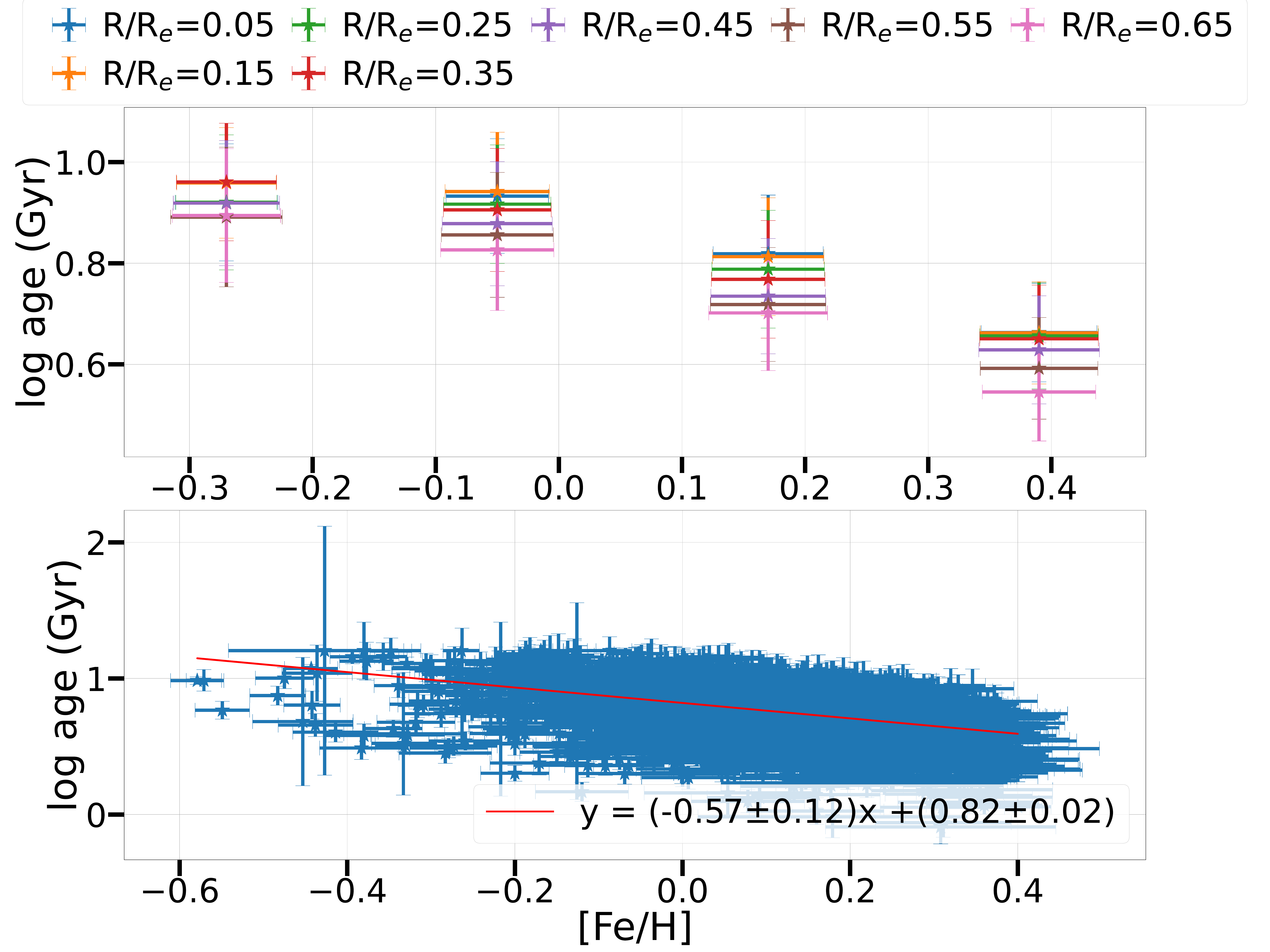}
\caption{The (log of) mean age is plotted against [Fe/H] with errorbars. The top panel shows the mean values for each [Fe/H] bin ([Fe/H] bins are described in Table \ref{table:bins_met}) color coded by annuli. The legend on top shows the middle value for each radial bin. The bottom panel shows the mean age of each galaxy's 3$^{\rm rd}$ annulus. The legend describes the best fit line with `y' as the mean age and `x' as the [Fe/H].
\label{fig:age_met_rad_with_3ann}}
\end{figure}

As is mentioned earlier, there have not been many studies on the topic of intrinsic scatter in stellar population parameters like the age. \cite{2005MNRAS.362...41G} showed that their age and [Fe/H] estimates contain significant intrinsic scatter at all masses of galaxies. The authors also note that there is a decreasing trend in the scatter with mass for [Fe/H] and very flat trend for age. In our work, both age and [Fe/H] show a decreasing trend in $\delta_i$ with $\sigma$. The values reported by \cite{2005MNRAS.362...41G} are close to the $\delta_i$ values in age and [Fe/H] that we report in this work. \cite{10.1111/j.1365-2966.2010.17862.x} also reported a decrease in intrinsic scatter in [Fe/H] with increasing $\sigma$.

\begin{table*}[]
\centering
\caption{Table showing the intrinsic scatter ($\delta_i$) in all of the measured astrophysical parameters and their radial gradients for all the $\sigma$ bins used in this work. The unit for $\delta_i$ in astrophysical parameters is dex and that for $\delta_i$ in their radial gradients is dex/log(R/R$_e$).}
\label{table:bins_delta}
\begin{tabular}{ccccccccccccc}
\hline\hline
log $\sigma$ & $\delta_i$ in & $\delta_i$ in & $\delta_i$ in & $\delta_i$ in & $\delta_i$ in & $\delta_i$ in & $\delta_i$ in & $\delta_i$ in & $\delta_i$ in & $\delta_i$ in & $\delta_i$ in & $\delta_i$ in \\[-3pt] 
Range & age & [Fe/H] & [C/R] & [N/R] & [Na/R] & [Mg/R] & $\nabla_r$(age) & $\nabla_r$[Fe/H] & $\nabla_r$[C/R] & $\nabla_r$[N/R] & $\nabla_r$[Na/R] & $\nabla_r$[Mg/R] \\[-3pt] 
(km/s) & & & & & & & & & & & & \\[5pt] \hline
{[}1.6, 2{]}     & 0.13 & 0.17 & 0.06 & 0.15 & 0.14 & 0.09 & 0.09 & 0.15 & 0.07 & 0.15 & 0.16 & 0.09 \\ \hline
{[}2, 2.2{]}     & 0.11 & 0.13 & 0.06 & 0.11 & 0.15 & 0.07 & 0.07 & 0.18 & 0.05 & 0.08 & 0.12 & 0.05 \\ \hline
{[}2.2, 2.34{]}  & 0.11 & 0.13 & 0.06 & 0.10 & 0.13 & 0.07 & 0.09 & 0.19 & 0.06 & 0.11 & 0.16 & 0.06 \\ \hline
{[}2.34, 2.45{]} & 0.14 & 0.13 & 0.07 & 0.11 & 0.12 & 0.07 & 0.17 & 0.23 & 0.09 & 0.16 & 0.21 & 0.17 \\ \hline
{[}2.45, 2.53{]} & 0.07 & 0.12 & 0.06 & 0.09 & 0.12 & 0.06 & 0.14 & 0.25 & 0.07 & 0.14 & 0.17 & 0.08 \\ \hline
\end{tabular}
\end{table*}

We now provide brief remarks on the age-[Fe/H] relation. Fig.~\ref{fig:age_met_rad_with_3ann} shows the distribution of age with [Fe/H] for each annulus grouped by [Fe/H] (top panel) and for the 3$^{rd}$ annulus of each individual galaxy (bottom panel). The $\delta_i$ in age and [Fe/H] is sufficient to explain the errorbars in top panel of Fig.~\ref{fig:age_met_rad_with_3ann}. We see a decreasing trend of age with [Fe/H] (at least for the last three [Fe/H] bins). The trendline shows that the age of ETGs decrease by as much as 0.57 dex (equivalent to 3.7 Gyr) for an increase in [Fe/H] by 1 dex. One of the first works on characterizing age-[Fe/H] relation was done by \cite{1995ASPC...86..203W}. They showed that E/S0 galaxies follow the direction of \textit{null spectral change} characterized by $\Delta$log$Z$/$\Delta$log age = -2/3. The slope of -0.57 dex/dex that we derived in log age - [Fe/H] plane (bottom panel of Fig.~\ref{fig:age_met_rad_with_3ann}) is not too far away from the numerical value of -2/3 ($\approx$-0.67). A very similar result was reported by other studies that looked into elliptical galaxies in Fornax cluster \citep{2000MNRAS.315..184K} and Coma cluster \citep{1999MNRAS.306..607J, 2001ApJ...562..689P}. \cite{2005MNRAS.362...41G} analyzed almost 26,000 ETGs from SDSS-DR2 and showed that (their Fig. 12) the age-[Fe/H] relation have an anti-correlation for all of their mass bins. Galaxies in different environments also show the trend from old and metal-poor to young and metal-rich in Coma supercluster \citep{2020NewA...8101417T}. Although any trend that flirts with being parallel to the age-metallicity degeneracy is fraught with modeling systematics, the variety of models employed to find the same result is now significant, so perhaps we should begin to accept it as a genuine constraint on galaxy evolution codes like TNG. 

This observed trend of decreasing metallicity with increasing age may be a straightforward result of chemical evolution. Elliptical galaxy precursors are thought to form most of their stars early in the Universe’s history through rapid, intense starbursts \citep{1967ApJ...147..868P, 2006MNRAS.366..499D}. At that early stage of Universe's history, the relatively metal-poor gas formed metal poor stars. Elliptical galaxies are often the product of major mergers of smaller galaxies \citep{1972ApJ...178..623T, 2015IAUGA..2256542T}. These mergers can trigger intense star formation early in a galaxy’s evolution but can also lead to the expulsion of gas through galactic winds, particularly if supernovae and active galactic nuclei (AGN) are involved \citep{1987ApJ...319..614L, 1990NASCP3084..185F}. However, if the gas does not escape altogether, it will only become more chemically enriched with time, form stars, and create an age-metallicity relation. The question then turns toward matching the amplitude of what is observed with galaxy evolution models.

\subsection{Comparison with TNG}\label{subsec:diss_illus}

\textit{The Next Generation} Illustris (TNG) project is a series of magnetohydrodynamical cosmological simulations designed to investigate galaxy formation and evolution within the framework of the $\Lambda$ Cold Dark Matter ($\Lambda$ -CDM) cosmological model \citep{2018MNRAS.480.5113M, 2018MNRAS.477.1206N, 2018MNRAS.475..624N, 2018MNRAS.475..648P, 2018MNRAS.475..676S}. It builds and improves on the scientific insights and achievements of the original TNG simulation \citep{2014Natur.509..177V, 2014MNRAS.445..175G}. Nine individual elements -- hydrogen (H), helium (He), carbon (C), nitrogen (N), oxygen (O), neon (Ne), magnesium (Mg), silicon (Si), and iron (Fe) -- were explicitly tracked throughout the TNG simulation \citep{2018MNRAS.477.1206N}. In this work, we compare our results with those of publicly available TNG data \citep{2019ComAC...6....2N}. 
Note that we use the TNG-100 simulation results for comparison with our observations. Only a couple of studies have so far compared the results of the TNG simulations with the observations \citep{2018MNRAS.477.1206N, 2020A&A...641A..60P}. 

Selection of theoretical data began with selecting 1000 halo IDs from the `deep learning morphologies' list \citep{2019MNRAS.489.1859H} that were ETGs at a probability of 60\% or greater. For computation of half light radii, we assumed the halos had spherical symmetry and iterated to the radius at which half the particles lay inside and half outside. We adopted this radius as R$_e$ for the computation of stellar particle velocity dispersion and radially-dependent quantities. Mean ages are $V$-band weighted averages of the star particle formation ages.

We compare our results with that of TNG in two different ways: (1) grouped by $\sigma$ following the binning scheme outlined in the Table \ref{table:bins_vd} and plotted versus $R/R_e$, and (2) ungrouped versus $\sigma$ using the 3$^{rd}$ annulus for the MaNGA galaxies. We provide all the comparison plots in Appendix \ref{appen:TNG_comp}.

\textit{Mean age comparison.} Fig.~\ref{fig:ill_man_age} shows the comparison for TNG and MaNGA ages. In the left-hand panel of the plot we see that the TNG galaxies (solid lines) in all $\sigma$ bins show a flat radial trend in age similar to our calculations (stars with errorbars) of age radial gradients within each $\sigma$ bin. The right-hand panel shows the age of each individual TNG galaxy (in red cross) plotted against their $\sigma$, overlaid on our findings (in blue star) for the age at 3$^{rd}$ annulus for each MaNGA galaxy. The TNG simulations show an old age ``ceiling'' that the MaNGA observations do not mimic except for high-$\sigma$ galaxies. The scatter in TNG measurements is less across all $\sigma$ values compared to MaNGA. However, scatter decreases with increasing $\sigma$ for TNG simulation as well. 

\textit{Mean [Fe/H] comparison.} Two panels in Fig.~\ref{fig:ill_man_met} shows comparison between TNG and MaNGA results for [Fe/H]. The morphologies of the trends are remarkably similar. The spatial gradients are quantitatively similar (left-hand panel). The TNG results also show a decreasing trend in [Fe/H] for galaxies with log($\sigma$)$>$2.0 (right-hand panel). The existence of this inflection is a remarkable success for TNG. Because {\sc compfit} will not return values of [Fe/H] higher than 0.4, we cannot comment on the small number of TNG galaxies that scatter to higher [Fe/H] values. We also observe a $\sim$1\% fraction of galaxies that scatter to low abundance. These are not reflected the theoretical simulation.

\textit{Mean [X/Fe] comparison.} While considering the individual elements (Fig.~\ref{fig:ill_man_C} for C, Fig.~\ref{fig:ill_man_N} for N, and Fig.~\ref{fig:ill_man_Mg} for Mg), we see zeropoint offsets along with much less scatter in TNG simulations compared to MaNGA results. We include magnesium amplification for the first time, so our abundance results should approach actual [X/Fe] values for the first time in the literature. TNG, on the other hand, will have adjusted their assumptions of stellar element yields to follow earlier estimates. Therefore, the existence of zeropoint offsets should be no surprise. The offset is strongest for magnesium. Parenthetically, \cite{2018MNRAS.477.1206N} also reported an overabundance of magnesium for Milky Way-like galaxies in the TNG simulation.  
A hint of abundance inflection at log $(\sigma)=2$ is seen in all elements in both theory and observation.

\textit{The inflection at log $(\sigma)\approx 2$.} The fact that TNG simulations qualitatively reproduce the chemical ``phase change'' at log $(\sigma)\approx 2$ is a success of the theory. To our knowledge, no author has pointed out this particular success before. Low-$\sigma$ galaxies are effective at Fe enrichment, but high-$\sigma$ galaxies favor light metal enrichment to such a degree that [Fe/H] falls with increasing $\sigma$. By itself, this implies that high=$\sigma$ galaxies cannot simply be dry mergers of $\sim$100 km s$^{-1}$ subunits: the inherited abundance pattern does not persist. It is also hard to imagine a monolithic collapse scenario in which galaxies with shallower potential wells retain \textit{more} iron, since supernova winds should be more effective in low-$\sigma$ galaxies. Despite these considerations, TNG demonstrates that the hierarchical merger picture can thread the needle between extremes and reproduce the age-$\sigma$ trend along with the chemical inflection at log $(\sigma)\approx 2$.

\textit{Nitrogen.} The large scatter in N, especially compared to TNG, is remarkable. Because the scatter in [N/Fe] is astrophysical, it is clear that much information is being carried in this tracer element regarding the chemical histories of galaxies. Nitrogen of some quantity is produced (1) instantly in supernovae, (2) at young ages in WR stars and other massive stars, (3) at intermediate ages via CNO processed material, (4) and in old red giants by mass loss after first dredge-up. Due to the unsolved status of mass loss in stars, the balancing of yields from these sources is not predictable from first principles. However, our results here hint at the importance of CNO processed material. Since increased N comes at the expense of C in the CNO cycle, one might expect an anticorrelation between the two abundances, other things being equal. In fact, we might observe this because the correlation between [N/R] and [C/R] gradients is the weakest of any element gradient combination. In the annulus 3 plots, [N/R] correlates least tightly with [C/R] and [Na/R]. The possibility that we see the effects of CNO processing in these data is real.

\textit{ADF comparison.} We compared the Abundance Distribution Function (ADF) used in this study with six halos taken from TNG, as shown in Fig.~\ref{fig:ill_man_adf}. The analytical ADF formula was taken from \cite{2005ApJ...631..820W} and \cite{2014MNRAS.445.1538T}, and we used the `normal' width ADF to derive peak [Fe/H] from MaNGA galaxies. The `normal' ADF has FWHM width of 0.65 dex. The six TNG ADFs average to FWHM = 0.67 dex, and the overall shapes are similar. Our ADF assumption, which is based on observations in the local universe, seem consistent with TNG calculations. Therefore, our composite models should be free of bias as compared to TNG, at least along the dimension of abundance-compositeness. The overall chemical evolution assumptions in TNG give ADFs that resemble observations, except for a small parcel of relic stars at very low metallicity (Population III stars, not illustrated in Fig.~\ref{fig:ill_man_adf}) that are so far unseen in the local universe and therefore may not exist.  

\section{Conclusion} \label{sec:conclusion}

We use updated \cite{1994ApJS...95..107W} stellar population models to investigate stellar population age and abundance parameters and their radial gradients for 2968 SDSS MaNGA ETGs. The models are composite in metallicity with an ADF that resembles nearby galaxies. They also include an improvement, magnesium amplification, that allows them to approach true [X/H] and [X/R] values for the first time. Thus, these results supersede previous results by models without abundance-sensitive isochrone shifts, which, as of this writing, is all past literature. That said, the improved results are not dramatically different, albeit with interesting new behavior in N and Na. The models are inverted to infer light-weighted mean age, ADF peak [Fe/H], and mean elemental abundances for C, N, Na, and Mg for each elliptical annulus within each of the 2968 galaxies. We investigate primarily trends with $\sigma$ and [Fe/H] in this paper. 

We find positive correlation between the age and size of the galaxies implying more massive ETGs have older age. But for [Fe/H], this trend reverses such that galaxies starting at log$\sigma\approx 2$ trend toward lower [Fe/H], while galaxies at all $\sigma$ trend toward higher C, N, Na, and Mg with increased $\sigma$. The light element abundances strengthen faster than [Fe/H] falls, so the most massive galaxies are enriched in heavy elements ($Z$) overall. These two findings can be merged together to conclude that older galaxies have lower [Fe/H] but higher light metals and $Z$. The light metals show a negative correlation with [Fe/H], presumably implying a spread of Type \Romannum{1}a/Type \Romannum{2} supernova yield contributions.

Among 7 annuli within each galaxy, a nearly flat average radial gradient in age is accompanied by large individual galaxy scatter. That is, nearly as many galaxies show a younger center as show a younger portion at 0.65 R$_e$. Gradients in abundances are negative but nearly flat for low-$\sigma$ galaxies. For high-$\sigma$ galaxies, the gradients are more strongly negative, with the most emphatic gradients in Fe and Na. 

In terms of galaxy formation and evolution, our findings are more consistent with formation of ETGs via \textit{merging} (citations in $\S$\ref{sec:intro}) rather than strong \textit{monolithic collapse} (citations in $\S$\ref{sec:intro}). \cite{1993MNRAS.262..650D} proposes a scenario where smaller galactic subunits are formed by dissipational collapse which underwent dissipationless major mergers in subsequent times. This theory is supported in other works on stellar population parameter gradients \citep{2004MNRAS.347..740K, 2017MNRAS.466.4731G}. In our work, the strong dependence of [Fe/H] radial gradient values on $\sigma$ poses a problem to such theory. A merging dominant galactic evolution pathway is more apt in explaining our results. The radial gradients in C, N, and Mg also correspond well with the negative radial gradient in [Fe/H] pointing to the fact that overall element enrichment ($Z$) varies with galactocentric radius, with metal-rich centers and relatively metal-poor outer parts \citep{2003A&A...407..423M}. While some cosmological simulations of galaxy formation support a two-phase formation model \citep{2010ApJ...725.2312O}, our findings suggest that merger events were the primary drivers of galaxy evolution, with a minor contribution from dissipational collapse.

We also discussed the observed trend of having more metal poor population in older elliptical galaxies and more metal rich stars in the younger ellipticals. This result agrees well with a number of previous studies and points to continued chemical enrichment in a substantial number of ETGs across much of cosmic time.  

The astrophysical scatter ($\delta_i$) calculated in this work is an inference only rarely exploited in the literature. We found a decreasing trend for $\delta_i$ with $\sigma$ in both age and [Fe/H]. The $\delta_i$ in those two quantities are sufficient to explain the scatter in the radial gradients calculated for them. The $\delta_i$ in elemental abundances and their gradients show in general lower $\delta_i$ values compared to the age and [Fe/H]. The only exception to this rule is Na. For the future, it is worth investigating the standalone behavior of Na. Also for future work, we intend to study the effect of environment on the intrinsic scatter of astrophysical parameters.

Comparing with TNG simulations, we find substantial zero point and scale shifts in the ages, though we note that our population models do not yet include blue straggler stars. The TNG simulation result of [Fe/H] decreasing with $\sigma$ (for log($\sigma$)$>$2.0) matches our results. Also, the light elements (C, N, and Mg) show an abundance slope inflection for log($\sigma$)$>$2.0 both in TNG and our work. This potentially points to better and more efficient chemical enrichment in light elements in the subunits that eventually merge to form the largest ETGs. 

\section*{Acknowledgements}

Funding for the Sloan Digital Sky Survey IV has been provided by the Alfred P. Sloan Foundation, the U.S. Department of Energy Office of Science, and the Participating Institutions. 

SDSS-IV acknowledges support and resources from the Center for High Performance Computing  at the University of Utah. The SDSS website is www.sdss4.org.

SDSS-IV is managed by the Astrophysical Research Consortium 
for the Participating Institutions of the SDSS Collaboration including the Brazilian Participation Group, the Carnegie Institution for Science, Carnegie Mellon University, Center for Astrophysics | Harvard \& Smithsonian, the Chilean Participation Group, the French Participation Group, Instituto de Astrof\'isica de Canarias, The Johns Hopkins 
University, Kavli Institute for the Physics and Mathematics of the Universe (IPMU) / University of Tokyo, the Korean Participation Group, Lawrence Berkeley National Laboratory, 
Leibniz Institut f\"ur Astrophysik Potsdam (AIP),  Max-Planck-Institut f\"ur Astronomie (MPIA Heidelberg), Max-Planck-Institut f\"ur Astrophysik (MPA Garching), Max-Planck-Institut f\"ur Extraterrestrische Physik (MPE), National Astronomical Observatories of China, New Mexico State University, New York University, University of Notre Dame, Observat\'ario Nacional / MCTI, The Ohio State University, Pennsylvania State University, Shanghai Astronomical Observatory, United Kingdom Participation Group, Universidad Nacional Aut\'onoma de M\'exico, University of Arizona, University of Colorado Boulder, University of Oxford, University of Portsmouth, University of Utah, University of Virginia, University of Washington, University of Wisconsin, Vanderbilt University, and Yale University.

\twocolumngrid
\appendix

\section{Plots that compare with TNG}\label{appen:TNG_comp}

This section of the appendix contains figures \ref{fig:ill_man_age}, \ref{fig:ill_man_met}, \ref{fig:ill_man_C}, \ref{fig:ill_man_N}, \ref{fig:ill_man_Mg}, and \ref{fig:ill_man_adf} elucidating comparison between our results and TNG results for age, [Fe/H], [C/R], [N/R], [Mg/R], and ADF respectively.

\begin{figure}[h!]
\plottwo{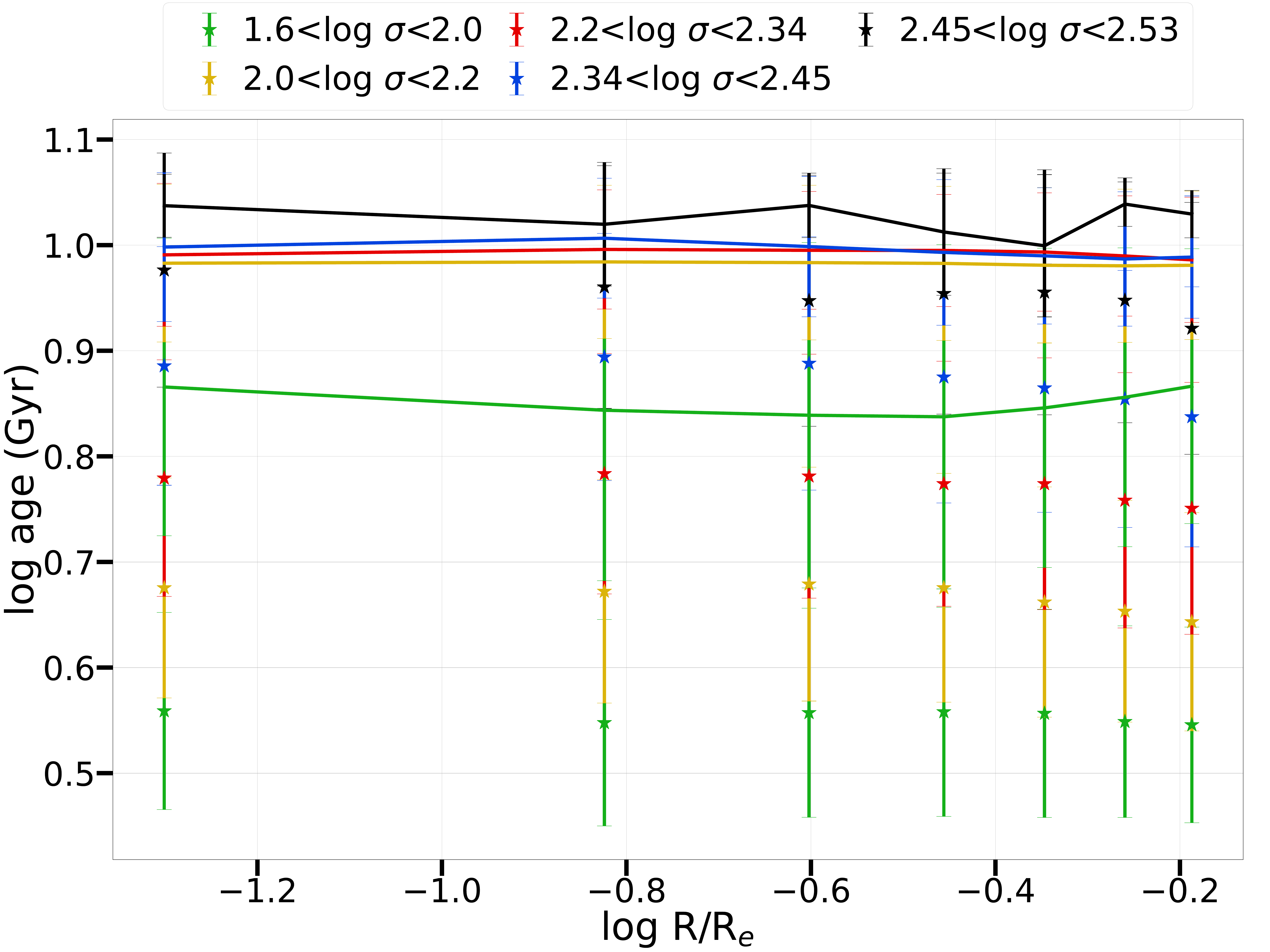}{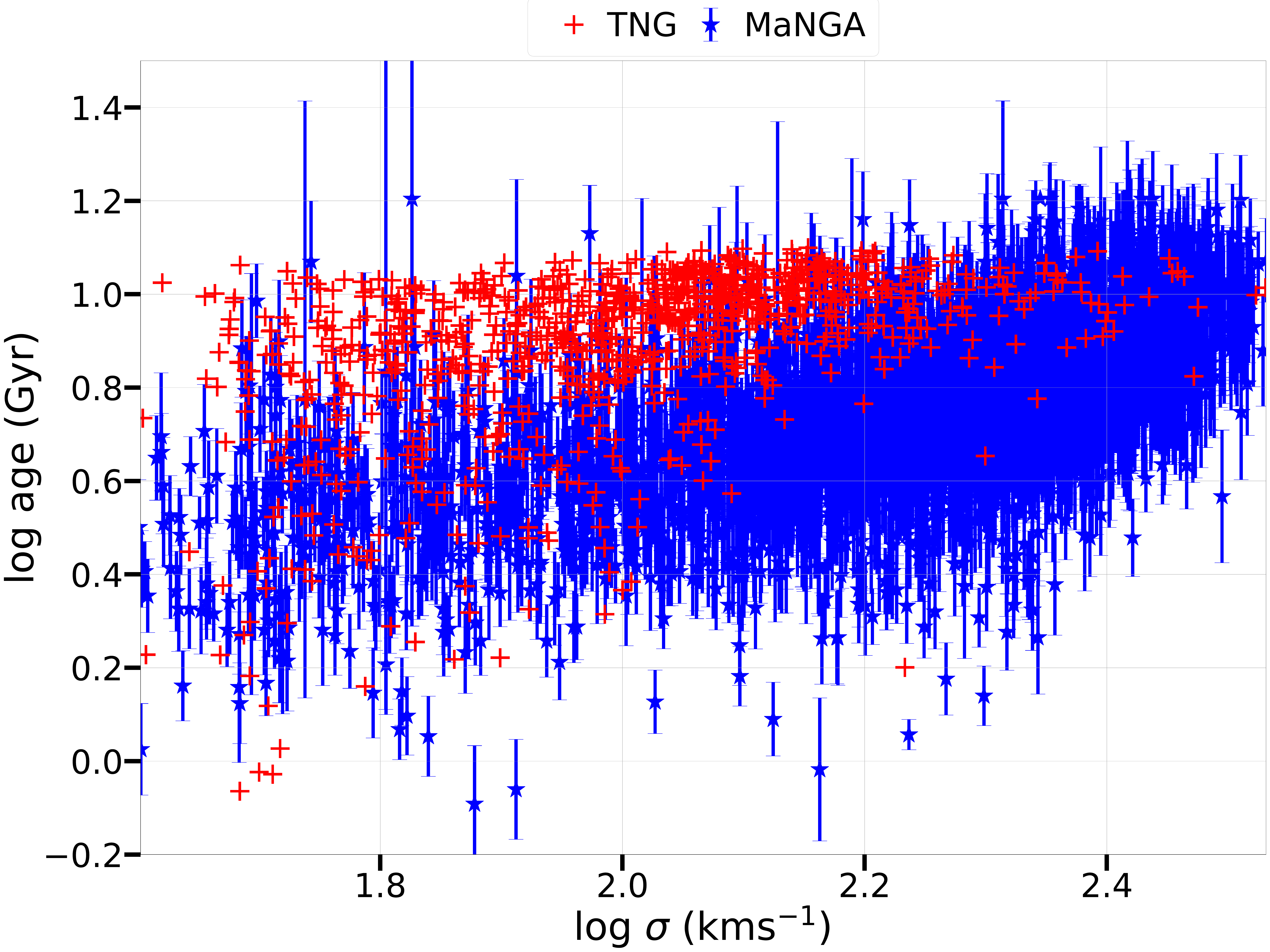}
\caption{The left panel shows how the mean age for different $\sigma$ bins are changing with distance from the center of the galaxy. The solid lines show results from the TNG and the stars with errorbars show results from our work. Both the lines and symbols use same color codes for different $\sigma$ bins. The right panel shows how the age of each individual TNG galaxy (in red plus) and age from 3$^{rd}$ annulus of each MaNGA galaxy (in blue star) are changing with $\sigma$. MaNGA galaxies have errors associated with each data point.
\label{fig:ill_man_age}}
\end{figure}

\begin{figure}[h!]
\plottwo{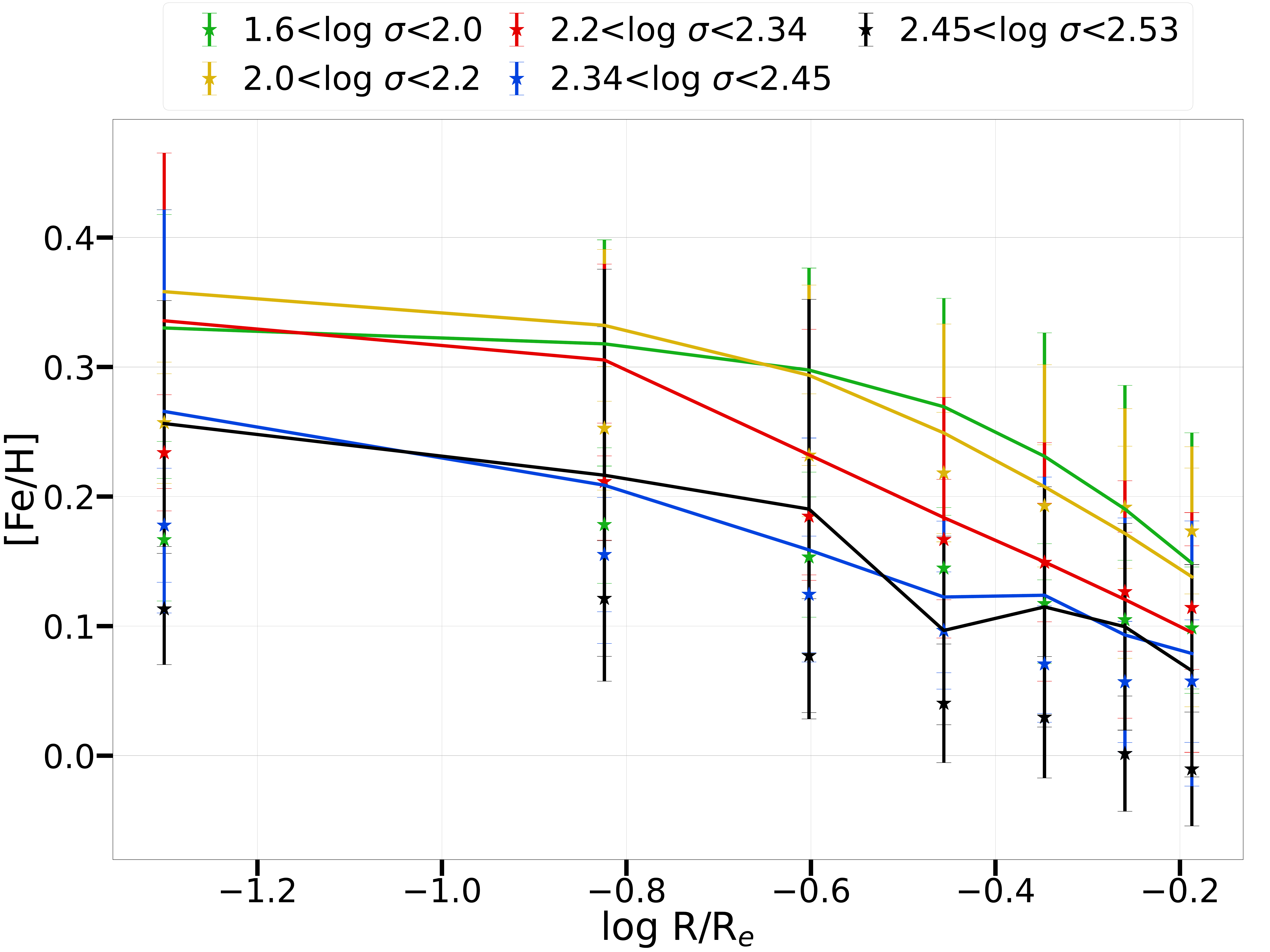}{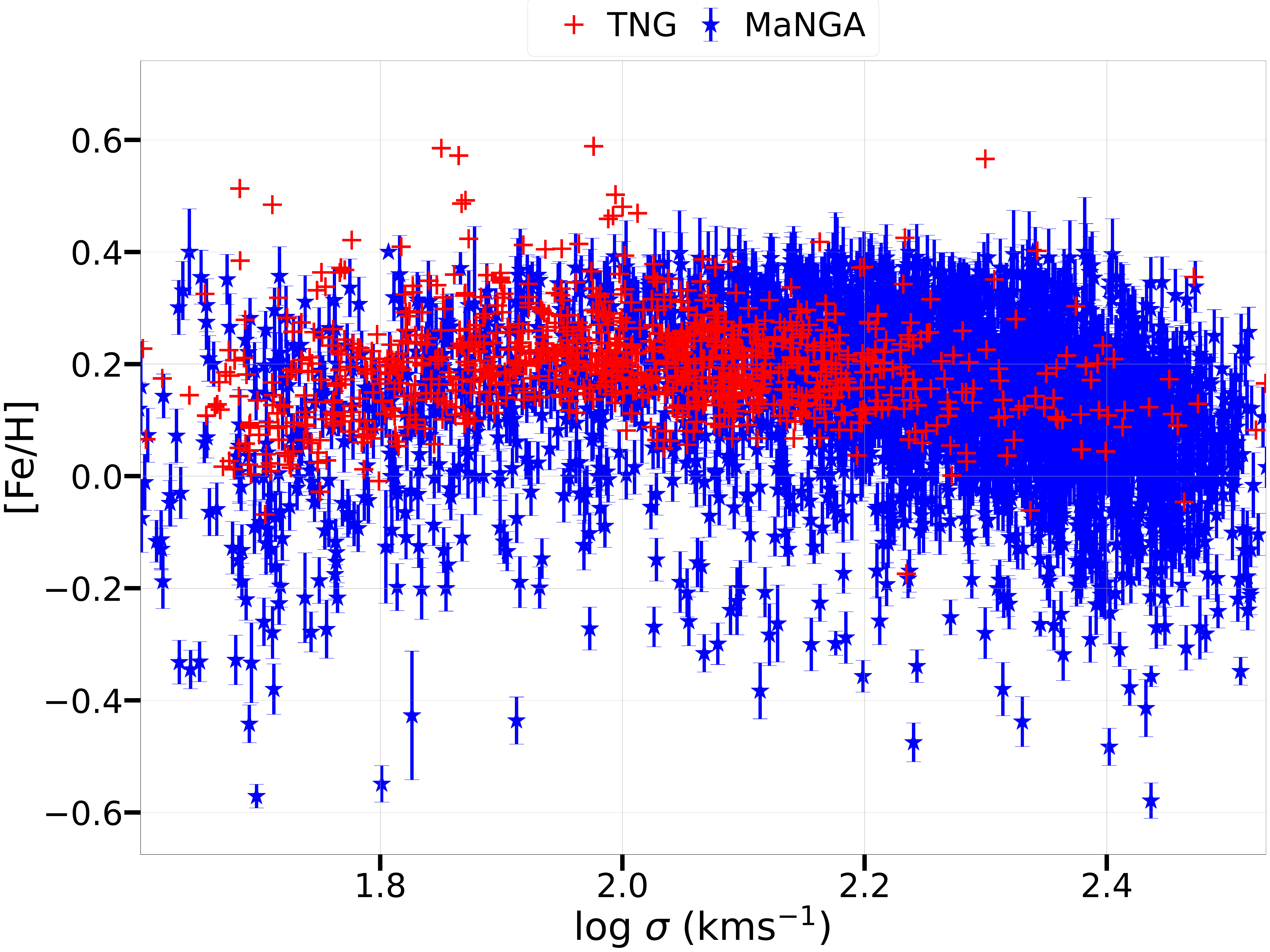}
\caption{The left panel shows how the mean [Fe/H] for different $\sigma$ bins are changing with distance from the center of the galaxy. The solid lines show results from the TNG and the stars with errorbars show results from our work. Both the lines and symbols use same color codes for different $\sigma$ bins. The right panel shows how the [Fe/H] of each individual TNG galaxy (in red plus) and age from 3$^{rd}$ annulus of each MaNGA galaxy (in blue star) are changing with $\sigma$. MaNGA galaxies have errors associated with each data point.
\label{fig:ill_man_met}}
\end{figure}

\begin{figure}[h!]
\plottwo{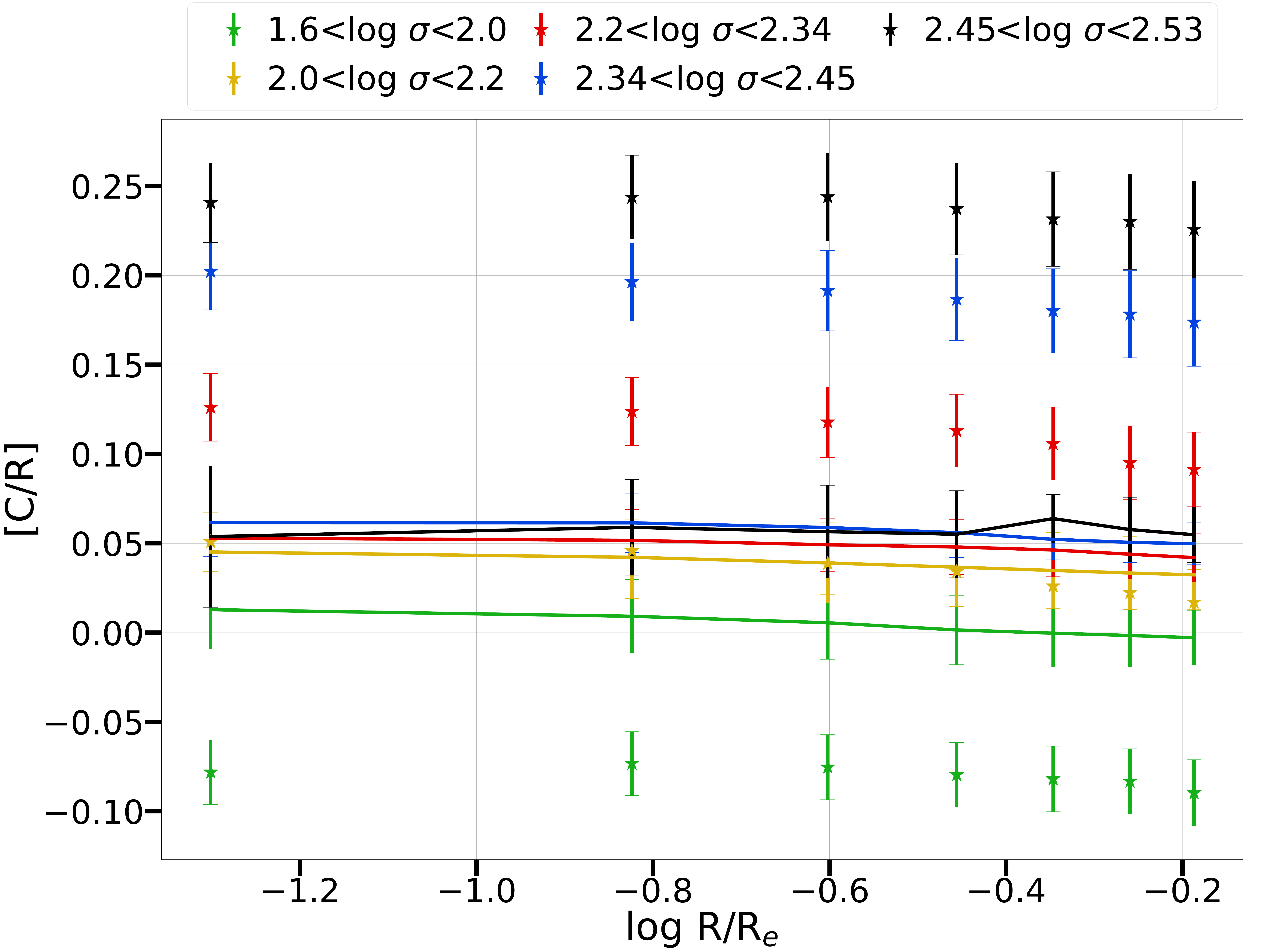}{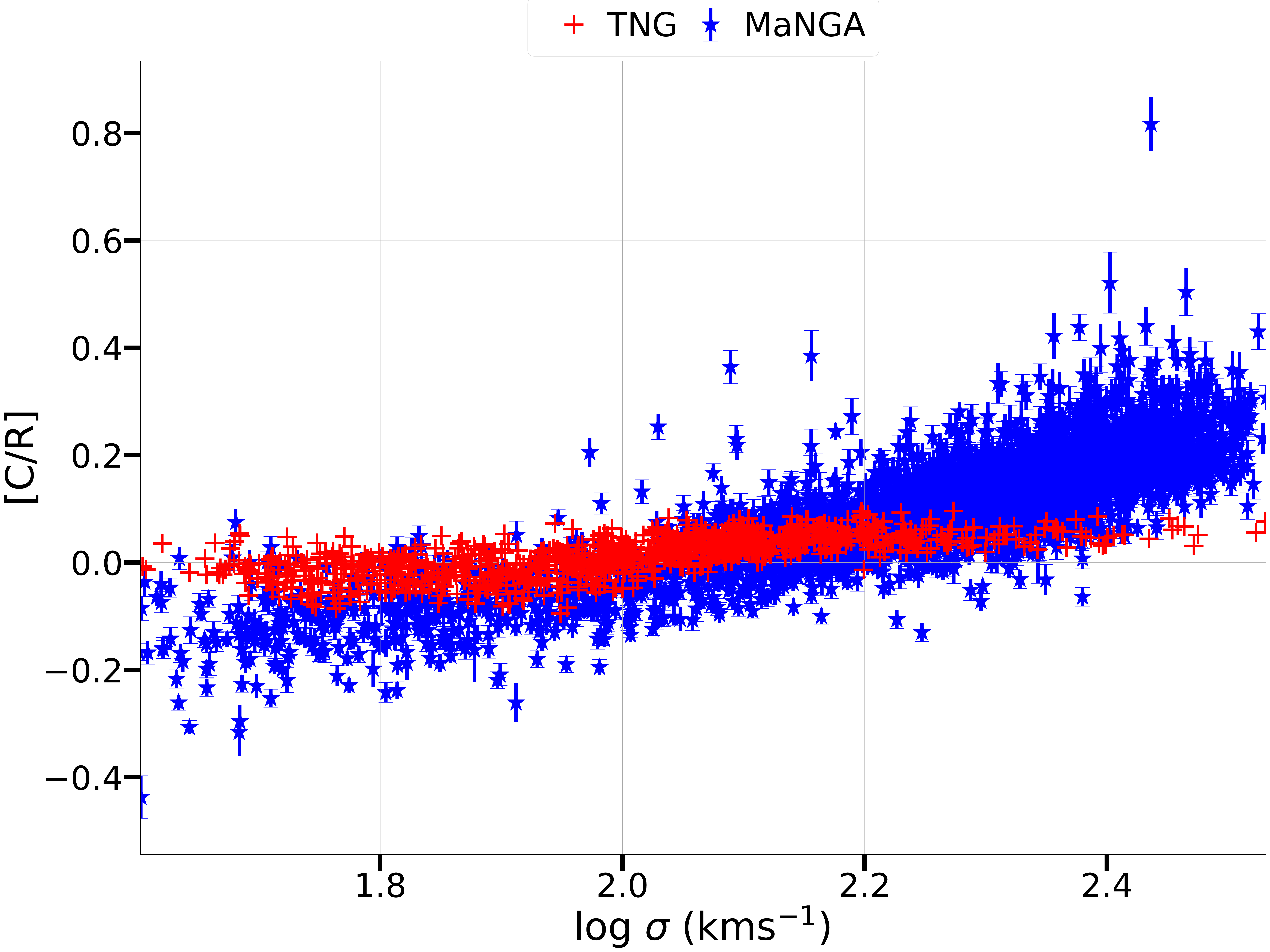}
\caption{The left panel shows how the mean carbon abundance for different $\sigma$ bins are changing with distance from the center of the galaxy. The solid lines show results from the TNG and the stars with errorbars show results from our work. Both the lines and symbols use same color codes for different $\sigma$ bins. The right panel shows how the [C/R] of each individual TNG galaxy (in red plus) and age from 3$^{rd}$ annulus of each MaNGA galaxy (in blue star) are changing with $\sigma$. MaNGA galaxies have errors associated with each data point.
\label{fig:ill_man_C}}
\end{figure}

\begin{figure}[h!]
\plottwo{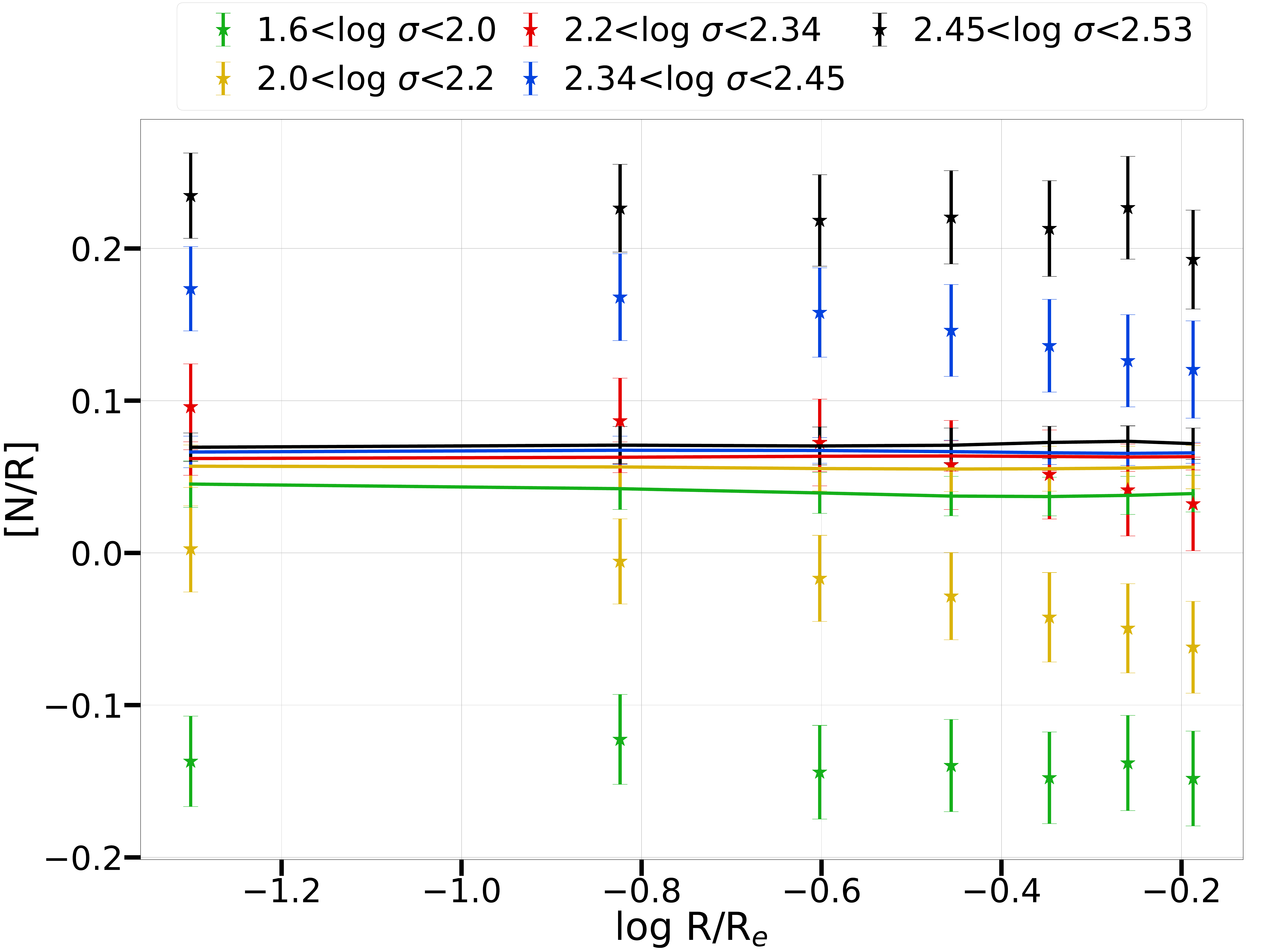}{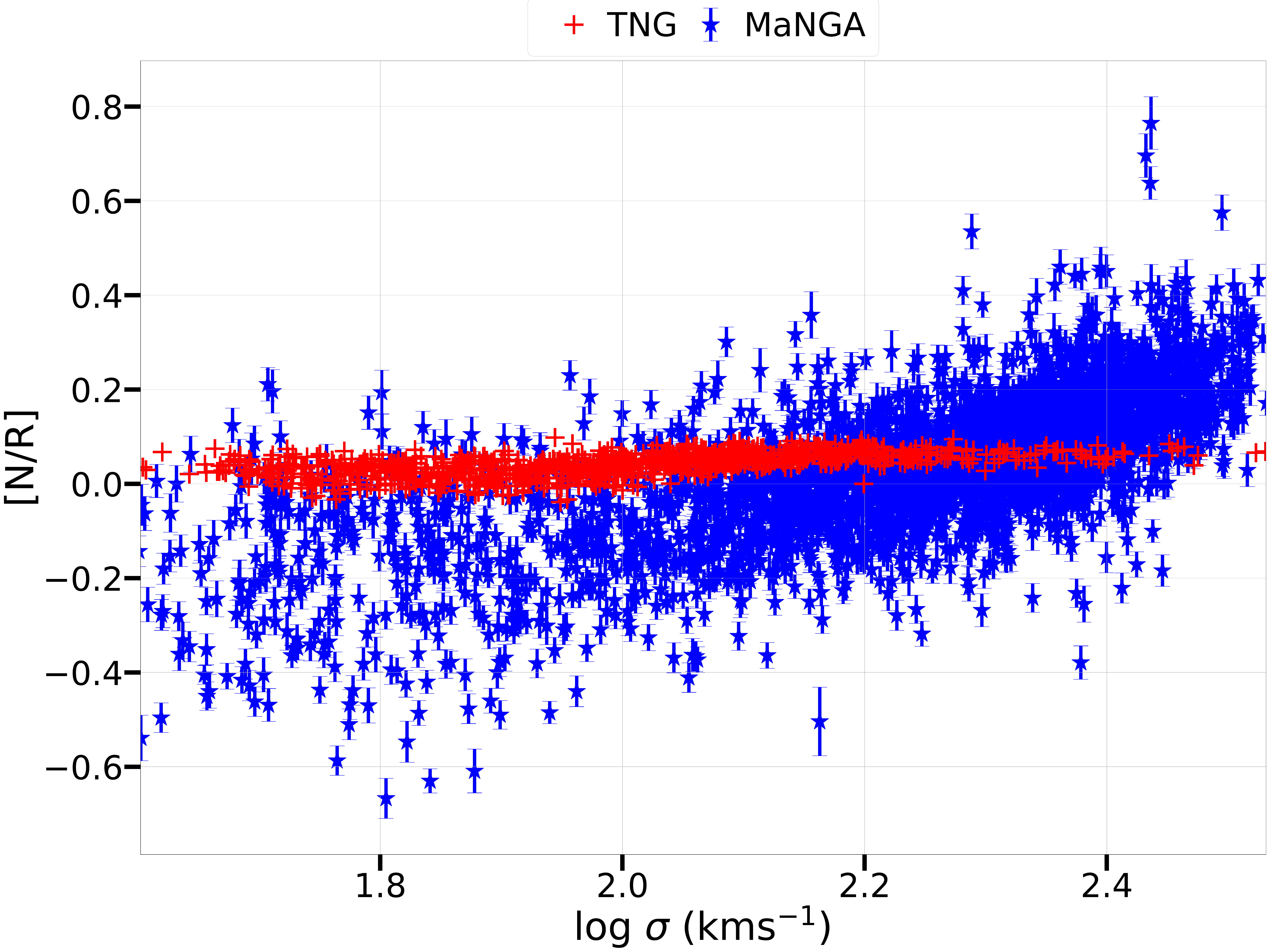}
\caption{The left panel shows how the mean nitrogen abundance for different $\sigma$ bins are changing with distance from the center of the galaxy. The solid lines show results from the TNG and the stars with errorbars show results from our work. Both the lines and symbols use same color codes for different $\sigma$ bins. The right panel shows how the [N/R] of each individual TNG galaxy (in red plus) and age from 3$^{rd}$ annulus of each MaNGA galaxy (in blue star) are changing with $\sigma$. MaNGA galaxies have errors associated with each data point.
\label{fig:ill_man_N}}
\end{figure}

\begin{figure}[h!]
\plottwo{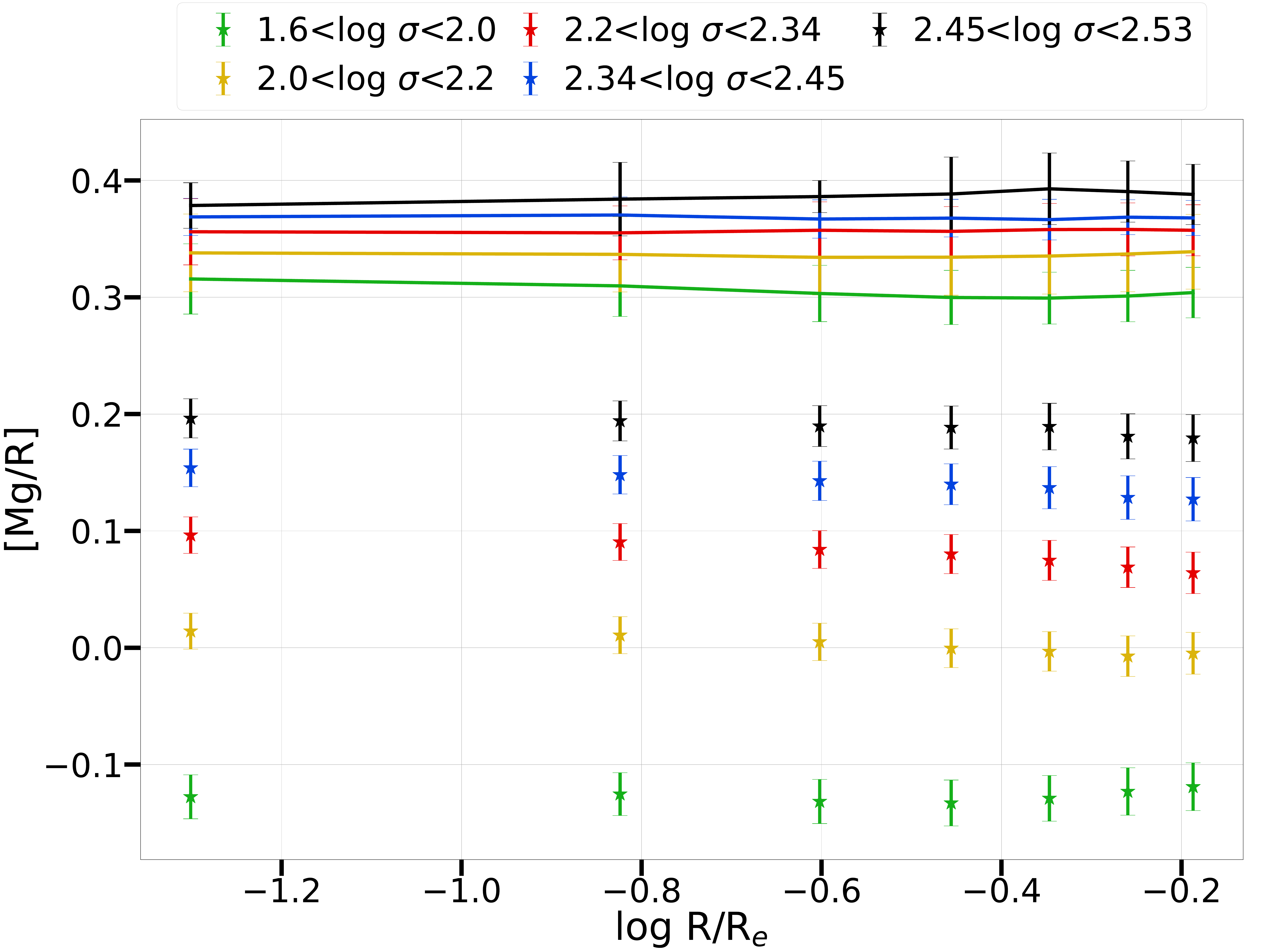}{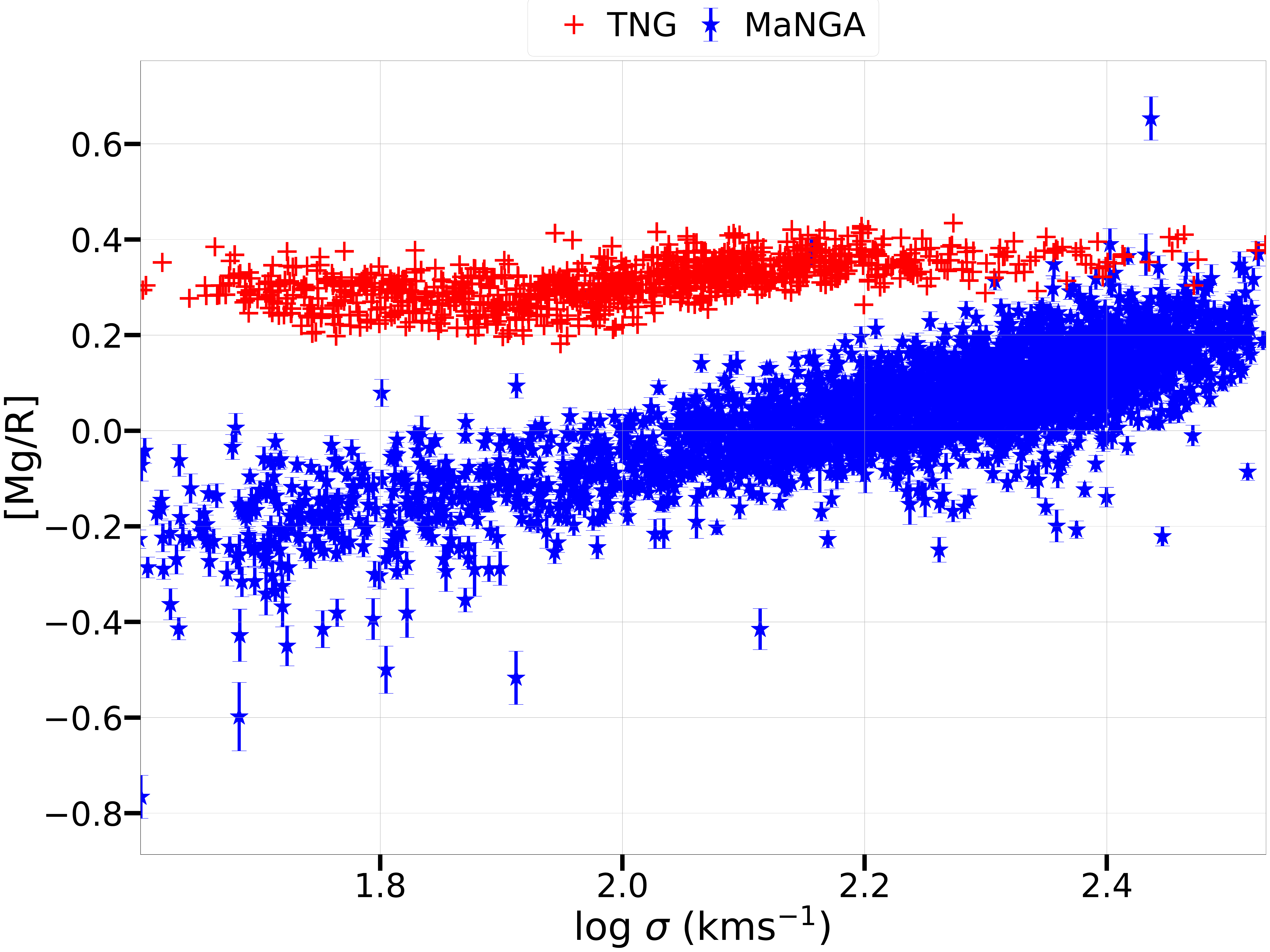}
\caption{The left panel shows how the mean magnesium abundance for different $\sigma$ bins are changing with distance from the center of the galaxy. The solid lines show results from the TNG and the stars with errorbars show results from our work. Both the lines and symbols use same color codes for different $\sigma$ bins. The right panel shows how the [Mg/R] of each individual TNG galaxy (in red plus) and age from 3$^{rd}$ annulus of each MaNGA galaxy (in blue star) are changing with $\sigma$. MaNGA galaxies have errors associated with each data point.
\label{fig:ill_man_Mg}}
\end{figure}

\begin{figure}[h!]
\plotone{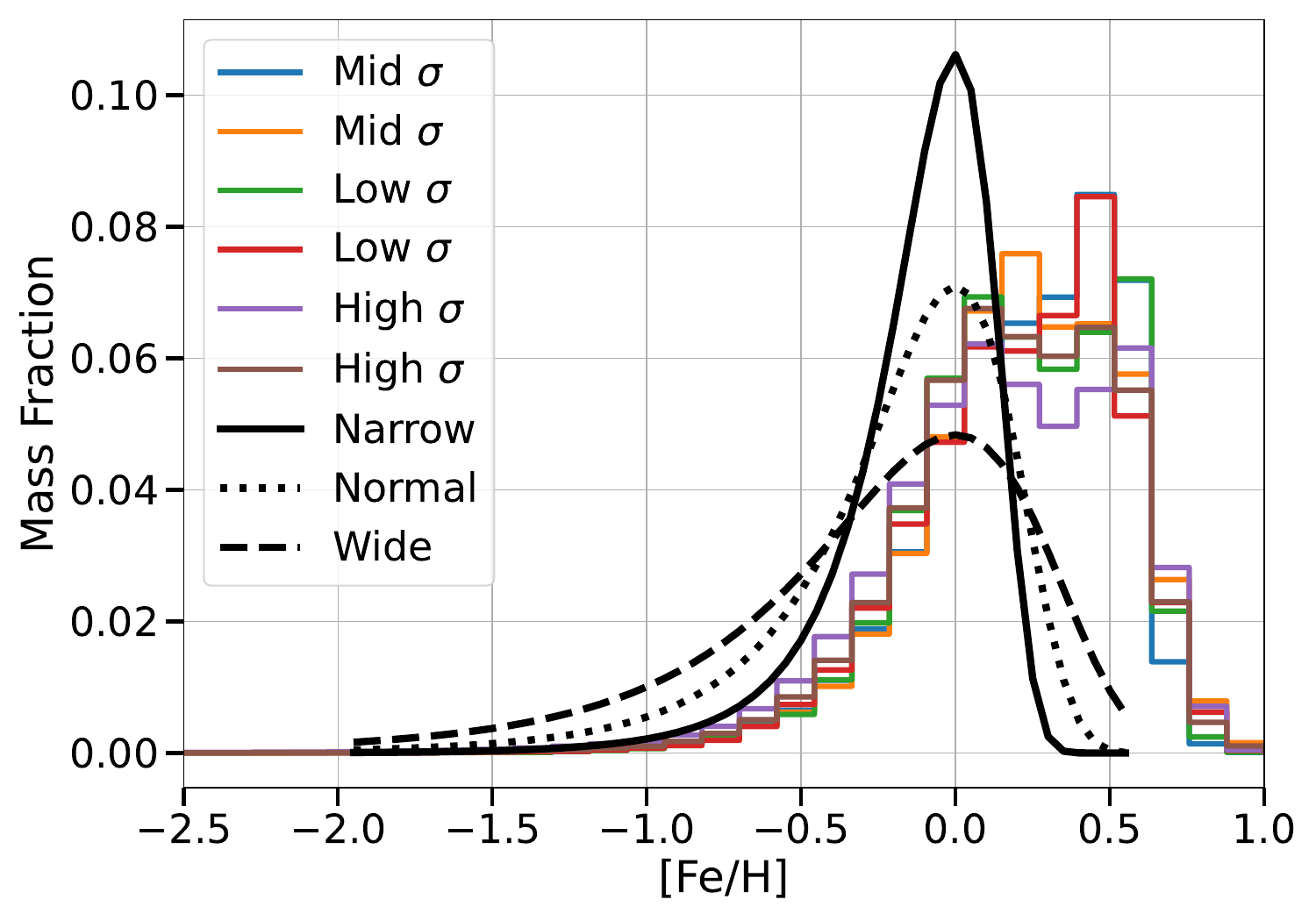}
\caption{The Abundance Distribution Function (ADF) for individual TNG galaxies (in colored solid lines) plotted along with three ADFs from an analytic function that is used in the Worthey CSPs. In this work, the `Normal' ADF (dotted black) is used. {\sc compfit} centers the ADF peak anew for each galaxy annulus, but for illustration purposes the model ADF peaks are drawn centered on zero.
\label{fig:ill_man_adf}}
\end{figure}

\clearpage

\section{Spectral Feature Table}\label{appen:feature_table}

This section of the appendix contains a part of the table showing the details of the spectral features used in conjunction with the \cite{1994ApJS...95..107W} model to calculate age, [Fe/H], and elemental abundances. The full table is available upon request to the authors.

\begin{table*}[h!]
\centering
\caption{List of all the spectral features used in the \textsc{compfit} program (full table is available online at [link to be inserted]). The acronyms are as follows: IB=Index Band, BC=Blue Continuum, RC=Red Continuum. The ``Start" and ``End" phrases are used to denote the start and end wavelengths of the passbands in \AA. The ``Unit" column specifies if the index used is an atomic feature (denoted by 0; units: \AA) or a molecular feature (denoted by 1; units: mag).}
\begin{tabular}{cccccccc}
\hline\hline
Feature Name & IB Start & IB End & BC Start & BC End & RC Start & RC End & Unit\\ \hline
CN$_1$	&	4142.125	&	4177.125	&	4080.125	&	4117.625	&	4284.125	&	4244.125	&	1	\\ \hline
CN$_2$	&	4142.125	&	4177.125	&	4083.875	&	4096.375	&	4284.125	&	4244.125	&	1	\\ \hline
Ca4227	&	4222.25	&	4234.75	&	4211	&	4219.75	&	4251	&	4241	&	0	\\ \hline
\end{tabular}
\end{table*}

\section{Monte Carlo (MC) Results}\label{appen:MC_results}

This section of the appendix shows a part of the table with the results from our MC simulation. The full table is available upon request to the authors.

\begin{splitdeluxetable*}{ccccccccccccccBcccccccccccccc}
\tabletypesize{\scriptsize}
\tablecaption{Table showing the values of our derived astrophysical parameters, their radial gradients, and corresponding errors. This is a part of the table and the full table is available upon request to the authors. }
\tablehead{
\colhead{Name}	&	\colhead{R/R$_e$}	&	\colhead{$\sigma_{cen}$}	&	\colhead{$\sigma_{ann}$}	&	\colhead{log age}	&	\colhead{Error in}	&	\colhead{$\nabla_r$log age}	&	\colhead{Error in}	&	\colhead{[Fe/H]}	&	\colhead{Error in}	&	\colhead{$\nabla_r$[Fe/H]}	&	\colhead{Error in}	&	\colhead{[C/R]}	&	\colhead{Error in}	&	\colhead{$\nabla_r$[C/R]}	&	\colhead{Error in}	&	\colhead{[N/R]}	&	\colhead{Error in}	&	\colhead{$\nabla_r$[N/R]}	&	\colhead{Error in}	&	\colhead{[Na/R]}	&	\colhead{Error in}	&	\colhead{$\nabla_r$[Na/R]}	&	\colhead{Error in}	&	\colhead{[Mg/R]}	&	\colhead{Error in}	&	\colhead{$\nabla_r$[Mg/R]}	&	\colhead{Error in} \\[-9pt]
\colhead{}	&	\colhead{}	&	\colhead{$(kms^{-1})$}	&	\colhead{$(kms^{-1})$}	&	\colhead{}	&	\colhead{log age}	&	\colhead{}	&	\colhead{$\nabla_r$log age}	&	\colhead{}	&	\colhead{[Fe/H]}	&	\colhead{}	&	\colhead{$\nabla_r$[Fe/H]}	&	\colhead{}	&	\colhead{[C/R]}	&	\colhead{}	&	\colhead{$\nabla_r$[C/R]}	&	\colhead{}	&	\colhead{[N/R]}	&	\colhead{}	&	\colhead{$\nabla_r$[N/R]}	&	\colhead{}	&	\colhead{[Na/R]}	&	\colhead{}	&	\colhead{$\nabla_r$[Na/R]}	&	\colhead{}	&	\colhead{[Mg/R]}	&	\colhead{}	&	\colhead{$\nabla_r$[Mg/R]}
} 
\startdata 
11946-1901\_ann\_0 & 0.050 & 183.460 & 217.250 & 0.850 & 0.120 & -0.080 & 0.180 & 0.250 & 0.040 & 0.050 & 0.040 & 0.200 & 0.020 & -0.060 & 0.020 & 0.040 & 0.020 & 0.050 & 0.030 & 0.200 & 0.030 & -0.110 & 0.030 & 0.100 & 0.010 & 0.020 & 0.020 \\
11946-1901\_ann\_1 & 0.150 & 183.460 & 201.060 & 0.780 & 0.080 & -0.080 & 0.180 & 0.300 & 0.030 & 0.050 & 0.040 & 0.220 & 0.020 & -0.060 & 0.020 & 0.030 & 0.020 & 0.050 & 0.030 & 0.180 & 0.030 & -0.110 & 0.030 & 0.100 & 0.010 & 0.020 & 0.020 \\
11946-1901\_ann\_2 & 0.250 & 183.460 & 193.940 & 0.720 & 0.100 & -0.080 & 0.180 & 0.340 & 0.040 & 0.050 & 0.040 & 0.160 & 0.020 & -0.060 & 0.020 & 0.100 & 0.030 & 0.050 & 0.030 & 0.170 & 0.040 & -0.110 & 0.030 & 0.120 & 0.010 & 0.020 & 0.020 \\
11946-1901\_ann\_3 & 0.350 & 183.460 & 189.400 & 0.700 & 0.090 & -0.080 & 0.180 & 0.350 & 0.030 & 0.050 & 0.040 & 0.200 & 0.020 & -0.060 & 0.020 & 0.040 & 0.020 & 0.050 & 0.030 & 0.150 & 0.030 & -0.110 & 0.030 & 0.100 & 0.010 & 0.020 & 0.020 \\
11946-1901\_ann\_4 & 0.450 & 183.460 & 186.070 & 0.720 & 0.100 & -0.080 & 0.180 & 0.330 & 0.040 & 0.050 & 0.040 & 0.140 & 0.020 & -0.060 & 0.020 & 0.090 & 0.020 & 0.050 & 0.030 & 0.120 & 0.030 & -0.110 & 0.030 & 0.110 & 0.010 & 0.020 & 0.020 \\
11946-1901\_ann\_5 & 0.550 & 183.460 & 183.460 & 0.720 & 0.120 & -0.080 & 0.180 & 0.320 & 0.050 & 0.050 & 0.040 & 0.150 & 0.020 & -0.060 & 0.020 & 0.070 & 0.020 & 0.050 & 0.030 & 0.070 & 0.040 & -0.110 & 0.030 & 0.120 & 0.020 & 0.020 & 0.020 \\
11946-1901\_ann\_6 & 0.650 & 183.460 & 181.310 & 0.760 & 0.120 & -0.080 & 0.180 & 0.270 & 0.050 & 0.050 & 0.040 & 0.130 & 0.020 & -0.060 & 0.020 & 0.100 & 0.030 & 0.050 & 0.030 & 0.080 & 0.030 & -0.110 & 0.030 & 0.140 & 0.020 & 0.020 & 0.020 \\
8336-3703\_ann\_0 & 0.050 & 154.490 & 182.950 & 0.780 & 0.100 & 0.090 & 0.110 & 0.120 & 0.040 & -0.130 & 0.040 & 0.080 & 0.020 & -0.020 & 0.020 & 0.140 & 0.030 & -0.070 & 0.030 & 0.080 & 0.030 & -0.010 & 0.050 & 0.090 & 0.020 & -0.030 & 0.020 \\
8336-3703\_ann\_1 & 0.150 & 154.490 & 169.310 & 0.790 & 0.130 & 0.090 & 0.110 & 0.100 & 0.050 & -0.130 & 0.040 & 0.080 & 0.020 & -0.020 & 0.020 & 0.130 & 0.030 & -0.070 & 0.030 & 0.080 & 0.040 & -0.010 & 0.050 & 0.100 & 0.010 & -0.030 & 0.020 \\
8336-3703\_ann\_2 & 0.250 & 154.490 & 163.320 & 0.810 & 0.120 & 0.090 & 0.110 & 0.090 & 0.040 & -0.130 & 0.040 & 0.080 & 0.020 & -0.020 & 0.020 & 0.110 & 0.030 & -0.070 & 0.030 & 0.080 & 0.030 & -0.010 & 0.050 & 0.090 & 0.010 & -0.030 & 0.020 \\
8336-3703\_ann\_3 & 0.350 & 154.490 & 159.500 & 0.810 & 0.120 & 0.090 & 0.110 & 0.080 & 0.050 & -0.130 & 0.040 & 0.080 & 0.020 & -0.020 & 0.020 & 0.100 & 0.030 & -0.070 & 0.030 & 0.070 & 0.040 & -0.010 & 0.050 & 0.080 & 0.020 & -0.030 & 0.020 \\
8336-3703\_ann\_4 & 0.450 & 154.490 & 156.700 & 0.850 & 0.150 & 0.090 & 0.110 & 0.030 & 0.050 & -0.130 & 0.040 & 0.060 & 0.020 & -0.020 & 0.020 & 0.090 & 0.030 & -0.070 & 0.030 & 0.070 & 0.030 & -0.010 & 0.050 & 0.070 & 0.020 & -0.030 & 0.020 \\
\enddata
\tablecomments{$\sigma_{cen}$ is the $\sigma$ at the position of the 5$^{th}$ annulus from the center of the galaxy and $\sigma_{ann}$ is the individual $\sigma$ for each annulus.}
\end{splitdeluxetable*}

\newpage

\bibliography{ref}{}
\bibliographystyle{aasjournal}



\end{document}